\documentclass[12pt]{article}
\usepackage{color,rotate,graphics,epsf,psfrag,amssymb}

\newcommand{\insertfig}[2]{\mbox{\epsfxsize=#1cm \epsfbox{#2.eps}}}

\newcommand{\Bx}{x_{\rm B}}
\newcommand{\GeV}{{\rm GeV}}
\font\cmss=cmss12 
\def\1{\hbox{{1}\kern-.25em\hbox{l}}}
\def\bfZ{\relax{\hbox{\cmss Z\kern-.4em Z}}}

\begin{document}

\begin{titlepage}

\centerline{\large \bf Complex conformal spin partial wave
expansion  of} \centerline{\large \bf generalized parton
distributions and distribution amplitudes}

\vspace{15mm}

\centerline{\bf  D.~M\"uller$^{a,b}$ and A. Sch\"afer$^a$ }

\vspace{10mm}

\centerline{\it $^a$ Institut f\"ur Theoretische Physik, Universit\"at
Regensburg} \centerline{\it D-93040 Regensburg, Germany}

\vspace{10mm}

\centerline{\it $^b$Department of Physics and Astronomy, Arizona State University}
\centerline{\it Tempe, AZ 85287-1504, USA}

\vspace{3mm}

\vspace{5mm}

\centerline{\bf Abstract}

\vspace{0.5cm}

\noindent We introduce a new representation of generalized parton
distributions and generalized distribution amplitudes that is
based on the partial wave decomposition with respect to the {\em
complex} collinear conformal spin. This decomposition leads us to
a versatile parameterization of these non-perturbative functions
in terms of conformal moments, which are measurable for integer
value on the lattice. This new representation has several
advantages: basic properties and crossing relations are
automatically implemented, a rather flexible parameterization is
possible, the numerical treatment  of evolution is simple  and
analytic approximation of scattering amplitudes can be given. We
demonstrate this for simple examples. In particular,
phenomenological considerations indicate that the $t$-dependence
of Mellin moments is governed by  Regge trajectories.  The new
representation is vital to push the analysis of deeply virtual
Compton scattering to next-to-next-to-leading order.

\end{titlepage}

\newpage

\pagestyle{empty} {\small\tableofcontents}

\newpage

\pagestyle{plain}  \setcounter{page} 1

\section{Introduction}

Generalized parton distributions (GPDs)
\cite{GeyDitHorMueRob88,MueRobGeyDitHor94,Ji96,Rad96,ColFraStr96}
and their analog, obtained by crossing, generalized distribution
amplitudes (GDAs) \cite{MueRobGeyDitHor94,DieGouPirTer98} are
non-perturbative functions that are accessible in certain hard
exclusive  processes such as the hard electroproduction of photons
and mesons off a nucleon or nucleus or hadron pair production by
two photon fusion in $e^+ e^- $ colliders. These functions are
related to parton densities, form factors, and distribution
amplitudes but contain additional non-perturbative information
about the internal structure of hadrons and nuclei. Some of this
information can even not be obtained in any other way than through
a global fit of GPDs. This has been widely realized for the first
time in connection with the proton spin puzzle. Here the second
moment of a certain combination of GPDs provides the orbital
angular momentum fraction carried by  quarks of a given flavour
\cite{Ji96}. Moreover, GPDs simultaneously possess a longitudinal
and transverse momentum dependence and so it has been pointed out
that they encode a three dimensional femto-holographic picture of
the probed hadron or nucleus \cite{RalPir01}. Indeed, it could be
shown that a partonic density interpretation holds in the infinite
momentum frame as long as the longitudinal momentum fraction in
the $t$-channel is vanishing \cite{Bur00,Bur02}, see also Refs.\
\cite{Die02,BelMue02}. More precisely, in the impact parameter
space GPDs are interpreted as the probability to find a parton
species $i$ with  momentum fraction $x$ at a relative distance
$\mbox{\boldmath{$b$}}_\perp$ from the proton center. Even an
interpretation of the three dimensional Fourier transform of GPDs
in the rest frame has been suggested within the concept of phase
space (Wigner) distributions \cite{Bel03,Ji03}. For further
details we refer to the comprehensive reviews in Ref.\
\cite{Die03a,BelRad05}.

At present generalized parton distributions are one of the  main
topics of collider and fixed target experiments at DESY and JLAB.
Further experiments are planed or proposed for COMPASS  and ERIC.
Also it should be mentioned that information on generalized
distribution amplitudes has, e.g., been extracted from  LEP data.
Unfortunately, the wealth on information encoded in GPDs and GDAs
goes along with their functional complexity. For instance, GPDs
depend on both the momentum fractions in the $s-$ and $t-$channel,
$x$ and $\eta$, the momentum transfer squared $t$, the resolution
scale $\cal Q$, and the quantum numbers of the target and the
probed parton. This multitude of functional dependencies is,
however, very strongly  constrained by their relation to parton
densities, form factors, and distribution amplitudes and even more
so by crossing relations, positivity bounds  and  Lorentz
invariance in general. The latter implies in particular that the
Mellin $x$-moments of GPDs must be polynomials of given order in
the skewness parameter $\eta$.

Unfortunately, there is still another complication. Typically
experimental observables allow only to determine convolutions
including  GPDs or GDAs and a formal deconvolution can practically
not be done for most of the processes\footnote{This is actually
only possible if the hadron is probed with two virtual photons and
the virtuality of both photons can be independently varied, which
is experimentally an extremely challenging task
\cite{GuiVan03,BelMue02a,BelMue03}.}. To determine GPDs or GDAs
from experimental data one therefore needs ans\"atze for them
which involve only a minimal number of parameters. Although, on
the theoretical side GPDs have been intensively studied in the
last few years, only a few of such parameterizations have been
proposed and used in phenomenology. Perhaps the most popular
parameterization is based on the ansatz suggested by Radyushkin,
in which by construction the relation to parton densities, form
factors, and polynomiality is assured. However, this advantages
arise from the simplicity of this ansatz which also might
implement a certain rigidity. Especially, it is widely used in
combination with a factorized $t$-dependence although the latter
is known to be wrong. This factorization does not respect the
disappearance of the $t$-dependence for $x\to 1$
\cite{Yua03,Bur04} and is also basically ruled out by lattice
results \cite{Hagetal03,Gocetal03,Hagetal04}. How far the employed
versions of this ansatz are suited for the kinematics accessible
in present experiments remains an open question. Obviously, it is
highly desired to have a versatile parameterization of GPDs and
GDAs, which respects all of their formal properties.

The main idea guiding the search for a more appropriate
parameterization of GPDs and GDAs is that the relevant kinematic
variables should be separated in this new representation. Let us
remind how helpful the representation of parton densities by
Mellin moments has proven to be for the analysis of hard inclusive
processes. Mellin moments are given by the analytic continuation
of the forward expectation values of leading twist-two operators
with given spin $J$. The main advantage of the Mellin space is
that operators with different spin $J$ do not mix under evolution
and so the solution of the evolution equations is trivial. In the
case of GPDs and GDAs leading twist-two operators can contain
total derivatives and so the operator basis has to be chosen
differently, in such a way that the operators again do not mix
under evolution. The appropriate operator basis
is given in terms of collinear conformal operators that are
labelled by the (collinear) conformal spin and the normal spin of
the operator. The former quantum number characterize the
irreducible multiplets or conformal towers of the collinear
conformal algebra while the latter denotes the members of a given
multiplet. Group theoretically we are dealing with the
representation of the so-called collinear conformal algebra
$so(2,1)$ which is a subalgebra of the full conformal algebra
$so(4,2)$. Let us remark that except for the trace anomaly, which
is proportional to the renormalization group coefficient
$\beta(\alpha_s)$, conformal symmetry is preserved in perturbative
QCD. Even in the case of a non-vanishing $\beta$ function the
conformal representations can be changed in such a way that the
evolution operator is diagonal. The evolution equation can then be
solved trivially. The conformal moments at the input scale depend
on the skewness parameter and can be expanded in an appropriate
orthogonal polynomial basis where the expansion coefficients
depend on the momentum transfer squared. In other words, GPDs and
GDAs can be represented by a conformal partial wave expansion,
where the expansion coefficients are characterized by form factors
that are labelled by the conformal spin and  by  an appropriate
second quantum number, e.g., the angular momentum. Most
importantly, it has been demonstrated that the first few form
factors are measurable on the lattice. Moreover, the crossing
relation between GPDs and GDAs are reduced in this representation
of the continuation of these form factors from the space- to the
time-like region and reverse.

Group theoretical discussions based on  discrete conformal spin
have a long history in QCD. However, in practice, they seemed to
be only useful for distribution amplitudes and GDAs, where the
resulting series convergence \cite{BroFriLepSac80,EfrRad80}.
Combined with conformal symmetry predictions, the perturbative
corrections for (virtual) two photon processes in the generalized
Bjorken limit can be worked out, e.g., for the photon-to-pion
transition form factor, to next-to-next-to-leading (NNLO) order
accuracy \cite{Mue97a,MelMuePas02}.

For GPDs the conformal partial wave expansion, where the conformal
spin is a non-negative integer, is represented by a series in
terms of mathematical distributions, which only converges if it is
convoluted with suitable test functions. An appropriate
resummation of this series  has been proposed in Ref.\
\cite{Mue05} and the details are presented here. There are also
several other suggestions in the literature to define out of this
divergent series. One might insert the identity expanded in terms
of polynomials, which is however only applicable for a certain
kinematical region \cite{BelGeyMueSch97,ManPilWei97a}, and for a
next-to-leading (NLO)  analysis see Refs.\
\cite{BelMueNieSch98,BelMueNieSch98a}. One can also  represent the
identity by its Fourier transform which makes contact to the group
theoretical representation with complex valued conformal spin
\cite{BalBra89}, later adopted for GPDs in Ref.\ \cite{KivMan99b}
and more recently in Ref.\ \cite{ManKirSch05}. Also a resummation
by an integral transformation has been suggested \cite{Shu99},
which, however, under close scrutiny turned  out to be
unpracticable or at least rather complicated \cite{Nor00}. An
attempt to approximately resum the conformal partial wave
expansion within a Taylor expansion of conformal moments has been
suggested in Ref.\ \cite{PolShu02}. Of course, all these proposals
can be related to each other. However, because of the  intricate
mathematics involved a correct, complete  and efficient
resummation of conformal partial waves has not yet been worked
out.

In this paper we employ the Sommerfeld-Watson transformation to
resum the conformal partial wave expansion of GPDs, finally
yielding a Mellin-Barnes integral for GPDs. The resulting
representation is similar to the one recently proposed in Ref.
\cite{ManKirSch05}, however, not identical. The  Sommerfeld-Watson
transformation requires the analytic continuation of the conformal
spin, which plays here the analogous role to the complex spin $J$
in the inverse Mellin representation of parton densities. While
the analytic continuation of Mellin moments for parton
distribution functions is a rather simple task, it is a highly
non-trivial one for the conformal moments of GPDs. This central
problem is solved by us to an extent such that the framework can
be applied as soon as a GPD ansatz is given. Although the final
GPD representation as a Mellin-Barnes integral over the complex
conformal spin seems to be rather complicated  it has several
advantages. The dependence on kinematic variables is separated in
this representation, it allows a simple and stable numerical
treatment of GPDs and their convolution with hard-scattering
amplitudes, and can be used for the analytic approximation of
scattering amplitudes. Moreover, the evolution equations to
leading order (LO) accuracy are trivially solved and the conformal
approach in Ref.\ \cite{Mue97a,MelMuePas02}   can be adopted for
the study of higher order corrections in perturbative QCD.
Especially, the NNLO corrections to deeply virtual Compton
scattering (DVCS) are calculable in a rather economic manner
\cite{MueSch05a}. Also the use of the crossing relation between
GDAs and GPDs is possible in our representation and should be very
valuable for phenomenology.

The paper is organized as following. Sect.\ \ref{Sec-FeaParGPDs}
is devoted to the conformal partial wave  decomposition of GPDs
for complex conformal spin. We start with a review on the anatomy
of GPDs in Sect.\ \ref{SubSec-AnaGPDs} and discuss the extension
of the GPD support \cite{Mue05}. To the best of our knowledge this issue has
not been presented in detail so far. Here it will guide us to find the correct
treatment of partial waves with complex conformal spin. We
consider then the crossing relation between GDAs and GPDs and
derive in Sect.\ \ref{SubSec-ConColSpi} the new GPD and GDA
representations in terms of Mellin--Barnes integrals. In Sect.\
\ref{Sec-CS-Sch} we consider evolution kernels and their
convolution with GPDs in a scheme that preserves conformal
symmetry.  Moreover, we present the Mellin-Barnes integral for the
scattering amplitude in  DVCS and discuss its analytic
approximation. In Sect.\ \ref{Sec-ExaGPDs} we have a closer look
to the analytic continuation procedure of conformal moments for a
simple GPD toy ansatz. Then we address the issue of  ans\"atze for
conformal moments and explore the features of the resulting GPDs
and GDAs. Furthermore, for vanishing longitudinal momentum
fraction in the $t$ channel we have a short look at valence quark
GPDs. Motivated by lattice results
\cite{Hagetal03,Gocetal03,Hagetal04}, we introduce a
parameterization for which the experimental constraints on GPDs
indicate that leading Regge trajectories are present in conformal
moments. Finally, we summarize and conclude. In Appendix
\ref{App-Int} integrals, which are used in the main text, and the
rotation from ordinary Mellin moments to conformal ones are
presented. Appendices \ref{App-GluCas} and \ref{App-MelBar-Ker}
contain the Mellin-Barnes integrals for gluonic GPDs and conformal
evolution kernels, respectively.

\section{Features and parameterization of GPDs
\label{Sec-FeaParGPDs}}

\subsection{The anatomy of GPDs}
\label{SubSec-AnaGPDs}

GPDs are defined as Fourier transform of light-ray operators,
sandwiched between the initial and final hadronic states. There is
a whole compendium of GPDs for each hadron. In addition the
initial and final states can have different quantum numbers
(transition GPDs), and one even can replace the hadrons by nuclei
(nucleus GPDs). Once the initial and final states are specified,
GPDs are classified with respect to the twist of the operators and
the spin content of fields. At leading twist-two level three
different types of quark and gluon GPDs can be defined (here the
gauge link is omitted):
\begin{eqnarray}
\label{Def-GPD-q} \left\{\!\!\!
\begin{array}{c}
 {^q\!F}^V \\ {{^q\!F}^A }  \\ {{^q\!F}^T }
\end{array}
\!\!\!\right\}(x,\eta,\Delta^2,\mu^2) \!\!\! &=&\!\!\! \int\!
\frac{d\kappa}{2\pi}\; e^{i \kappa { x} P_+}
     \langle P_2, S_2 \big|\bar{\psi}_q^r(-\kappa n)
\left\{\!\!\!
\begin{array}{c}
{ \gamma_+}\\ {\gamma_+\gamma_5} \\ {i\sigma_{+\perp}}
\end{array}
\!\!\!\right\}\psi^r_q(\kappa n) \big|P_1, S_1 \rangle ,
\\
\label{Def-GPD-g} \left\{\!\!\!
\begin{array}{c}
{ {^G\!F}^V} \\ {{^G\!F}^A} \\   { {^G\!F}^T}
\end{array}
\!\!\!\right\}(x,\eta,\Delta^2,\mu^2) \!\!\! &=&\!\!\!  2
\int\!\frac{d\kappa}{\pi P_+}\;  e^{i \kappa {x} P_+}
         \langle P_2,S_2 \big| G^a_{+\mu}(\!-\kappa n)\!\!
\left\{\!\!\!
\begin{array}{c}
{ g_{\mu\nu}}\\ { i\epsilon_{\mu\nu-+}} \\ {
\tau_{\mu\nu;\rho\sigma}}
\end{array}
 \!\!\!\right\}\!\! G^a_{\nu+}(\kappa n)\big| P_1,S_1 \rangle ,
\end{eqnarray}
with $P_+ = n\cdot (P_1+P_2), V_- = n^\ast \cdot V,\ n^2=
(n^\ast)^2=0, n\cdot n^\ast=1$. In the first (vector) and second
(axial-vector) entry the in- and outgoing partons have the same
helicities, and the  sum (vector) and  difference (axial-vector)
of left- and right-handed partons is taken, respectively. For the
third entry, called transversity, a helicity flip appears. GPDs
depend on the momentum fraction $x$, conjugated to the light-cone
distance $2\kappa$, the longitudinal momentum fraction $\eta =
(P_1 -P_2)^+/(P_1 +P_2)^+$ in the $t$-channel\footnote{ In the
literature $\eta$ is now denoted as $\xi$, which also is the
Bjorken like scaling variable in hard inelastic exclusive
processes. To be precise, we distinguish between both variables.
The sign convention for $\xi$ is fixed, for $\eta$ it is changing.
For quantities which are even under reflection, i.e., $\eta \to
-\eta$, the sign convention is irrelevant. Here we define $\eta$
in such a way that it corresponds to the variable $\xi$, commonly
used in the definition of GPDs, too.}, the momentum transfer
$\Delta^2\equiv t=(P_2 -P_1)^2$, and the renormalization scale
$\mu^2$. The latter is induced by the renormalization prescription
of the operators, which is part of the GPD definition. To deal
with the polarization of the hadronic states, one might introduces
a form factor decomposition \cite{Ji98,BerCanDiePir01}. For
instance, for the nucleon GPD Dirac and Pauli-like form factors
appear in the vector case \cite{Ji98}:
\begin{eqnarray}
\label{Def-ForFacDec} {^i\!F}^V= \overline{U}(P_2,S_2)  \gamma_+
U(P_1,S_1) {H_i}(x,\eta,\Delta^2) + \overline{U}(P_2,S_2)
\frac{i\sigma_{+\nu} \Delta^\nu}{2 M} U(P_1,S_1)
{E_i}(x,\eta,\Delta^2)\,,
\end{eqnarray}
where $i=u,d,s,\cdots, G$. To avoid confusion, let us note that
for the process $\gamma^*(q_1)+p(P_1)\to\gamma^*(q_2)+p(P_2)$ two
scaling  variables  exist. They are denoted as $\xi$ and $\eta$
and are defined by  \cite{MueRobGeyDitHor94}:
\begin{equation}
\xi=\frac{-q^2}{P\cdot q}\;,~~~~~~ \eta=-\frac{\Delta\cdot
q}{P\cdot q}\quad\mbox{with}\quad q=\frac{1}{2}(q_1+q_2)\,.
\end{equation}
Both variable coincide, up to power suppressed corrections ${\cal
O}(\Delta^2/Q^2)$, when the outgoing photon is real, i.e., for
DVCS one can simply replace $\eta$ by $\xi$. In this paper we will
treat the general case.

The definitions
(\ref{Def-GPD-q})--(\ref{Def-ForFacDec}) imply the basic
properties of GPDs:
\begin{itemize}
\item In the forward limit $\Delta\!\to\! 0$
helicity non-flip GPDs reduce  to parton densities
\cite{MueRobGeyDitHor94,Ji96,Rad96,ColFraStr96}, e.g.,
\begin{eqnarray}
q_i(x,\mu^2) = \lim_{\Delta\to 0} {H_i}(x,\eta,\Delta^2, \mu^2)
\end{eqnarray}
and the helicity flip GPDs ${E_i}$ decouple, but
\begin{eqnarray}
 \lim_{\Delta\to 0} {E_i}(x,\eta,\Delta^2,\mu^2) \not= 0\,.
\end{eqnarray}
 \item
 The $\mu^2$-dependence is governed by linear evolution equations
\cite{GeyDitHorMueRob88,MueRobGeyDitHor94},
 which can be derived from the renormalization group equation of
the light-ray operators \cite{BorRob80,Bal83}.
\item Hermiticity \cite{MueRobGeyDitHor94}
together with time reversal invariance \cite{Ji98} leads to a definite
symmetry with respect to the skewness parameter $\eta$, e.g.,
${H_i}(x,\eta)={H_i}(x,-\eta).$ \item The Mellin moments of GPDs
are expectation values of local twist-two operators:
\begin{eqnarray}
\label{Def-MellMom} \int\! dx\; x^n\; {^q\!F}^V(x,\eta,\Delta^2,
Q^2) =\frac{1}{P_+^{n+1}} n^{\mu_0}\cdots n^{\mu_n}
\langle P_2, S_2 \big| {\mbox{\bf
S}}\, \bar{\psi}_q^r { \gamma_{\mu_0}}\, i\!
\stackrel{\leftrightarrow}{D}_{\mu_{1}}\cdots  i\!
\stackrel{\leftrightarrow}{D}_{\mu_{n}} \psi^r_q  \big|P_1, S_1
\rangle\,,
\end{eqnarray}
where $\stackrel{\leftrightarrow}{D}_{\mu} =
\stackrel{\to}{D}_{\mu}-\stackrel{\gets}{D}_{\mu}$ is the
covariant derivative, acting as indicated by the arrows, and the
operator ${\mbox{\bf S}}$ symmetrizes all indices and subtracts
the traces. Lorentz covariance enforces  that this moments are
polynomials in $\eta$.
\end{itemize}
Furthermore,  GPDs are constrained in the region $x\geq |\eta|$ by
the positivity of the norm in the Hilbert space of states. The
most general form of such positivity bounds
\cite{MarRys97,PirSofTer98,Rad98a}, known so far, are given as an
infinite set of constraints \cite{Pob02,Pob03}.  Such constraints
can be alternatively understood within the  representation of GPDs
as overlap of light-cone wave functions \cite{DieFelJakKro00}.

Let us consider the support  of a GPD in more detail\footnote{The
GPD support might be directly derived by means of a partonic Fock
state decomposition and the so-called $\alpha$-representation for
Feynman diagrams, see for instance Ref.\ \cite{GeyRobBorHor85}.
Equivalently, a GPD can be expressed in terms of a DD
\cite{MueRobGeyDitHor94,Rad97}, which has a simpler structure, and
one might consider it as more convenient to derive the support of
the former from that of the latter.}. A generic quark GPD
$F(x,\eta,\Delta^2)$, e.g., in the vector case, is related to a
double distribution (DD) $D(y,z,\Delta^2)$ by the integral
transformation \cite{MueRobGeyDitHor94,Rad97}
\begin{eqnarray}
\label{RepKerDD} F(x,\eta,\Delta^2) = \int_{-1}^1\! dy\!
\int_{-1+|y|}^{1-|y|}\! dz \;  x^p \delta(x- y - \eta\, z)\;
D(y,z,\Delta^2)\,, \quad p=\{0,1\}\,.
\end{eqnarray}
Here  $D(y,z,\Delta^2)$ is an even function in $z$ so that
$F(x,\eta,\Delta^2)$ has the proper symmetric behavior under the
exchange $\eta\to-\eta$. Obviously, its Mellin moments, i.e.,
$\int\! dx\, x^n F(x,\eta,\Delta^2)$ are even polynomials in
$\eta$, since the support of $D(y,z,\Delta^2)$ is restricted.
Depending on the form factors appearing in the decomposition of
the GPDs (\ref{Def-GPD-q}) and (\ref{Def-GPD-g}), see, for
instance, Eq.\ (\ref{Def-ForFacDec}), the order of the polynomial
is $n$ or $n+1$. To treat both cases in a convenient and generic
manner, we have included in Eq.\ (\ref{RepKerDD}) the factor $x^p$
with $p=0$ ($p=1$) in the former (latter) case
\cite{BelMueKirSch00}. This restores the correct order of the
polynomials\footnote{Note that within $p=0$ an  additive so-called
$D$-term was proposed to generate $\eta^{n+1}$ terms
\cite{PolWei99}.  It is only non-zero in the restricted region
$|x| \le |\eta|$  and is contained in our representation with
$p=1$ as an additive term of $D(y,z,\Delta^2)$ that is
proportional to $\delta(y)$. Our parameterization offers the
possibility that the $\eta^{n+1}$ terms arise from an uniform
GPD.}. We can now fix $\eta$ to be positive and decompose the
integration with respect to $y$ into $y>0$ and $y<0$. This results
into a decomposition of $F(x,\eta,\Delta^2)$ in its quark $q$ and
anti-quark $\overline{q}$ part:
\begin{equation}\label{Dec-GPDs}
F(x,\eta,\Delta^2)= q(x,\eta,\Delta^2) \mp
\overline{q}(-x,\eta,\Delta^2)\, .
\end{equation}
Here both functions separately satisfy the polynomiality
condition:
\begin{eqnarray}
\label{RepKerDDq} q(x,\eta,\Delta^2) = \int_{0}^1\! dy\!
\int_{-1+y}^{1-y}\! dz \;  x^p\, \delta(x- y - \eta\, z)\;
D(y,z,\Delta^2)\,, \quad p=\{0,1\}
\end{eqnarray}
and analogous for the anti-quark GPD
$\overline{q}(x,\eta,\Delta^2)$, where $D(y,z,\Delta^2)$ is
replaced by $\overline{D}(-y,z,\Delta^2)= \pm (-1)^p
D(-y,z,\Delta^2)$
\begin{eqnarray}
\label{RepKerDDbq} \overline{q}(x,\eta,\Delta^2) =
 \int_{0}^1\! dy\! \int_{-1+y}^{1-y}\! dz \;  x^p \delta(x- y -
\eta\, z)\; \overline{D}(y,z,\Delta^2)\,, \quad p=\{0,1\}\,.
\end{eqnarray}
Obviously, both quark and anti-quark GPDs have the same
mathematical representation and so we will in the following mainly
deal with the quark one. The results for anti-quark GPDs are
easily obtained by replacements $D\to \overline{D}$.

\begin{figure}[t]
\begin{center}
\mbox{
\begin{picture}(600,85)(0,0)
\put(20,0){\insertfig{6}{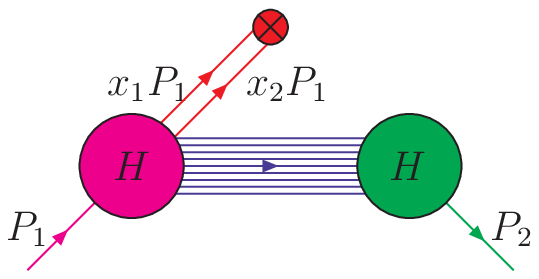}}
\put(240,0){\insertfig{6}{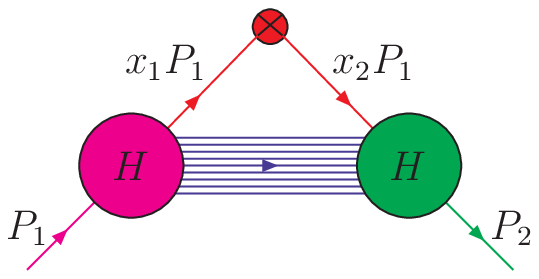}}
\end{picture}
}
\end{center}
\caption{\label{FigParInt} Partonic interpretation of GPDs in the
central (left)  and outer (right) region.}
\end{figure}
In the central (exclusive or ER-BL) region $-\eta \leq x \leq
\eta$, $q(x,\eta,\Delta^2)$ might be interpreted as probability
amplitude to have a meson like configuration inside the hadron,
while the outer (inclusive or DGLAP) region $\eta \leq x \leq 1$
can be viewed as probability amplitude for emission  and
absorbtion of a quark with momentum fraction $x_1 P_1 =
\frac{x+\eta}{1+\eta} P_1$ and $x_2 P_1 = \frac{x-\eta}{1+\eta}
P_1$, respectively, see Fig.\ \ref{FigParInt}. Remarkably, both
regions have a dual interpretation, namely,  as meson and parton
exchange in the $t$ and $s$ channel, respectively.

Lorentz invariance ties both dual regions, which can be read off
from the representation (\ref{RepKerDDq}) that ensures
polynomiality. Suppose $\eta \geq 0$, the $z$  integration in the
double distribution representation (\ref{RepKerDDq}) can be
trivially performed\footnote{\label{FooNot-sig} Since $\eta > 0$,
we have set $\delta(x-y-z\eta) = 1/\eta \delta(x/\eta-y/\eta-z)$
rather than to indicate the modulus $1/|\eta|$. The sign
convention of $\omega(x,-\eta)$ and its transformation under
reflection $\eta \to -\eta$  avoids an overall ${\rm sign}(\eta)$
factor in Eq.\ (\ref{Dec-GPD-1}).  This allows us to treat
$\omega(x,\eta)$ as a holomorphic function in the complex $\eta$
plane.} and leads to the support
\begin{eqnarray}
\label{Dec-GPD-1} q(x,\eta,\Delta^2)= \theta\left(-\eta \leq x \leq
1\right) \omega\left(x,\eta,\Delta^2\right) + \theta\left(\eta \leq
x \leq 1\right) \omega\left(x,-\eta,\Delta^2\right)\,.
\end{eqnarray}
The function $\omega$ follows from the $y$ integration in Eq.\
(\ref{RepKerDDq})
\begin{eqnarray}
\label{Def-omega}
 \omega\left(x,\eta,\Delta^2\right)=
 \frac{1}{\eta} \int_0^{\frac{x+\eta}{1+\eta}}\!dy\, x^p
 D(y,(x-y)/\eta,\Delta^2)\,.
\end{eqnarray} The GPD representation (\ref{Dec-GPD-1}) is
manifestly invariant under the transformation $\eta\to - \eta$,
especially, the support $-\eta \leq x \leq 1$ remains untouched.

\begin{figure}[t]
\begin{center}
\mbox{
\begin{picture}(600,180)(0,0)
\put(-5,163){$\displaystyle\frac{x_1}{\eta}$}
\put(10,166){\line(1,0){15}}
\put(-5,196){$\displaystyle\frac{x_2}{\eta}$}
\put(10,199){\line(1,0){15}} \put(0,0){\insertfig{7.5}{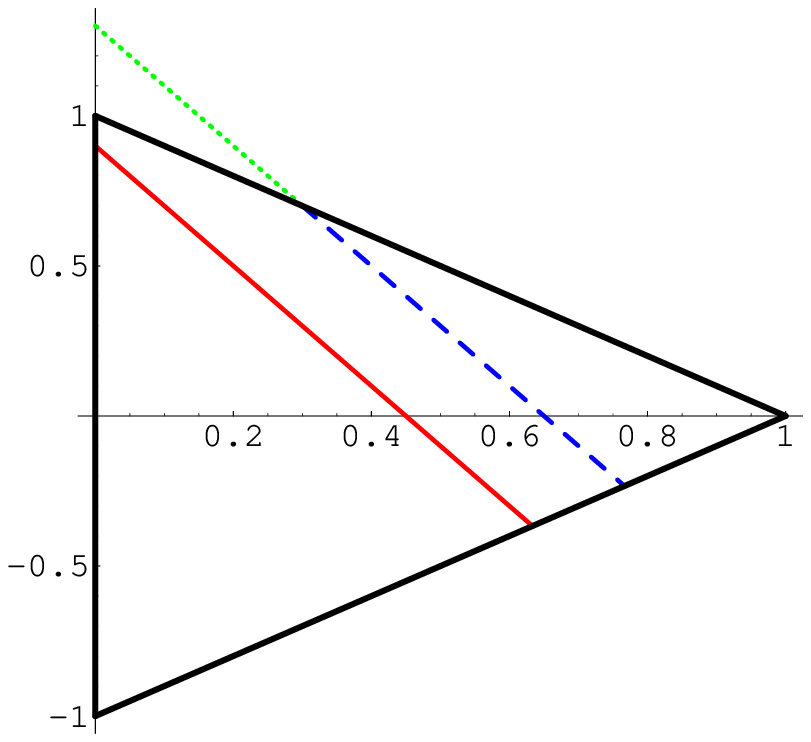}}
\put(105,105){$x_1$} \put(106,105){\line(0,-1){10}}
\put(142,105){$x_2$} \put(143,105){\line(0,-1){10}}
\put(210,75){$y$} \put(17,202){$z$} \put(80,-10){(a)}
\put(260,0){\insertfig{7}{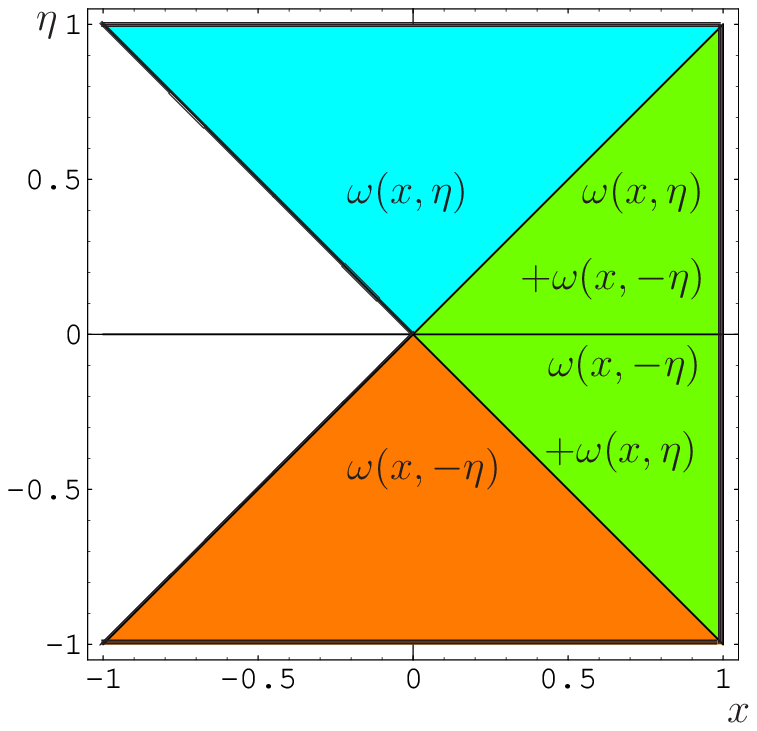}} \put(360,-10){(b)}
\end{picture}
}
\end{center}
\caption{\label{Fig-SupPro} Support of a DD (the area surrounded
by thick lines) and integration path for the calculation of
$\omega(x,\eta)$, see Eq.\ (\ref{Def-omega}), for the central
region $x_1 < \eta $ (solid) and the outer region $\eta < x_2$
(dashed) are shown in Fig.\ (a). The continuation of the
integration path over the support boundary is indicated as dotted
line. In Fig.\ (b) the support of the resulting GPD
(\ref{Dec-GPD-1}) is depicted. }
\end{figure}
In the central region the GPD is given by
$\omega\left(x,\eta,\Delta^2\right)$ from which the outer region,
determined by the symmetrized function
$\omega\left(x,\eta,\Delta^2\right)+
\omega\left(x,-\eta,\Delta^2\right)$, can be restored, see Fig.\
\ref{Fig-SupPro}. Let us have a closer look at this continuation.
In the central region  the integration variable in the integral
(\ref{Def-omega}) takes the values $0\leq y\leq
\frac{x+\eta}{1+\eta}\leq 1 $. The restriction of the second
argument in the double distribution $|z| = |(x-y)/\eta| \leq 1-y$
is ensured by the values of the lower and upper limit, see solid
line in Fig.\ \ref{Fig-SupPro}(a). The integration path, starting
at $y=0$ and $z=x/\eta$, lies inside of the DD support as it must
be. At the cross-over point $x=\eta$ it starts at the support edge
$y=0$ and $z=1$ and in the outer region the lower limit  is
$y=\frac{x-\eta}{1-\eta}$ rather than zero. We can now define an
(ambiguous) continuation  of the  DD support for $|z| > 1-y$ by
any smooth function $\tilde D(y,z,\Delta^2)$ symmetric in $z$.
\begin{equation}
 D(y,(x-y)/\eta,\Delta^2)~\to~
D(y,(x-y)/\eta,\Delta^2)\,\theta\!\left(\!
y-\frac{x-\eta}{1-\eta}\!\right) + \tilde
D(y,(x-y)/\eta,\Delta^2)\,\theta\!\left(\!
\frac{x-\eta}{1-\eta}-y\!\right).
\end{equation}
This provides the (ambiguous) continuation of
$\omega\left(x,\eta,\Delta^2\right)$ into the outer region.
$\omega\left(x,-\eta,\Delta^2\right)$ is obtained by reflection
symmetry $\eta\to -\eta$ and adding both contributions leads to
the integral representation
\begin{eqnarray}
\label{Sum-ome}
 \omega\left(x,\eta,\Delta^2\right) + \omega\left(x,-\eta,\Delta^2\right)=
 \frac{1}{\eta}
 \int_{\frac{x-\eta}{1-\eta}}^{\frac{x+\eta}{1+\eta}}\!dy\, x^p
 D(y,(x-y)/\eta,\Delta^2),
\end{eqnarray}
in which the integration  runs only over the original support of
the DD. Hence, the ambiguity in the  continuation of
$\omega\left(x,\eta,\Delta^2\right)$ and
$\omega\left(x,-\eta,\Delta^2\right)$  drops out in their sum.

Let us suppose that $D(y,z,\Delta^2)$ can be viewed as a
holomorphic function of $y$ and $z$ inside its support and has
branch cuts outside of it. Then the integration path in Fig.\
(\ref{Fig-SupPro}) (a) can cross or go along such branch cuts. To
deal with a unique definition of $\omega\left(x,\eta\right)$ for
all values of $x$ we might define its  value within its integral
representation (\ref{Def-omega}) for $\eta \leq x$ by the
principal value prescription
\begin{eqnarray}
\label{Pre-Con}
\frac{1}{2}\omega(x+i\epsilon,\eta) +
\frac{1}{2} \omega(x-i\epsilon,\eta)\quad \mbox{for} \quad \eta < x\,.
\end{eqnarray}

For illustration we give a simple example for the Radyushkin
ansatz
\begin{eqnarray}
\label{RadAns}
q(x,\eta) = \int_{0}^1\! dy\! \int_{-1+y}^{1-y}\! dz \; \delta(x-
y - \eta\, z)\; \frac{q(y)}{1-y}\Pi(|z|/(1-y))\,,
\end{eqnarray}
where the $\Delta^2$ dependence is disregarded. The parton density
is parameterized by a toy  ansatz $q(y)= y^\alpha
(1-y)^{\beta}/B(\alpha+1,\beta+1)$, which can be considered as
building block for realistic parameterizations. A popular ansatz
for the  profile function is
\begin{eqnarray}
\Pi(z) =\Pi(z|b) = \frac{ (1-z^2)^b }{B(b+1,1/2)}\,, \qquad B(x,y)
= \frac{\Gamma(x)\Gamma(y)}{\Gamma(x+y)}\,,
\end{eqnarray}
where the parameter $b$ controls the strength of the skewness
effect. This GPD ansatz can be evaluated in an analytically form
in terms of hypergeometric functions for non-negative integer
value of the parameter $b$. We might choose here for simplicity
$b=0$, i.e., $\Pi(z,0)=1/2$. In the central region this toy GPD
reads
\begin{eqnarray}
q(x,\eta) =  \frac{\Gamma(2+\alpha+\beta)}
{2\eta\Gamma(2+\alpha)\Gamma(1 + \beta)} \left( \frac{x + \eta }{1
+ \eta } \right)^{1+\alpha } {_2F_1}\left({1 + \alpha , 1 - \beta
\atop 2 + \alpha}\Bigg|\frac{x + \eta }{1 + \eta }\right) \quad
\mbox{for}\quad -\eta \leq x \leq \eta\,.
\end{eqnarray}
The extension of the DD support
corresponds to the analytic continuation of $\omega(x,\eta)$ into
the outer region and so we find:
\begin{eqnarray}
q(x,\eta) &\!\!\! =&\!\!\!  \frac{\Gamma(2+\alpha+\beta)}
{2\eta\Gamma(2+\alpha)\Gamma(1 + \beta)} \Bigg[ \left( \frac{x +
\eta }{1 + \eta } \right)^{1+\alpha } {_2F_1}\left({1 + \alpha , 1
- \beta \atop 2 + \alpha}\Bigg|\frac{x + \eta }{1 + \eta }\right)
\\
&&\hspace{3.3cm} -
 \left( \frac{x -
\eta }{1 - \eta } \right)^{1+\alpha } {_2F_1}\left({1 + \alpha , 1
- \beta \atop 2 + \alpha}\Bigg|\frac{x - \eta }{1 - \eta }\right)
 \Bigg]\quad
\mbox{for}\quad x \geq \eta\,.
 \nonumber
\end{eqnarray}

We remark that  the rescaled distribution
\begin{eqnarray}
\label{Def-resc-GPD}
  Y^{p-1} q\left(\frac{X}{Y},\frac{1}{Y},\Delta^2\right)  =
  \int_{0}^1\! dy\!
\int_{-1+y}^{1-y}\! dz \;  X^p\, \delta(X- y Y -  z)\;
D(y,z,\Delta^2)\,, \quad p=\{0,1\}
\end{eqnarray}
has technically the same support as it appears in evolution kernels.
Defining $\varpi$ by
\begin{equation}\label{Def-varpi}
  \varpi(X,Y,\Delta^2) =
 \int_0^{\frac{1+X}{1+Y}}\!dy\, X^p D(y,X-Y y,\Delta^2)\,,
 \quad p=\{0,1\}\,,
\end{equation}
we have
\begin{eqnarray}
\label{DefCroGPD}
 Y^{p-1} q\left(\frac{X}{Y},\frac{1}{Y},\Delta^2\right) = {\rm sign}(1+Y)
\theta\left(\frac{1+X}{1+Y}\right)
\theta\left(\frac{Y-X}{1+Y}\right) \varpi\left(X,Y,\Delta^2\right)
+ \left\{ { X\to -X\atop Y\to -Y} \right\}\, .
\end{eqnarray}
The evolution kernels have the form (\ref{DefCroGPD}) in the
collinear limit $\Delta^2=0$. This enables us to adopt results for
the representation of GPDs to the representation of GDAs (see next
section) and of kernels. Needless to say, the GPD
(\ref{Dec-GPD-1}) follows from  $X= x/\eta$ and $Y= 1/\eta$, where
$\omega(x,\eta,\Delta^2)$, see Eq.\ (\ref{Def-omega}), is related
to $\varpi(X,Y,\Delta^2)$ by the formula
\begin{eqnarray}
\label{Rel-ome-varome}
\omega\left(x,\eta,\Delta^2\right)  =
  \eta^{p-1}
  \varpi\left(\frac{x}{\eta},\frac{1}{\eta},\Delta^2\right).
\end{eqnarray}

\subsection{Generalized distribution amplitudes and crossing}
\label{SubSec-Cro}

Another ingredient we need for our derivation of a Mellin-Barnes
representation of GPDs is their extension to $1< \eta$. This
immediately leads us to further non-perturbative distributions,
the so-called GDAs \cite{MueRobGeyDitHor94,DieGouPirTer98}, which
are related to GPDs by crossing \cite{Ter01}. The GDAs, denoted as
$\Phi(z,\zeta,W^2)$, are defined in analogy to the GPDs in Eqs.\
(\ref{Def-GPD-q}) and (\ref{Def-GPD-g}), however, the initial
state is replaced by the vacuum and the final one contains two
hadrons. For instance, the crossing analog of ${^q\!F}^V$ for a
spin-zero target reads
\begin{eqnarray}
 {^q\!\Phi}^V(z,\zeta,W^2,\mu^2)
 &=&\!\!\! \int\!
\frac{d\kappa}{2\pi}\; e^{i \kappa (1-2z) P_+}
     \langle 0 \big|\bar{\psi}_q^r(-\kappa n)\,
 \gamma_+ \,\psi^r_q(\kappa n) \big|P_1, P_2\rangle
\end{eqnarray}
with $P_+ = n\!\cdot\!(P_1 + P_2)$.
Here $0\leq z\leq 1$ and $1-z$ are the momentum
fractions of the quark and anti-quark, respectively, which produce
the hadron pair with invariant mass squared $W^2$. $0\leq \zeta
\leq 1$ is the momentum fraction of one of the hadrons.

In the following we give a rather generic discussion of the
crossing relation to GPDs, which neglects details about form
factor decomposition or quantum numbers. We consider also only a
quark GPD and its analog. For anti-quarks the only additional
aspect,  that one has to implement, is the  sign convention in the
decomposition (\ref{Dec-GPDs}). The GDA reads in terms of the
rescaled distribution (\ref{Def-resc-GPD}) with $X=
1-2z,Y=1-2\zeta$, and $\Delta^2=W^2$ as
\begin{eqnarray}
\label{DefGPDtoDA} \Phi(z,\zeta,W^2) =
Y^{p-1} q\left(\frac{X}{Y},\frac{1}{Y},\Delta^2\right)\Big|_{X=
1-2z,Y=1-2\zeta, \Delta^2=W^2}\,.
\end{eqnarray}
 Consequently, from Eq.\ (\ref{DefCroGPD}) we read off its representation
\begin{eqnarray}
\label{SupGDA}
 \Phi(z,\zeta,W^2) = \theta\left(z-\zeta\right)
\varpi\left(1-2z,1-2\zeta,W^2\right)+\theta\left(\zeta-z\right)
\varpi\left(2z-1,2\zeta-1,W^2\right) \,,
\end{eqnarray}
where the function $\varpi$ is given by the integral
(\ref{Def-varpi}). Having in mind that in Eq.\ (\ref{DefGPDtoDA})
a rescaled GPD appears, it remains a trivial exercise to directly
relate GPDs and GDAs\footnote{Note that the upper line in Eq.\
(\ref{CroRel}) is only valid for $1< \eta$, i.e., $\zeta < 1/2$.
For $ 1/2 < \zeta $ one must replace $\eta^{1-p}$ by ${\rm
sign}(\eta) \eta^{1-p}$. It is more convenient to use the second
line, valid for all values of $\eta$, together with the explicit
representations (\ref{Dec-GPD-1}) and (\ref{SupGDA}), see also
footnote \ref{FooNot-sig}.}:
\begin{eqnarray}
\label{CroRel} \left\{{
 \Phi(z,\zeta,W^2)\,,
 \atop
  \varpi\left(1-2z,1-2\zeta,W^2\right)}\right\}
\leftrightarrow
 \left\{ {
 \eta^{1-p}\, q(x,\eta,\Delta^2)
 \atop
 \eta^{1-p}\, \omega\left(x,\eta,\Delta^2\right)
 }
 \right\},
 \quad
1-2 z  \leftrightarrow  \frac{x}{\eta}\,, 1-2 \zeta
\leftrightarrow \frac{1}{\eta}\,, W^2 \leftrightarrow \Delta^2\,,
\end{eqnarray}
where the lower line is in fact the relation (\ref{Rel-ome-varome}).

Let us shortly discuss this crossing relation. A GDA
is obtained from the GPD analog by
\begin{eqnarray}
\label{CroGPDtoGDA} \varpi\left(1-2z,1-2\zeta,W^2\right) =
(1-2\zeta)^{p-1}\,
\omega\left(\frac{1-2z}{1-2\zeta},\frac{1}{1-2\zeta},W^2\right)\,.
\end{eqnarray}
Since $0\leq \zeta\leq 1$, the second argument of $\omega$ covers
the region $ |1/(2\zeta-1)|\geq 1$. Thus, the crossing relation
requires to enter a kinematic region which is unphysical for GPDs,
except for the point $\eta=1$, i.e., $\zeta=0$. However, for a
given functional form of a GPD, the  relation (\ref{CroGPDtoGDA})
allows to fix its phenomenological parameters. So for instance,
the same function appears after factorization in hard exclusive
electroproduction of a photon or mesons on a hadron target as GPD
and in the production of a hadron pair due to two photon fusion as
GDA. The knowledge of the analytic form of the GPD in the central
region is sufficient to perform the symmetry transformation
(\ref{CroGPDtoGDA}) to obtain the corresponding GDA.  Reversely, a
GPD follows from a given GDA using the symmetry transformation
(\ref{Rel-ome-varome}). As expected from general reasons, the
physical and unphysical regions are again connected by crossing.
Moreover, as we realized above, $\omega(x,\eta,\Delta^2)$ is not
uniquely defined in the outer region. As explained above this
problem is artificial and does not affect the net contribution in
this region.

Let us stress ones more that the extension procedure of the
support is unique, which was shown in connection with the support
extension of evolution kernels
\cite{GeyDitHorMueRob88,MueRobGeyDitHor94}. Here we adopt the same
arguments. Suppose  we know the  function
$\varpi\left(1-2z,1-2\zeta,\Delta^2\right)$ in the region $0\leq
\zeta \leq z\leq 1$, which is equivalently to the knowledge of the
GDA $\Phi(z,\zeta,W^2)$, see Eq.\ (\ref{SupGDA}). Next
representing the GDA (\ref{DefGPDtoDA}) in terms of the DD
(\ref{Def-resc-GPD}), we realize that its convolution with any
holomorphic test function $\tau(z)$ yields an holomorphic function
in $\zeta$. For instance,  one finds for $p=0$
\begin{equation}\label{AnaConGPD}
\int_{0}^1\!dz\, \tau(z) \Phi(z,\zeta,W^2)  =\frac{1}{2}\int_{0}^1\!
dy\! \int_{-1+y}^{1-y}\! dz \;
\tau\left(\!\frac{1-y(1-2\zeta)-z}{2}\!\right)\;
D(y,z,W^2)\,.
\end{equation}
 Hence, also the Fourier transform  of the GDA
with respect to $z$ is a holomorphic function in the conjugate
variables $\lambda$ and $\zeta$. Consequently, we can employ
analytic continuation.  Then the inverse Fourier transform
together with the crossing relation (\ref{CroRel}) yields the
result we desire,
\begin{eqnarray}
\label{ExtPro}
 q(x,\eta,\Delta^2)
 =  {\rm sign}(\eta)\; \eta^{p-1}
 \int_{-\infty}^\infty\! \frac{d\lambda}{2\pi}
e^{-i \lambda x/\eta} {\rm AC}\left[2  \int_{0}^{1}\! dz\, e^{i
\lambda (1-2 z)} \Phi(z,\zeta,W^2) \right]\,,
\end{eqnarray}
where $\zeta=(\eta-1)/2\eta$, $W^2=\Delta^2$ and ${\rm AC}$
denotes the analytic continuation of both variables  $\lambda$ and
$\zeta$. (Actually, the variables stays real and analytic
continuation is only used to extend their  numerical values on the
real axis, e.g., for $\lambda$ up to $\pm$ infinity.)

\subsection{Complex collinear conformal spin partial wave expansion}
\label{SubSec-ConColSpi}

We derive now  a new representation for GPDs which has several
advantages, already mentioned in the introduction.  In fact we
will deal with a partial wave decomposition of GPDs, where the
partial waves  are labelled by the complex conformal spin, the
quantum number which characterize the multiplets (towers) of
conformal operators. This is rather analogous to the partial wave
expansion of scattering amplitudes with respect to the complex
angular momentum, however, requires a more attentive
consideration. Irrespectively, of whether the symmetry is
preserved or not, one can introduce such an expansion. Certainly,
conformal symmetry is broken in the non-perturbative QCD sector.
Fortunately, up to calculable corrections proportional the
non-vanishing $\beta$ function, it holds true in the perturbative
sector and thus has nevertheless still predictive power.

In the following we  consider only quark GPDs, the gluon case can
be treated analogously and the results are collected in Appendix
\ref{App-GluCas}. To derive the partial wave decomposition of
GPDs, the following steps will be performed:
\begin{itemize}
\item
First the conformal moments and partial wave expansion of GPDs are
introduced for discrete conformal spin. Here GPDs  are represented
as  a divergent series  in terms of mathematical distributions.
\item
This series will be summed in the unphysical region, where it
can be viewed as an ordinary expansion in terms of orthogonal
polynomials, by means of the Sommerfeld--Watson transformation,
which requires the analytic continuation of the conformal spin.
\item
Finally, we complete the Sommerfeld-Watson transformation and
derive a representation of GPDs in terms of a Mellin--Barnes
integral for the central region. The outer region follows from an
suitable continuation in $x$ that arises from the analytic
structure of GPDs.
\end{itemize}

\subsubsection{Conformal partial wave expansion}
\label{SubSubSec-ConParWavExp}

So-called conformal moments of quark GPDs are formed with respect
to Gegenbauer polynomials $\eta^n C_n^{3/2}\left(x/\eta\right)$
with index 3/2 and order $n$. These moments are given by the
expectation value of local conformal operators, see below Eq.\
(\ref{Def-ConMomVec}). Relevant group theoretical aspects can be
found in  Appendix \ref{App-GluCas} and in the review
\cite{BraKorMue03}. These moments can be viewed as an appropriate
generalization of the ordinary forward Mellin moments, used in the
analysis of deep inelastic scattering. The Gegenbauer polynomials
are orthogonal polynomials, possess a definite reflection symmetry
$C^{3/2}_n(x)= (-1)^n C^{3/2}_n(-x)$, and are the only solution of
the second order differential equation
\begin{eqnarray}
\frac{d^2}{dx^2} (1-x^2) C^{3/2}_n(x) = -(n+1)(n+2) C^{3/2}_n(x)
\end{eqnarray}
that is finite at the singular points $x=\pm 1$. These polynomials
form a complete basis in the interval $[-1,1]$. Here we rescale
the polynomials and choose the normalization
\begin{eqnarray}
\label{Def-c}
 c_n(x,\eta) =  \eta^n c_n\!\left(\!\frac{x}{\eta}\!\right)
\quad\mbox{with}\quad
c_n(x) =
\frac{\Gamma(3/2)\Gamma(1+n)}{2^{n} \Gamma(3/2+n)}
C_n^{3/2}\left(x\right)
\end{eqnarray}
in such a way
that in the forward case the ordinary Mellin moments appear:
\begin{eqnarray}
\label{Def-csca} \lim_{\eta\to 0} c_n(x,\eta) = x^n\,.
\end{eqnarray}
There are several possibilities to express $c_n(x,\eta)$ in terms
of hypergeometric functions, which might provide different
prescriptions for the analytic continuation of the discrete
variable $n$. Below  we will use
\begin{eqnarray}
\label{Def-cHyp} c_j(x,\eta) = \frac{\Gamma(3/2) \Gamma(3 + j)}{2^{1+j}
\Gamma(3/2 + j)}\; \eta^j {_2\!F}_1\!\left({-j,j+3\atop
2}\Big|\frac{\eta-x}{2\eta}\right)
\end{eqnarray}
for complex valued $j$. Equivalently, we can express it in terms of
(associated) Legendre functions of the first kind.

The conformal moments of a GPD, separated into quark and anti-quark
ones, are defined as
\begin{eqnarray}
\label{Def-ConMom} m_n(\eta,\Delta^2)=\int_{-\eta}^1\! dx\,
c_n(x,\eta) q(x,\eta,\Delta^2) \,, \quad
\overline{m}_n(\eta,\Delta^2)= \int_{-\eta}^1\! dx\, c_n(x,\eta)
\overline{q}(x,\eta,\Delta^2) \,.
\end{eqnarray}
They are given by the expectation values of collinear conformal
operators, e.g., in the vector case,
\begin{eqnarray}
\label{Def-ConMomVec}
m_n(\eta,\Delta^2) - (-1)^n \overline{m}_n(\eta,\Delta^2)  =
 \frac{\Gamma(3/2)\Gamma(1+n)\, \eta^n}{2^{n}  \Gamma(3/2+n)\, P_+ } \langle
P_2, S_2 \big|  \bar{\psi}_q^r { \gamma_+}\,
C_n^{3/2}\!\left(\!
\frac{i\!\stackrel{\leftrightarrow}{D}_{+}}{\eta P_+}\!
\right)\psi^r_q \big|P_1, S_1 \rangle\,.
\end{eqnarray}
The  rotation  to the ordinary Mellin moments (\ref{Def-MellMom})
and its inversion are given in Appendix \ref{App-Int} by Eqs.\
(\ref{Exp-Sim2ConMom}) and (\ref{Exp-2Con2SimMom}). The operators
are characterized by the conformal spin, which in our case is
$n+2$. The (conformal) moments can be either calculated on the
lattice\footnote{The separation of valence quarks and sea quarks
is done by measuring even and odd moments. For instance, in the
vector case: if the quark and anti-quark sea is equivalent, in
even moments the complete sea drops out, while for odd moments
valence and sea quarks are added.}, can be directly modelled, or
evaluated from a given GPD ansatz. In the latter case they are
naturally decomposed as, e.g., for quarks,
\begin{eqnarray}
\label{DefConMomGPD}
 m_n(\eta,\Delta^2) = \mu_n(\eta,\Delta^2) + \mu_n(-\eta,\Delta^2)\,,
 \quad
 \mu_n(\eta,\Delta^2) = \int_{-\eta}^1\!dx\, c_n(x,\eta)\,
 \omega(x,\eta,\Delta^2)\,.
\end{eqnarray}
This is a simple consequence of the symmetry relations
$c_n(x,\eta)= c_n(x,-\eta)$ together with the representation
(\ref{Dec-GPD-1}).  The $\mu_n(\eta,\Delta^2)$ are only
defined in terms of $\omega(x,\eta,\Delta^2)$ and, thus, they can
be quite general functions of $\eta$. However, after
symmetrization with respect to $\eta$, see  first formula in Eq.\
(\ref{DefConMomGPD}), one obtains the polynomial $
m_n(\eta,\Delta^2)$.

Now we would like to invert the transformation (\ref{Def-ConMom}). As
mentioned above the polynomials $c_n(x,\eta)$, see Eqs.\
(\ref{Def-c}) and (\ref{Def-csca}), form only a complete basis  in
the central region $[-\eta,\eta]$. Let us denote by $p_n(x,\eta)$
the polynomials that include the weight $(1-x^2)$ and an
appropriate normalization
\begin{eqnarray}
\label{Def-pn}
 p_n(x,\eta) = \eta^{-n-1}
 p_n\left(\frac{x}{\eta}\right)\,,
 \quad
p_n(x) &\!\!\!=\!\!\!& \theta(1-|x|)\frac{2^{n} \Gamma(5/2 +
n)}{\Gamma(3/2)\Gamma(3 + n)} \left( 1 - x^2 \right)
C_n^{3/2}(-x)\,.
\end{eqnarray}
The orthogonality relation for Gegenbauer polynomials reads in our
notation
\begin{eqnarray}
\label{OrtRel} \int_{-1}^1 dx\, p_n(x,\eta)  c_m(x,\eta) = (-1)^{n}
\delta_{nm}\,.
\end{eqnarray}
The minus sign in the argument of Gegenbauer polynomials in Eq.\
(\ref{Def-pn}) is conventionally and induces the factor $(-1)^{n}$
in the orthogonality relation. This sign convention is appropriate
to perform the steps which follow. Note also that the support
restriction is explicitly contained in the definition
(\ref{Def-pn}) and so the integration region in the integral
(\ref{OrtRel}) is restricted to the central region. We might now
expand a GPD in terms of such polynomials (\ref{Def-pn})
\begin{equation}\label{SumConMom1}
 q(x,\eta,\Delta^2) = \sum_{n=0}^\infty (-1)^{n} p_n(x,\eta)
 m_n(\eta,\Delta^2)\,.
\end{equation}
It is easily to see that the conformal moments (\ref{Def-ConMom})
of a GPD are reproduced by this series (\ref{SumConMom1}).
However, it is divergent as expansion in terms of
polyonomials\footnote{This is also the case  in the central region
$[-\eta,\eta]$, since the  coefficients in front or the Gegenbauer
polynomials are enhanced by the factor $\eta^{-j-1}$, which
divergences for $|\eta| < 1$ at $j\to \infty$.}. Especially, the
restricted support property of each individual term does not
imply that the GPD vanishes in the outer region. Rather one
should understand this expansion as an ill-convergent
sum of distributions (in the mathematical sense) that yields
a result which is non-zero in the outer region. Indeed,
$p_n(x,\eta)$  can be considered as the $n$th derivative of a
smeared $\delta$ function:
\begin{eqnarray}
p_n(x,\eta) = \frac{ \Gamma(5/2 + n)}{n! \Gamma(1/2) \Gamma(2 +
n)} \int_{-1}^1\! du\, (1-u^2)^{n+1}
 \delta^{(n)}(x-u \eta)\,.
\end{eqnarray}
 Taking now the
forward limit $\lim_{\eta\to 0}p_n(x,\eta) = \delta^{(n)}(x)/n!$,
the series (\ref{SumConMom1}) turns  out to be the expansion of
parton densities in terms of derivatives of $\delta(x)$.

On the other hand, Eq.\ (\ref{SumConMom1}) might be a convergent
series for $\eta > 1$ and by means of the crossing relation
(\ref{CroRel}) we find for a GDA the partial wave decomposition:
\begin{eqnarray}
\label{ConParWavGDA} \Phi(z,\zeta,W^2) = \sum_{n=0}^\infty (-1)^n
p_n(1-2z,1) M_n(\zeta,W^2)\,,
\end{eqnarray}
where the polynomials $M_n(\zeta,W^2) =
 (1-2\zeta)^n m_n(1/(1-2\zeta),W^2)$ are of order $n$. If
  $\Phi(z,\zeta,W^2)$ is a smooth function that vanishes
 at the end-points $z=\{0,1\}$, the series converges and the
 moments $M_n(\zeta,W^2)$ behave for $n\to
 \infty$ as $\sim 2^{-n} n^{-1/2-\epsilon}$ with $\epsilon>0$ .
Here the exponential suppression by $2^{-n}$ is a consequence of
our normalization, which has been adopted from the Mellin moments
of GPDs respectively parton densities.

The series (\ref{SumConMom1}) and (\ref{ConParWavGDA}) are the
conformal partial wave expansions with respect to the conformal
spin $n+2$. They have the advantage that in perturbative QCD the
conformal spin is,  to some extend, a good quantum number. So for
instance the LO evolution kernels are diagonal with respect to
Gegenbauer polynomials. Furthermore, in this expansion the $x$ and
$\Delta^2$ (respectively $z$ and $W^2$) dependence factorizes. In
the case of GPDs the ($x$,$\eta$) dependence is decomposed in an
intrinsic $x/\eta$ and a remaining $(\eta,\Delta^2)$  dependence
contained in the conformal moments. The $(\eta,\Delta^2)$ or
($\zeta,W^2$) dependencies can be separated by an expansion  of
the moments with respect to an appropriate set of orthogonal
polynomials. It has been proposed to expand $M_n(\zeta,W^2)$ in
terms of Legendre polynomials $P_l(\cos(\theta))$ \cite{PolShu02},
the eigenfunctions of the rotation group $SO(3)$ for spinless
states. Here $\cos(\theta)\simeq 1-2\zeta =1/\eta$ and $\theta$ is
the scattering angle in the center-of-mass system.  Certainly, it
is appealing to have such an expansion with respect to angular
momentum $l$, which makes contact to the $SO(3)$ partial wave
expansion. When spin is involved, the expansion is rather given in
terms of Wigner functions, expressible in terms of associated
Legendre polynomials. To be not specific on the spin content and
in view of the definition in terms of conformal operators, it
appears also natural to expand conformal moments
with respect to Gegenbauer polynomials with index $3/2$:
\begin{eqnarray}
\label{Def-Exp-Coe}
m_n(\eta,\Delta^2) = \sum_{k=0}^n F_{nk}(\Delta^2)\, \eta^n
c_k(1/\eta)\,,\quad M_n(\zeta,W^2) = \sum_{k=0}^n F_{nk}(W^2)\,
c_k(1-2\zeta)\,,
\end{eqnarray}
where all $\Delta^2$ and $W^2$ dependence is absorbed into  the
form factors\footnote{We note that in this expansion the scaling
invariance and so the conformal symmetry is  broken, since in
general the form factors contain now a massive parameter for
dimensional reason. This is not surprising, since in
non-perturbative QCD this symmetry does not hold -- otherwise
there would exist only massless hadrons.} $F_{nk}$. Obviously, the
crossing of GPDs and GDAs, i.e., the transfer from the space-like
to the time-like region, concerns now only these form factors. The
form factors, appearing in a specific partial wave expansion with
respect to the angular momentum, are simply obtained  by a
rotation from the conformal ones, i.e., $F_{nk}(\Delta^2)$.

\subsubsection{Sommerfeld-Watson transformation}
\label{SubSubSec-SomWatTra}

As already mentioned, the series (\ref{SumConMom1}) can not be
directly used in practical calculation. Rather, it must be either
resummed or the individual terms must be smeared by inserting the
identity, expanded with respect to an appropriate basis. However,
the latter method is only applicable in a restricted kinematical
region and has been performed only  approximately. So we chose to
resum the conformal partial waves series for GPDs instead.
\begin{figure}[t]
\begin{center}
\mbox{
\begin{picture}(600,130)(0,0)
\put(50,0){\insertfig{4}{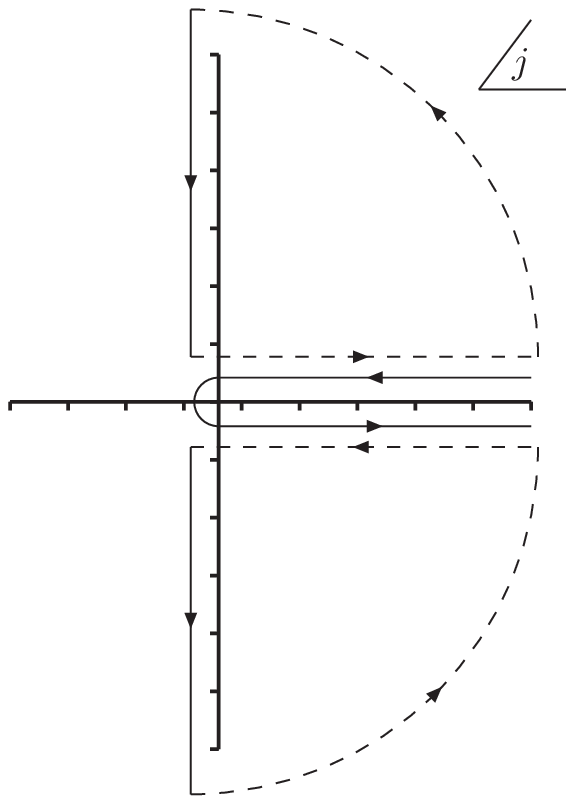}}
\put(90,-15){$(a)$}
\put(260,23){\insertfig{4}{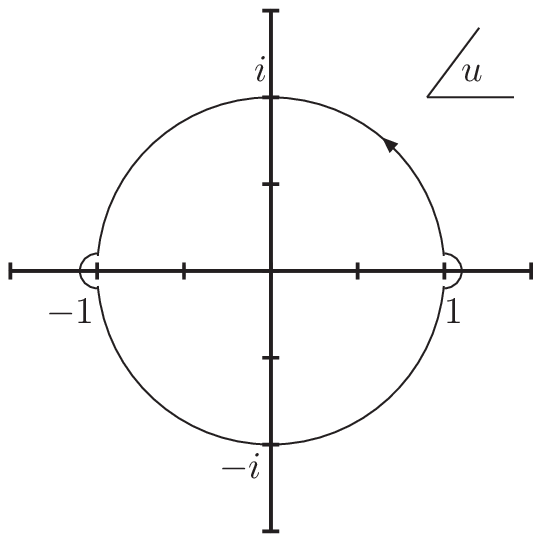}}
\put(315,-15){$(b)$}
\end{picture}
}
\end{center}
\caption{\label{FigCon1} $(a)$ The integration contour in Eq.
(\ref{Def-IntParWav}) enclosing the real axis in the complex
$j$-plane. Adding semicircles results in an integration path which
is parallel to the imaginary axis. $(b)$ The contour of the
Schl\"afli integral (\ref{Def-pn-SchInt}) in the complex
$u$-plane.}
\end{figure}
We consider it first in the unphysical region $\eta>1$ and rewrite
$q(x,\eta,\Delta^2)$ as a contour integral in the complex plane
that includes the positive real axis, see Fig. \ref{FigCon1}:
\begin{eqnarray}
\label{Def-IntParWav} q(x,\eta,\Delta^2) = \frac{1}{2
i}\oint_{(0)}^{(\infty)}\!dj\,  \frac{1}{\sin(\pi j)}
p_j(x,\eta)\; m_j(\eta,\Delta^2)\,.
\end{eqnarray}
Here we included a factor $1/\sin(\pi j)$, which has the residue
$\mbox{Res}_{j=n}\, 1/\sin(\pi j) = (-1)^n/\pi$ for
$n={0,1,2,\cdots}$. Thus, if no other singularities are present
inside the integration contour, the residue theorem leads to the
conformal partial wave expansion (\ref{SumConMom1}). The main
difficulty is to find an appropriate analytic
continuation\footnote{We remind that the analytic continuation of
a function that depends on a discrete variable is not unique. For
instance, a term  proportional to $\sin(\pi j)$ could be added,
which drops out for $j=n=0,1,2,\cdots$.} of both functions
$p_j(x,\eta)$ and $m_j(\eta,\Delta^2)$ with respect to the
conformal spin $n+2$.

First let us  define the analytic continuation of $p_j(x,\eta)$.
This can be done using its definition (\ref{Def-pn}) in
terms of hypergeometric functions. To include also the support
restriction, we represent the analytic continuation of the
Gegenbauer polynomials by the Schl\"afli integral
\begin{eqnarray}
\label{Def-pn-SchInt}
 p_j(x,\eta) =
- \frac{\Gamma(5/2 + j)}{\Gamma(1/2) \Gamma(2 + j)}
 \frac{1}{2\pi i} \oint_{(-1+\epsilon)}^{(+1-\epsilon)}\! du\,
 \frac{(u^2-1)^{j+1}}{(x+u\,
 \eta)^{j+1}}\,.
\end{eqnarray}
Here the integration contour is essentially the unit circle in the
complex $u$-plane where the points $-1$ and $+1$ are included, see
Fig.\ \ref{FigCon1} (b).  The integrand has four branch points in
the complex $u$-plane, namely at $\{-\infty, -1,-x/\eta,1\}$.
These points will be connected by a single branch cut that goes
along the real axis from $-\infty$ to ${\rm Max}(-x/\eta,1)$. It
is easy to see that for non-negative integer $j=n$ the  Schl\"afli
integral is equivalent to the definition (\ref{Def-pn}). The
integrand possesses now only a pole of order $n+1$ at $u=-x/\eta$
(and at infinity). For $|x| < \eta $ the pole is inside the
integration contour  and the residue theorem gives $p_n(x,\eta)$.
On the other hand  for $ |x| > \eta$ the pole is moved out of the
contour and so the integral vanishes.

Before we evaluate the integral (\ref{Def-pn-SchInt}) in general,
let us consider the forward case $\eta=0$. Here the integrand is
essentially reduced to the function $(u^2-1)^{j+1}$ and possesses
for non-integer $j$ a discontinuity on the real axis in the
interval $-1\leq u\leq 1$. We might now deform the contour so that
the real axis is pinched. For $|\Re{\rm e}\, u|\leq 1$  we pick up
a phase factor $e^{\pm i \pi (j+1)}$ for $\Im{\rm m}\,u \gtrless
0$ and so we can write
\begin{eqnarray}
p_j(x,\eta=0) =  x^{-j-1} \frac{\Gamma(5/2 + j)}{\Gamma(1/2) \Gamma(2 + j)}
\frac{1}{2\pi i} \left(e^{i \pi (j+1)} -e^{- i \pi (j+1)} \right)
                 \int_{-1}^1\! du\,(1-u^2)^{j+1}\,.
\end{eqnarray}
The remaining integral  represents just $\Gamma(1/2) \Gamma(2 +
j)/\Gamma(5/2 + j)$, which results in
\begin{eqnarray}
p_j(x,\eta=0) =\frac{\sin(\pi[j+1])}{\pi} x^{-j-1}\,.
\end{eqnarray}
This is, up to the  conventional  factor
$\sin(\pi[1+j])/\pi$, nothing else but the integral kernel of the
inverse Mellin transform, widely used in deep inelastic
scattering.

{F}or $\eta>0$ and non integer values $j$ the integrand in Eq.\
(\ref{Def-pn-SchInt}) has no discontinuity for $x \leq -\eta$. For
$x\geq -\eta$ we will pick up phase factors by surrounding  the
branch points at $-x/\eta$ and $x=1$. They appear in the following
interval above or below the real axis:
\begin{eqnarray}
\left\{
\begin{array}{c}
\left[-x/\eta,1\right] \\
\left[-1,1\right]
\end{array}
\right\}
 \quad&\mbox{for}&\quad
\left\{
\begin{array}{c}
 -\eta < x \leq \eta\;\;\; \mbox{and}\;\;\; \eta > 0 \\
 \eta < x
\end{array}
\right. .
\end{eqnarray}
Consequently, at the endpoint $x=-\eta$ and for $x<-\eta$ the
integral vanishes
\begin{equation} \label{Valpj-End}
p_j(x\leq-\eta,\eta)=0\,.
\end{equation}
For both the central and the non-vanishing
outer region the complex integral can be evaluated by deforming
the contour as before so that  the real axis is pinched. Taking the
discontinuity  yield the following integrals along the real axis
\begin{eqnarray}
\label{Def-PjInt} p_j(x,\eta) &\!\!\!=\!\!\!& \frac{\Gamma(5/2 +
j)}{\Gamma(1/2) \Gamma(2 + j)} \frac{ \sin(\pi [j+1])}{\pi}
\int_{-x/\eta}^{1} \! du\,\frac{(1-u^2)^{j+1}}{(x+u\,
 \eta)^{j+1}}\,,\quad -\eta \leq x \leq \eta\,, \;\; \eta>0\,
 \nonumber\\
 &\!\!\!=\!\!\!& \frac{\Gamma(5/2 + j)}{\Gamma(1/2)
\Gamma(2 + j)} \frac{ \sin(\pi [j+1])}{\pi}  \int_{-1}^{1} \!
du\,\frac{(1-u^2)^{j+1}}{(x+u\,
 \eta)^{j+1}}\,,\;\;\qquad  0 \leq\eta \leq x \,.
\end{eqnarray}
At the cross over point $x=\eta$ both  integrals have the same
value and represent a Beta function. So  $p_j(x=\eta,\eta)$  is
smooth in the vicinity of this point and  takes the value
\begin{equation}
\label{Valpj-Cros} p_j(x=\eta,\eta) = \frac{2^{1 + j}}{\eta^{j+1}}
\frac{\Gamma(5/2 + j)}{\Gamma(3/2)\Gamma(3 + j)} \frac{\sin (\pi
[j+1] )}{\pi} \,.
\end{equation}

The integrals in Eq.\ (\ref{Def-PjInt}) define hypergeometric
functions. The analytic continuation of the mathematical
distributions $p_n(x,\eta)$ with respect to the conformal spin is
expressed by them as following:
\begin{eqnarray}
\label{Def-p-all}
 p_j(x,\eta)  = \theta(\eta-|x|) \eta^{-j-1}
{\cal P}_j\left(\frac{x}{\eta}\right) +\theta(x-\eta) \eta^{-j-1}
{\cal Q}_j\left(\frac{x}{\eta}\right)\,
\end{eqnarray}
where
\begin{eqnarray}
\label{Def-p-P}
{\cal P}_j(x)&\!\!\!=\!\!\!&
 \frac{2^{j+1}\Gamma(5/2 + j)}{\Gamma(1/2)\Gamma(1 +j)}
(1+x) \, {_2\!F}_1\!\left({-j-1,j+2\atop
2}\Big|\frac{1+x}{2}\right)\,,
\\
\label{Def-p-Q}
{\cal Q}_j(x) &\!\!\!=\!\!\!&  -\frac{\sin(\pi j)}{\pi}\;
x^{-j-1}\; {_2\!F}_1\!\left({(j+1)/2,(j+2)/2\atop
5/2+j}\Big|\frac{1}{x^2}\right) \,.
\end{eqnarray}

Here, a few comments are in order. First, for $j=n=0,1,2,\cdots$
only the central region contributes and the relation
\begin{eqnarray}
{_2\!F}_1\!\left({-j,j+3\atop 2}\Big|\frac{1+x}{2}\right) =
\frac{2}{1-x}\;\; {_2\!F}_1\!\left({-j-1,j+2\atop
2}\Big|\frac{1+x}{2}\right)
\end{eqnarray}
establishes the definition (\ref{Def-pn}) of $p_n$ in terms of
Gegenbauer polynomials, cf.\  Eqs.\ (\ref{Def-c}) -
(\ref{Def-cHyp}). Obviously, in the central region  the analytic
continuation is based on the definition of  hypergeometric
functions. In the outer region, however, the result might be
surprising.  To clarify its meaning, we decompose the integral
(\ref{Def-PjInt}) for $\eta < x$ as $\int^{1}_{-1}\!du\cdots =
\int^{1}_{-x/\eta}\!du\cdots  - \int^{-1}_{-x/\eta}\!du\cdots$ and
realize that it can be expressed by the function ${\cal P}_j$.
Thus, we can write Eq.\ (\ref{Def-p-all}) as
\begin{eqnarray}
\label{Dec-p}
 p_j(x,\eta) =\theta(-\eta\leq x )\;
\eta^{-j-1} {\cal P}_j\left(\frac{x}{\eta}\right) + \theta(\eta\leq
x ) \cos(\pi [j+1])\; \eta^{-j-1} {\cal
P}_j\left(-\frac{x}{\eta}\right)\,.
\end{eqnarray}
Here it is  understood that the principal value is taken at the
branch cut, which starts at $x=\eta$, i.e.,  we insert $[{\cal
P}_j(x+i \epsilon)+{\cal P}_j(x-i \epsilon)]/2$ for $x\geq 1$. The $\cos(\pi
[j+1])$ term in the second expression on the r.h.s.\ arise from
the continuation of $\eta$ to $-\eta$, again by taking the
principal value. Hence, one realizes that this result precisely
fits the structure of the representation (\ref{Dec-GPD-1}) for
$q(x,\eta,\Delta^2)$ in terms of the functions $\omega$. We remark
that the identity
\begin{eqnarray}
{\cal Q}_j(x) = \frac{1}{2}{\cal P}_j(x+i \epsilon)+ \frac{1}{2} {\cal P}_j(x-i \epsilon)+
\cos(\pi [j+1]) {\cal P}_j(-x) \quad\mbox{for}\quad x\geq 1\,,
\end{eqnarray}
we derived here, is a known relation between associated Legendre functions
of the first and second kind.

\begin{figure}[t]
\begin{center}
\mbox{
\begin{picture}(600,100)(0,0)
\put(20,0){\insertfig{7}{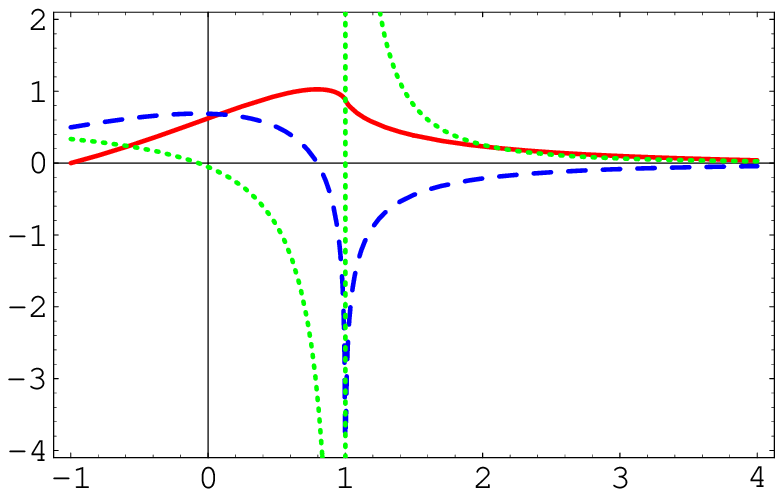}}
\put(280,0){\insertfig{7}{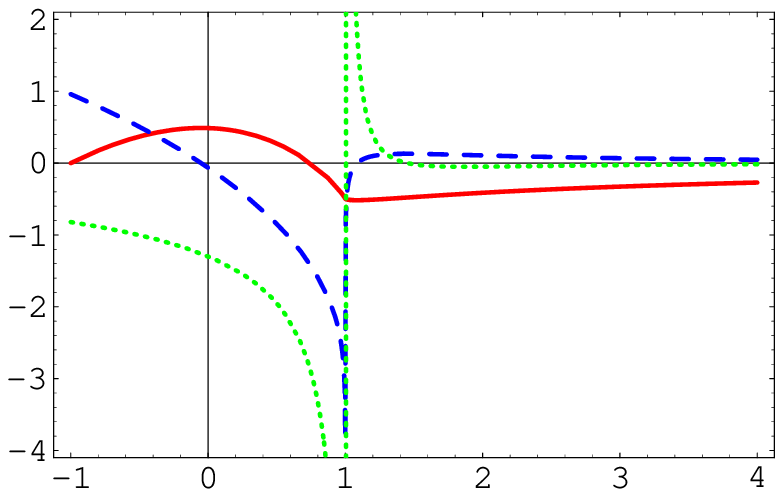}}
\put(0,40){\rotatebox{90}{$\Re{\rm e}\, p_{j}(x,1)$}}
\put(210,-10){$x$} \put(265,40){\rotatebox{90}{$\Im{\rm m}\,
p_{j}(x,1)$}} \put(470,-10){$x$}
\end{picture}
}
\end{center}
\caption{\label{Fig-Pro-p} The real (left) and imaginary (right)
part  of the conformal partial wave $p_{j}(x,\eta=1)$ (solid), its
first (dashed), and second (dotted) derivative for $j=-1/4+ i/2$.}
\end{figure}
The function $p_j(x,\eta)$ is continuous at the cross-over point
$x=\eta$, however, the imaginary part of the first derivative has
a jump, while the real part is still continuous, see Fig.\
\ref{Fig-Pro-p}. It satisfies the second order differential
equation for conformal partial waves with complex valued conformal
spin $j+2$:
\begin{equation}
\label{AnaConSpi} (1-x^2)\frac{d^2}{dx^2} p_j(x,\eta=1) =
-(j+1)(j+2)p_j(x,\eta=1)\,.
\end{equation}

The expressions for the conformal partial waves in Eqs.\
(\ref{Def-p-all}), (\ref{Def-p-P}), and (\ref{Def-p-Q}), turn out
to be in agreement with a representation, given recently in Ref.\
\cite{ManKirSch05}, in terms of associated Legendre functions for
the gluonic GPDs. In fact, up to some normalization factors and
shift in both the conformal spin and the index,  the same
functions appear in the central and outer region. For the former
one we, however, prefer to work with complex conformal spin, too.
This incorporates  the underlying duality between central and
outer region in a manifest manner and yields a uniform
representation of scattering amplitudes, see below Sect.
\ref{Sec-RepAmp}. This is essential for the perturbative QCD
analysis at higher orders.

The analytic continuation of the polynomials $m_n(\eta,\Delta^2)$
is denoted as $m_j(\eta,\Delta^2)$. These functions will be also
analytic in $\eta$, however, might have branch points at $\eta=0$,
$\eta=\pm 1$, and $\eta=\infty$. It would be desirable to have an
integral representation that makes this property transparent and
might allow  the continuation from $\eta\geq 1$ to
$\eta\leq 1$ or even to negative values. Moreover, we will also
require that the moments $m_j(\eta,\Delta^2)$ are bounded at large
$j$. It turned out that with these requirements the analytic or
numerical calculation of moments from a given GPD is a rather
intricate task. We have to admit that this mathematical problem is
also not solved here for any conceivable GPD. In the following we
give, however, some recipes to evaluate the conformal moments  for
complex conformal spin in the region $|\eta| \leq 1$.

To derive an appropriate integral representation, we decompose here,
in contrast to Eq.\ (\ref{DefConMomGPD}), the conformal moments
into contributions that arise from the outer and
the central region:
\begin{eqnarray}
\label{Def-m-j}
m_n(\eta,\Delta^2) = \mu^{\rm cen}_n(\eta,\Delta^2) +
\mu^{\rm out}_n(\eta,\Delta^2)\,.
\end{eqnarray}
Again polynomiality is only manifest for the sum but not for the
separate terms on the r.h.s. The analytic continuation of
$\mu^{\rm out}_n(\eta,\Delta^2)$ is defined in terms of
hypergeometric functions (\ref{Def-cHyp})
\begin{eqnarray}
\mu^{\rm out}_j(\eta,\Delta^2) = \int_{\eta}^1\! dx\, c_j(x,\eta)\,
\left[\omega(x,\eta,\Delta^2) +\omega(x,-\eta,\Delta^2)  \right]\,.
\end{eqnarray}
From the asymptotics of the hypergeometric functions
\begin{eqnarray}
c_j(x,\eta) \sim  \left(\frac{x+ \sqrt{x^2-\eta^2}}{2}\right)^j
\quad\mbox{for}\quad
j\to\infty\,, \quad |{\rm arg}(j)| \leq \pi/2\,,
\end{eqnarray}
we can estimate the large $j$-behavior of conformal moments
\begin{eqnarray}
\label{Est-larj-out}
\mu_j^{\rm out}(\eta,\Delta^2) \sim
\int_{\eta}^1\! dx\, \left(\frac{x+ \sqrt{x^2-\eta^2}}{2}\right)^j
\left[\omega(x,\eta,\Delta^2) +\omega(x,-\eta,\Delta^2) \right]\,.
\end{eqnarray}
{F}or $0 <\eta < 1 $  the integrand and thus  the conformal
moments are exponentially suppressed for $j\to\infty$ with  $|{\rm
arg}(j)| < \pi/2$. In the limit $\eta =0$ we arrive at the parton
densities and get the power like suppression factor $j^{-p}$.

The contribution from the central region reads for integer values $n$
\begin{eqnarray}
\label{Def-mu-cen}
\mu_n^{\rm cen}(\eta,\Delta^2) =
\int_{-\eta}^\eta\! dx\, c_n(x,\eta)\, \omega(x,\eta,\Delta^2) \,.
\end{eqnarray}
The analytic continuation of the conformal spin, as done for the
central region above, would yield terms proportional to $\sin(\pi
j)$ that exponentially grow at large $\Im{\rm m}\, j$. The
presence of such terms can also be read off from the integral
(\ref{Def-Ort-Cen}), in the appendix, together with our final GPD
representation (\ref{Rep-F-MelBar}), given below. In the case that
this integral can be evaluated in an analytic form for integer $n$
these terms drop out and the analytic continuation can be
performed by means of the substitution $n\to j$. We remark that
the explicit knowledge of $\mu^{\rm out}_j(\eta,\Delta^2)$ allows
to restore the contribution  from the central region by means of
the polynomiality condition. Obviously, within such a procedure
one might miss some terms that satisfy the  polynomiality
condition, but  contribute only in the central region.
Nevertheless, this procedure  offers in principle a way to restore
the polynomiality of a GPD that is only known in the outer region,
e.g., from an ansatz within the overlap representation
\cite{DieFelJakKro00}.

Another possibility for the analytic continuation of the conformal spin,
appropriate for numerical calculations, arises from the definition of the
conformal moments in the central region by the following contour integral
(see Fig \ref{FigCon10}):
\begin{figure}[t]
\begin{center}
\mbox{
\begin{picture}(600,90)(0,0)
\put(100,-15){\insertfig{4}{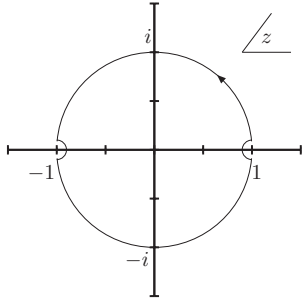}}
\end{picture}
}
\end{center}
\caption{\label{FigCon10}
The integration contour in Eq. (\ref{Def-mu-cen1}).}
\end{figure}
\begin{eqnarray}
\label{Def-mu-cen1}
\mu_n^{\rm cen}(\eta,\Delta^2) = \frac{\eta^{n+1} }{2  i}
 \oint_{-1+\epsilon}^{1-\epsilon}\! dz\, d_n(z)\,
 \omega(z \eta ,\eta,\Delta^2) \,.
\end{eqnarray}
Here the function $d_n(z)$ is defined in the complex plane by the integral
\begin{eqnarray}
\label{Def-d}
d_n(z) = \frac{1}{\pi} \frac{1}{(1-z^2)}
\int_{-1}^1\! dx\, \frac{(1-x^2) c_n(x,1)}{x-z}.
\end{eqnarray}
It has a branch cut on the real axis in the interval $[-1,1]$ and
in addition single poles at $x=\{-1,1\}$. Thus, we excluded in the
integration contour of the integral (\ref{Def-mu-cen1}) the points
$x=\{-1,1\}$. Obviously, inserting Eq.\ (\ref{Def-d}) into Eq.\
(\ref{Def-mu-cen1}) yield by means of the residue theorem and
taking  the limit $\epsilon \to 0$ to the original definition of
the conformal moments (\ref{Def-mu-cen}). The function $d_n(z)$
can be expressed through Legendre functions of the second kind and
an appropriate continuation of the conformal spin $n$ is given in
terms of a hypergeometric function
\begin{eqnarray}
\label{Def-d-com} d_j(x)= \frac{2^{-2j-2}
\Gamma(1+j)\Gamma(3+j)}{\Gamma(3/2+j)\Gamma(5/2+j)}
\frac{x^{-j-1}}{x^2-1}\; {_2\!F}_1\!\left({(j+1)/2,(j+2)/2\atop
5/2+j}\Big|\frac{1}{x^2}\right)\,.
\end{eqnarray}
The function $d_j(x)$ behaves at large $x$ as $x^{-j-3}$ and has a
pole at $x=\pm 1$. At large $j$ it is bound in the whole region
$1\leq x\leq\infty$ by
\begin{eqnarray}
d_j(x,\eta) \sim 2^{-j}  \frac{x^{-1-j}}{x^2-1}\,.
\end{eqnarray}
Let us suppose that the function $\omega(x\eta, \eta)$ is a
regular function that has only a branch cut along the real axis
$[-\infty,-1]$ and has a power like behavior at infinity. We can
expand the upper and lower part of the integration contour in Fig.
\ref{FigCon10} to semicircles encompassing the complete
half-planes such that the real axis is pinched in the intervals
$[-\infty,-1]$  and  $[1,\infty]$.  For a certain value of
$\Re{\rm e}\, j$, the contributions at infinity are negligible and
so we pick up the discontinuity along the negative real axis and
the pole at $x=1$. This result serves now for the definition of
the analytic continuation in $j$:
\begin{eqnarray}
\label{Def-mu-cen2} \mu_j^{\rm cen}(\eta,\Delta^2|\sigma)
&\!\!\!=\!\!\!& \sigma  \eta^{j+1} \int_{-\infty}^{-1}\! dx\,
d_j(-x)\, \frac{1}{2i}\left[\omega(x \eta - i \epsilon ,\eta,\Delta^2)+
\omega(x \eta + i \epsilon,\eta,\Delta^2)\right] \nonumber
\\
&&+ \left(\frac{\eta}{2}\right)^{j+1}
 \frac{\Gamma(1/2) \Gamma(1+j)}{\Gamma(3/2+j)} \omega(\eta ,\eta,\Delta^2)
\,.
\end{eqnarray}
Our choice of $\Re{\rm e}\, j$ guarantees the convergence of the
integral and, moreover, we assumed here that
$\omega(x=-\eta,\eta)$ vanishes. The choice of $\Re{\rm e}\, j$
can be relaxed by introducing appropriate subtraction terms that
improve the power behavior of $\omega(\eta x,\eta)$ at $x=\infty$.
On the r.h.s.\ of Eq.\ (\ref{Def-mu-cen2}) the last term arises
from the pole at $x=1$ and the first term contains a phase factor
$\sigma = e^{\pm i \pi j}$, that depends on the continuation
procedure from positive to negative real valued $x$. To get rid of
this factor we define for even and odd $n$ the analytic
continuation as
\begin{eqnarray}
\mu_n^{\rm cen}(\eta,\Delta^2)  \to
\left\{ \mu_j^{\rm cen}(\eta,\Delta^2|\sigma=1) \atop
\mu_j^{\rm cen}(\eta,\Delta^2|\sigma=-1)
\right\} \quad\mbox{for} \quad
n= \left\{ {\rm even} \atop {\rm odd} \right.\,.
\end{eqnarray}
Note that the appearance of the phase factor is strongly connected
to the analytic properties of GPDs and is only absent if the GPD
has  no branch point at $x=-\eta$.

Finally, we have for the analytic continuation of the conformal moments
\begin{eqnarray}
m_n(\eta, \Delta^2) \to\left\{ m^{\rm even}_j(\eta, \Delta^2)
\atop
 m^{\rm odd}_j(\eta, \Delta^2)
\right\}
= \mu_j^{\rm out}(\eta,\Delta^2) +
\left\{ \mu_j^{\rm cen}(\eta,\Delta^2|\sigma=1) \atop
\mu_j^{\rm cen}(\eta,\Delta^2|\sigma=-1)
\right\} \quad\mbox{for}
\quad n= \left\{ {\rm even} \atop {\rm odd} \right.\,.
\end{eqnarray}
It is obvious that $m_j^{\rm even}$  and $m_j^{\rm odd}$ lead only
for even and odd integer values of $j=n$ to polynomials, which are
even in $\eta$. The difference of even and odd moments arises only
from the central region and is expressed by a function in $\eta$
which does not degenerate into a polynomial for integer values of
$j=n$.

\subsubsection{GPDs represented as Mellin-Barnes integral}
\label{SubSubSec-MelBarInt}

Now  we like to change  the integration contour in Eq.\
(\ref{Def-IntParWav}), in such a way that it becomes parallel to
the imaginary axis in the $j$-plane, i.e., extends from $c-i\infty$
to $c+i\infty$. Here the constant $c <0$ is chosen such that all
singularities contained in the conformal moments
$m_j(\eta,\Delta^2)$ are on the left hand side of the integration
path. Let us assume for the moment that $m_j(\eta,\Delta^2)$
degenerates for all non-negative integer values of $j$ to
polynomials. As displayed in Fig.\ \ref{FigCon1} (a), we add two
quarters of circles in the first and fourth quadrant so that the
integration contour includes the imaginary axis and is closed by
the arc that includes the points $c+i\infty$, $\infty$, and
$c-i\infty$.

It remains to show that within our definition of
conformal moments the contribution from the arc vanishes.
Suppose that $\eta\geq 1$ and that the variable $x$ is rescaled by
$\eta$, i.e., $x=\eta X$. Hence, we must study the behavior of
$p_j(\eta X, \eta)$ for $j\to \infty$ in the interval $|X|\leq 1$.
The asymptotic expansion of the partial waves for large
$j$ with $|{\rm arg}(j)| \leq \pi/2 $ can be read off from the
behavior of hypergeometric functions \cite{Luk69} and  is
\begin{eqnarray}
\label{Est-Cir0}
p_j(X \eta,\eta)\sim
\left(\frac{2}{\eta}\right)^j \left[
 e^{j\; {\rm arccosh}(-X \pm i\epsilon)} \pm i\,
 e^{-(j+3)\; {\rm arccosh}(-X \pm i\epsilon)}\right)
\quad\mbox{for}\quad |X| \leq 1\,.
\end{eqnarray}
For $-1\leq X \leq 1$  the function ${\rm arccosh}(-X\pm i \epsilon)$ has
a monotonously  increasing or decreasing imaginary part that lies
in the interval $[0,\pi]$ and  $[-\pi,0]$ for the $+i\epsilon$ and $-i\epsilon$
prescription, respectively. Thus, within both prescriptions
\begin{eqnarray}
  \frac{1}{\sin(\pi j)}  p_j(\eta X,\eta) m_j(\eta,\Delta^2)
\end{eqnarray}
exponentially vanishes for $-1< X< 1$ on the arc, specified above,
as long as $m_j(\eta,\Delta^2)$ behaves for $\eta>1$ as
\begin{eqnarray}
\label{Bou-ConMom}
m_j(\eta,\Delta^2) \sim  \left(\frac{\eta}{2}\right)^j\quad
\mbox{for}\quad j\to \infty.
\end{eqnarray}
This remains true, if we replace $p_j(x,\eta)$ by the symmetric
and antisymmetric combinations
\begin{eqnarray}
\label{Def-pSym} p^{\rm even}_j(x,\eta) = \frac{1}{2}\left[
p_j(x,\eta) + p_j(-x,\eta)\right],\qquad p^{\rm odd}_j(x,\eta) =
\frac{1}{2}\left[ p_j(x,\eta) -  p_j(-x,\eta)\right]\,
\end{eqnarray}
and $m_j(\eta,\Delta^2)$ by the corresponding even and odd moments,
respectively. Since $p_n(x,\eta)=(-1)^n p_n(-x,\eta)$ has definite
symmetry, the  residues of
\begin{eqnarray}
\frac{1}{\sin(\pi j)} p^{\rm even }_j(x,\eta)  m^{\rm
even}_j(\eta,\Delta^2) \quad
  \mbox{and}\quad
 \frac{1}{\sin(\pi j)}  p^{\rm odd }_j(x,\eta) m^{\rm odd}_j(\eta,\Delta^2)
\end{eqnarray}
only contribute for even  and odd values of $j=n$, respectively,
and, thus, the polynomiality is implemented in a manifest manner.

Unfortunately, we did not give in the previous section a
representation for $m_j(\eta,\Delta^2)$ that allows the
analytic continuation in $\eta$ to the region $\eta > 1$ and
simultaneously satisfies the requirement that $m_j(\eta,\Delta^2)$
fulfills the  bound (\ref{Bou-ConMom}).
This can be read off from Eq.\ (\ref{Est-larj-out}), where obviously,
we will pick up an additional phase.
It seems that such a representation can not easily be found
rather we encounter here similar problems as for the continuation
to the region $\eta < 0$.

This difficulty can be avoided once we replace  in
$m_j(\eta,\Delta^2)$ the variable $\eta$ by $\eta^\ast$ with
$\eta^\ast \leq 1$. It is sufficient to consider the cross-over
point, since for all other $x$ values inside the central region we
will have an additional exponential suppression due to the phases
of $p_j(\eta,\eta)$, see Eq.\ (\ref{Est-Cir0}). For $X=1$ we find
from Eq.\ (\ref{Valpj-Cros}) that $p_j(\eta,\eta)$  behaves on the
cross-over point as $\sim (2/\eta)^j \sin(\pi j)/\sqrt{j}$. To get
rid of the exponential growth, induced by the real part of $j$, we
might choose $\eta >2$ and can arrange in this way even an
exponential suppression. If now  the conformal moments
$m_j(\eta^\ast,\Delta^2)$ for given $\eta^\ast$ do not grow faster
than $(\eta/2)^j/\sqrt{j}$ for $j\to \infty$, the  integral on the
infinite arc does not contribute for $ x= 1$, too. Thus, after
appropriate scaling of $x$ with $\eta$, analytic continuation to
the region $0\leq \eta \leq 1$, and setting $\eta^\ast=\eta$ we
find for all values $|x|\leq \eta$ the following Mellin--Barnes
integral representation for GPDs:
\begin{eqnarray}
\label{Rep-F-MelBar} q(x,\eta,\Delta^2) = \frac{i}{2}\int_{c-i
\infty}^{c+i \infty}\!dj\, \frac{1}{\sin( \pi j)} p_j(x,\eta)\,
m_j(\eta,\Delta^2) \,.
\end{eqnarray}
Finally, we extend this integral into the outer region. For
$p_j(x,\eta)$ this is done by means of  the definition
(\ref{Def-p-all}).

The Mellin--Barnes integral (\ref{Rep-F-MelBar}) can be used when
the analytic continuation of even and odd moments leads to the
same function $m_j(\eta,\Delta^2)$. If this is not the case, we
separately introduce the Mellin--Barnes integral for the even and
odd part of
\begin{eqnarray}
\label{Dec-GPD-EveOdd}
 q(x,\eta,\Delta^2) =
q^{\rm even}(x,\eta,\Delta^2)+q^{\rm odd}(x,\eta,\Delta^2)
\end{eqnarray}
Since the representation (\ref{Rep-F-MelBar}) remains valid if we
substitute $p_j(x,\eta)$ by the symmetrized partial waves $p^{\rm
even}_j(x,\eta)$ and $p^{\rm odd}_j(x,\eta)$, we have the same
form of the Mellin--Barnes integral as before:
\begin{eqnarray}
\label{Rep-F-MelBarSym} q^{\rm even/odd}(x,\eta,\Delta^2) =
\frac{i}{2}\int_{c-i \infty}^{c+i \infty}\!dj\, \frac{1}{\sin( \pi
j)} p^{\rm even/odd}_j(x,\eta)\, m^{\rm
even/odd}_j(\eta,\Delta^2)\,.
\end{eqnarray}
Bearing in mind that $p_j(x,\eta)$ is set to zero for $x < -\eta$,
the extension of $p^{\rm even}_j(x,\eta)$ and $p^{\rm
odd}_j(x,\eta)$, defined in Eq.\ (\ref{Def-pSym}) in terms of
$p_j(x,\eta)$ and $p_j(-x,\eta)$, is consistently done by the
procedure (\ref{Def-p-all}). The original GPD
(\ref{Dec-GPD-EveOdd}) is restored by adding its even and odd
parts:
\begin{eqnarray}
 \label{RepGPD-Even-Odd}
q(x,\eta,\Delta^2) = \frac{i}{2}\int_{c-i \infty}^{c+i
\infty}\!dj\, \frac{1}{\sin(\pi j)} \left[ p^{\rm
even}_j(x,\eta)\, m^{\rm even}_j(\eta,\Delta^2)  +  p^{\rm
odd}_j(x,\eta)\, m^{\rm odd}_j(\eta,\Delta^2)\right]\,.
\end{eqnarray}

The extension into the outer region  is not based on analytic
continuation and the reader might wonder whether  this
prescription (\ref{Def-p-all}) is indeed correct. Besides the
arguments we gave in Sect.\ \ref{SubSubSec-SomWatTra}, we provide
next some further support. Let us first verify that the GPD
representation (\ref{RepGPD-Even-Odd}) has the correct support
property, i.e., it contributes only for $-\eta < x$. As mentioned
above, the difference of odd and even moments arise only from the
central region and has thus the functional form $(\eta/2)^{j+1}
f_j(\eta,\Delta^2)$, where $f_j(\eta,\Delta^2)$ has a power-like
behavior at $j\to \infty$, see Eq.\ (\ref{Def-mu-cen2}). We expect
that such a term does not contribute to the outer region. Indeed
this can be read off from the Mellin--Barnes integral, which takes
the form
\begin{eqnarray}
\frac{i}{2}\int_{c-i
\infty}^{c+i \infty}\!dj\, \frac{2^{-j-1}}{\sin(\pi j)} Q_j(x/\eta)
f_j(\eta,\Delta^2)\,,
\nonumber
\end{eqnarray}
see definitions (\ref{Def-p-all}) and (\ref{Def-p-Q}). For $x
>\eta$ the term $2^{-j-1} Q_j(x/\eta)/\sin(\pi j)$ contains an
exponential damping factor $(\eta/x)^j$ for $j\to\infty$ with
$|\arg(j)| < \pi/2$, while the remaining factor in the integrand
is bounded. So we can close the integration contour by an infinite
arc that includes the first and forth quadrant. Since the whole
integrand is analytic in these two quadrants, Cauchy theorem gives
zero. Note that $Q_j(x/\eta)/\sin(\pi j)$ contains no
singularities on the real positive axis. Consequently, even and
odd contributions are the same in the outer region and will add
for $x> \eta $, while they cancel for $x < -\eta $.

Alternatively,  we can represent the GPD as
\begin{eqnarray}
\label{Rep-GPD} q(x,\eta,\Delta^2) = \frac{i}{2}\int_{c-i
\infty}^{c+i \infty}\!dj\, \frac{1}{\sin(\pi j)} \left[\theta(|x|
\leq \eta) p^{\rm}_j(-x,\eta)  m_j^\Delta(\eta,\Delta^2) +
p^{\rm}_j(x,\eta) m_j^\Sigma(\eta,\Delta^2) \right]\,,
\end{eqnarray}
where $ m_j^\Sigma= (m_j^{\rm even}+ m_j^{\rm odd})/2 $ and $
m_j^\Delta= (m_j^{\rm even}- m_j^{\rm odd})/2 $. Here the
polynomiality is not manifest in the conformal moments, rather it
is separately restored  for $m_{2n}^\Sigma + m_{2n}^\Delta $  and
$m_{2n+1}^\Sigma - m_{2n+1}^\Delta $.

Let us next show that the Mellin-Barnes integral vanishes for
$|x|>1$.  Here the expression $\eta^{-j-1} Q_j(x/\eta)/\sin(\pi
j)$ in the integrand vanishes on the infinite arc, due to the
damping factor $(1/x)^j$, see Eqs.\ (\ref{Def-p-all}) and
(\ref{Def-p-Q}). As in the previous paragraph, we can close the
integration contour and find that the integral gives zero. If we
would have allowed an additional term proportional to ${\cal
P}_j(-x,\eta)$ in the extension of the support for $x>\eta$, it
would in general contribute in the unphysical region $x>1$, too.

Finally, the correctness of the GPD representations as
Mellin-Barnes integral is deduced from the fact that for
non-negative integers the conformal moments are reproduced.
Employing  the symmetry with respect to $x\to -x$, we find  from
the representation (\ref{RepGPD-Even-Odd}), e.g.,
for even moments,
\begin{eqnarray}
\label{PerConMom} \int_{-\infty}^\infty\!dx\, c_n(x,\eta)
q(x,\eta,\Delta^2) =  \int_{-\eta}^\infty \!dx\, c_n(x,\eta)
\frac{i}{2 }\int_{c-i
\infty}^{c+i \infty}\!dj\, \frac{1}{\sin(\pi j)} p_j(x,\eta)\,
m^{\rm even}_j(\eta,\Delta^2)
\end{eqnarray}
for $n=0,2,\cdots$. {F}rom the moments separately calculated in
the outer and central region, given in Eqs.\ (\ref{Def-Ort-Out})
and (\ref{Def-Ort-Cen}), respectively, it follows\footnote{The
$x$-integral over the outer region only exist for $\Re{\rm e}\, j
> n$. Since here the integrand has no singularities on the
positive real axis, we can first shift the Mellin--Barnes
integration path to the imaginary axis to the right. Then with
$\Re{\rm e}\, j = n+\epsilon $ the $x$-integration is performed.
On the other hand the $x$-integration over the central region must
be performed within the original contour. It produces a $\sin(\pi
j)$ term, which removes the poles on the real axis and so we can
shift the Mellin--Barnes integration path along the positive real
axis, too. Both remaining integrals can then be combined in the
limit that is indicated in Eq.\ (\ref{PerConMom-fin}). }
\begin{eqnarray}
\label{PerConMom-fin} \int_{-\infty}^\infty\!dx\, c_n(x,\eta)
q(x,\eta,\Delta^2) = \lim_{\epsilon\to +0} \frac{1}{2 i
\pi}\int_{n-i \infty}^{n+i\infty}\!dj\,  \left(
\frac{{\cal N}_{nj}(\eta)}{n-j+\epsilon} +
\frac{{\cal N}_{nj}(\eta)}{j-n+\epsilon} \right)
 m_j^{\rm even}(\eta,\Delta^2) \,,
\end{eqnarray}
for $n=0,2,\cdots$. The integration path can be  parameterized  by
$j=n+ i \lambda$ and making use of the identity $1/(\lambda-i\epsilon ) -
1/(\lambda+i\epsilon )= 2 i \pi \delta(\lambda)$, the integral yields
finally  the conformal moment $m_n^{\rm even}(\eta,\Delta^2)$ for
$n=0,2,\cdots$. Analogously, the conformal moments $m_n^{\rm
odd}(\eta,\Delta^2)$ arise for odd values of $n$.

We complete this section with the Mellin--Barnes representation
for GDAs, which follows now by crossing from Eq.\
(\ref{Rep-F-MelBar}) with the constrain $z \geq \zeta$, which
arises from  $-1/\eta  \leq x/\eta \leq 1$. Employing the symmetry
transformation $z\to 1-z$ and $\zeta\to 1-\zeta$ we get the
remaining contribution, see Eq.\ (\ref{SupGDA}):
\begin{eqnarray}
\label{Rep-GDA-MelBar} \Phi(z,\zeta,W^2) = \frac{i}{2}\int_{c-i
\infty}^{c+i \infty}\!dj\, \frac{1}{\sin( \pi j)} \left[
\theta(z-\zeta ) p_j(1-2z,1)M_j(\zeta,W^2) + \left\{ {z\to 1-z
\atop \zeta\to 1-\zeta }      \right\} \right]  \,,
\end{eqnarray}
where $M_j(\zeta,W^2)$ is the analytic continuation of the
conformal GPD moments, obtained by crossing, see below Eq.\
(\ref{ConParWavGDA}). If $M_j(\zeta,W^2)$ is bounded along the
integration path for  $0\leq \zeta\leq 1$, the integral of the
first term in the square brackets in this formula exist for all
values of $0\leq z\leq 1$. So we can drop both the restriction
$\theta(z-\zeta )$ and the second term in the square brackets,
obtained by symmetry. This latter term follows now from analytic
continuation and so we arrive at the representation:
\begin{eqnarray}
\label{Rep-GDA-MelBar-uni} \Phi(z,\zeta,W^2) =
\frac{i}{2}\int_{c-i \infty}^{c+i \infty}\!dj\, \frac{1}{\sin( \pi
j)} p_j(1-2z,1)M_j(\zeta,W^2)  \,.
\end{eqnarray}
Suppose that $M_j(\zeta,W^2)$ vanishes for $j\to \infty$ with
$|{\rm arg} j| \leq \pi/2 $, it is straightforward to see that the
conformal moments for non-negative integer conformal spin are
reproduced. Employing Eq.\ (\ref{Def-Ort-Cen}) with $\eta=1$ and
$k=n=\{0,1,2,\cdots\}$, we can then close the contour so that it
now encircles the positive real axis. The residue theorem yields
then $M_n(\zeta,W^2)$.

\section{Evolution kernels and coefficient functions in a manifest conformal  scheme}
\label{Sec-CS-Sch}

To LO accuracy the evolution kernels and coefficient functions for
exclusive processes respect conformal symmetry, which is the
symmetry of the QCD Lagrangian at the classical level for massless
quarks. Conformal symmetry find, in fact,  many practical
applications in QCD. It allows for instance to solve the mixing
problem of light-ray operators, caused by renormalization, and by
means of the conformal operator product expansion it predicts the
hard-scattering amplitude. The breaking of conformal symmetry
beyond LO has been studied with perturbative methods in great
detail \cite{Mue94,Cre97,Mue97a,BelMue97a,BelMue98a,BelMue98c}.
The main lesson from these studies is that the only physical
contributions violating conformal symmetry are generated by the
non-zero $\beta$ function. All other terms violating conformal
symmetry (in the perturbative sector) are mainly artifacts
introduced  by the standard renormalization/factorization
prescription for the light-ray operators, which is based on some
version of minimal subtraction within the dimensional
regularization scheme. Within the standard scheme the conformal
symmetry is separately broken in the evolution kernels and the
hard-scattering amplitude for two-photon processes.
However, it can be restored by a finite renormalization providing
a scheme in which conformal symmetry is manifest, except for terms
proportional to $\beta$ that are induced by the trace anomaly of
the energy momentum tensor. Thus, we will restrict us  in the
following two subsections not to LO accuracy. Rather, the results
can be (at least partly) applied to all orders of perturbation
theory. Of course, terms proportional to $\beta$ need special
consideration. A first discussion of this issue can be found in
Ref.\ \cite{MelMuePas02}, see also Ref.\ \cite{BroGabKatLu95}.

\subsection{Convolution of GPDs with conformal kernels}
\label{SubSec-Con}

In this section we consider the convolution
\begin{eqnarray}
\label{DefCon}
K\otimes q(x,\eta)\equiv \int_{-\infty}^\infty\!\frac{dy}{|\eta|}\,
K\left(\frac{x}{\eta},\frac{y}{\eta}\right) q(y,\eta,\Delta^2)
\end{eqnarray}
of a GPD with a generic kernel $K(x,y)$ that respects conformal
symmetry. Here the integration region is defined by the combined
restrictions for the  GPDs and the kernels. The requirement of
conformal symmetry means in fact that the kernel has for $|x|,|y|
\leq 1$ the following spectral representation
\begin{eqnarray}
\label{SpeRepERBLK} K(x,y) = \sum_{n=0}^\infty    (-1)^n p_n(x) k_n
c_n(y) \quad\mbox{for}\quad  |x|,|y| \leq 1,
\end{eqnarray}
where $k_n$ are the eigenvalues of $K(x,y)$ and the polynomials
$p_n(x)$ and $c_n(y)$ are defined in Eqs.\ (\ref{Def-c}) and
(\ref{Def-pn}), respectively. Suppose that even and odd
eigenvalues have the same sign and provide after analytic
continuation with respect to the conformal spin the same
holomorphic function $k_j$ so that it satisfies a bound for all
values of $j$ with ${\rm arg}(j) \leq \pi/2$.  In fact the
eigenvalues $k_n$ coincide with the Mellin moments of the DGLAP
kernel, given by rational functions and harmonic sums, and so we
know their analytic continuation. We remark that the support in
the $(x,y)$-plane is defined in Eq.\ (\ref{Str-Ker-K}) and the
extension to the full region is unique, see Appendix
\ref{App-MelBar-Ker}. The convolution of a GPD confined to
$[-\eta,1]$ with this kernel leads to a function that has also the
support $-\eta \le x\le 1$.

To proceed in the simplest possible manner let us extend the
representation (\ref{SpeRepERBLK}) to the whole region by the
series
\begin{eqnarray}
\label{SpeRepERBLK-Dis}
\frac{1}{\eta}
K\left(\frac{x}{\eta},\frac{y}{\eta}\right)  = \sum_{n=0}^\infty
(-1)^n p_n(x,\eta) k_n c_n(y,\eta)\,.
\end{eqnarray}
Again, here we understand that $p_n(x,\eta)$ are mathematical
distributions with the restricted support $-\eta\leq x\leq \eta$
for integer values of $n$, while $c_n(y,\eta)$ are the polynomials
(\ref{Def-c}), which can be extended to the whole $y$ region.
Taking the Mellin-Barnes integral (\ref{Rep-F-MelBar}) and  the
spectral representation (\ref{SpeRepERBLK-Dis}), the convolution
(\ref{DefCon}) leads to a divergent series
\begin{eqnarray}
\label{DefConRes} K\otimes q(x,\eta) = \sum_{n=0}^\infty   (-1)^n
p_n(x,\eta) k_n m_n(\eta,\Delta^2)\,.
\end{eqnarray}
Here we  employed the integrals (\ref{Def-Ort-Out})
and (\ref{Def-Ort-Cen}) for discrete conformal spin, as has been
explained at the end of Sect.\ \ref{SubSubSec-MelBarInt}. Since
the analytic continuation of  $k_n$ does not spoil our assumptions
for the Sommerfeld-Watson transform, we can proceed  as in Sect.\
\ref{SubSubSec-MelBarInt}. As expected, for a conformal kernel the
convolution
\begin{eqnarray}
K\otimes q(x,\eta) = \frac{i}{2}\int_{c-i \infty}^{c+i
\infty}\!dj\, \frac{1}{\sin(\pi j)} p_j(x,\eta)\, k_j\,
m_j(\eta,\Delta^2)
\end{eqnarray}
is in the Mellin momentum space given by a multiplication of the
conformal GPD moments with the corresponding eigenvalues.

In those cases in which we must distinguish between even and odd
eigenvalues of the kernel, we write $K$ as
\begin{eqnarray}
\label{SpeRepERBLK-EO}
\frac{1}{\eta} K\left(\frac{x}{\eta},\frac{y}{\eta}\right) =
\sum_{n=0}^\infty  (-1)^n \left[ p_n^{\rm even}(x,\eta) k^{\rm even}_n
c_n(y,\eta) +  p_n^{\rm odd}(x,\eta) k^{\rm odd}_n c_n(y,\eta)\right],
\end{eqnarray}
where $p_n^{\rm even}(x)$ and $p_n^{\rm odd}(x)$ are defined as in
Eq.\ (\ref{Def-pSym}), and employ the GPD representation
(\ref{RepGPD-Even-Odd}). Since even and odd Gegenbauer polynomials
and partial waves have definite symmetry under reflection, they
can not mix. Hence, the convolution leads to
\begin{eqnarray}
K\otimes q(x,\eta) =
 \frac{i}{2}\int_{c-i
\infty}^{c+i \infty}\!dj\, \frac{1}{\sin(\pi j)} \left[ p^{\rm
even}_j(x,\eta)\, k_j^{\rm even}  m^{\rm even}_j(\eta,\Delta^2)  +
\left\{ {\rm even} \to {\rm odd} \right\}
 \right]\,.
\end{eqnarray}
If the support of the GPD was in the interval $[-\eta,1]$ and the
eigenvalues  are different in the even and odd sector, the
convolution certainly gives us a function that lives now in the
whole region $[-1,1]$.  The support extension is caused by the
mixing with an anti-quark GPD. With $k_j^{\rm even} = k_j^{\Sigma}
+ k_j^{\Delta}$ and $k_j^{\rm odd} = k_j^{\Sigma} - k_j^{\Delta}$,
we can decompose the convolution as
\begin{eqnarray}
K\otimes q(x,\eta) = \delta q(x,\eta) + \delta \overline{q}(-x,\eta)\,.
\end{eqnarray}
Based on the representation (\ref{Rep-GPD}), we interpret the
terms on the r.h.s.\ as contributions to a quark GPD, with  $-\eta
\leq x \leq 1$,
\begin{eqnarray}
\delta q(x,\eta)=
 \frac{i}{2}\int_{c-i
\infty}^{c+i \infty}\!dj\, \frac{k_j^{\Sigma}}{\sin(\pi j)}
\left[\theta(|x| \leq \eta) p^{\rm}_j(-x,\eta)
m_j^\Delta(\eta,\Delta^2) + p^{\rm}_j(x,\eta)
m_j^\Sigma(\eta,\Delta^2) \right]\,
\end{eqnarray}
and an anti-quark GPD, with $-1 \leq -x \leq \eta$,
\begin{eqnarray}
\delta \overline{q}(-x,\eta)=
 \frac{i}{2}\int_{c-i
\infty}^{c+i \infty}\!dj\, \frac{k_j^{\Delta}}{\sin(\pi j)}
\left[\theta(|x| \leq \eta) p^{\rm}_j(x,\eta)
m_j^\Delta(\eta,\Delta^2) +  p^{\rm}_j(-x,\eta)
 m_j^\Sigma(\eta,\Delta^2) \right] \,.
\end{eqnarray}

For the convolution of a conformal kernel with a GDA we would
obtain the analogous results. The convolution is conventionally
defined as
\begin{eqnarray}
\label{Def-ConParWavGDA}
K \displaystyle{ \mathop{\otimes}^{\rm e}}
\Phi(z,\zeta,W^2) = \int_{0}^1\! dy\,
K(1-2z, 1-2y) \Phi(y,\zeta,W^2)\,.
\end{eqnarray}
Note that the change of variable $(1-2y)\to y$, induces a factor
$2$ in comparison to the definition (\ref{DefCon}). We can now
represent the GDA by the convergent series (\ref{ConParWavGDA}),
and the convolution immediately leads to
\begin{eqnarray}
\label{Res-ConParWavGDA}
K \displaystyle{ \mathop{\otimes}^{\rm e}}\Phi(z,\zeta,W^2) =
\sum_{n=0}^\infty (-1)^n
p_n(1-2z,1) \frac{k_n}{2}  M_n(\zeta,W^2)\,.
\end{eqnarray}

Within the Mellin-Barnes representation  the solution of the
evolution equation is a trivial task to LO accuracy, where the
conformal symmetry is manifest in any scheme. The restoration of
this symmetry is well understood to NLO and even the terms
proportional to $\beta$ can be diagonalized with respect to
conformal partial waves. Relying on this symmetry and borrowing
the eigenvalues of the evolution kernels from the Mellin moments
of the DGLAP kernel from Ref.\ \cite{MocVerVog04,VogMocVer04} one
can even proceed to NNLO.

The simplest example is given by the flavor non-singlet sector and
LO accuracy. Here the evolution equation reads
\begin{eqnarray}
\mu \frac{d}{d\mu} q(x,\eta,\Delta^2,\mu^2) =
 -\frac{\alpha_s(\mu)}{2\pi} \gamma^{(0)} \otimes
 q(x,\eta,\Delta^2,\mu^2)\,,
\end{eqnarray}
where the evolution kernel is
\begin{eqnarray}
\gamma^{(0)}(x,y) = \left[\Theta(x,y) \frac{1+x}{1+y}
\left(1+ \frac{2}{y-x}\right) +
\Theta(-x,-y) \frac{1-x}{1-y}
\left(1+ \frac{2}{x-y}\right)\right]_+\,,
\end{eqnarray}
with the +-prescription $[K(x,y)]_+ = K(x,y) - \delta(x-y) \int dz
K(z,y) $ and the shorthand notation
\begin{eqnarray*}
\Theta(x,y) = {\rm sign}(1+y) \theta\left(\frac{1+x}{1+y}\right)
\theta\left(\frac{y-x}{1+y}\right).
\end{eqnarray*}
Its eigenvalues can be simply calculated for discrete conformal
spin $n$. If one did this for complex conformal spin $j$, one
would encounter the same problems as for moments of GPDs. Namely,
terms proportional to $\sin(\pi j)$ would appear. The analytic
continuation of the discrete eigenvalues can, however, be defined
without such terms:
\begin{eqnarray}
\gamma_j^{(0)} = C_F \left(
4\psi(j+2)- 4\psi(1)- 3 -\frac{2}{(j+1)(j+2)}
\right)\,,
 \quad C_F= \frac{4}{3}\,,
\quad \psi(z) = \frac{d}{d z}\ln\Gamma(z)\,.
\end{eqnarray}
This is just the forward anomalous dimensions of twist-two
operators, well-known from deep inelastic scattering. The
evolution of the conformal moments is thus governed by the
ordinary differential equation
\begin{eqnarray}
\mu \frac{d}{d\mu} m_j(\eta,\Delta^2, \mu^2) =
 - \frac{\alpha_s(\mu)}{2\pi} \gamma_j^{(0)} m_j(\eta,\Delta^2, \mu^2)\,,
\end{eqnarray}
which is easily solved. Equating the renormalization scale $\mu$
with the resolution scale ${\cal Q}$ and inserting the solution
into the Mellin-Barnes integral (\ref{Rep-GPD}) leads to
 \begin{eqnarray}
 \label{Sol-EvoEqu-LO}
q(x,\eta,\Delta^2,{\cal Q}^2) = \frac{i}{2}\int_{c-i \infty}^{c+i
\infty}\!dj\, \frac{1}{\sin(\pi j)} p_j(x,\eta)\,
\exp\left\{-\frac{\gamma_j^{(0)}}{2} \int_{{\cal Q}_0^2}^{{\cal
Q}^2}\frac{d\sigma}{\sigma} \frac{\alpha_s(\sigma)}{2\pi} \right\}
m_j(\eta,\Delta^2,{\cal Q}^2_0)\,,
 \end{eqnarray}
where the moments $m_j(\eta,\Delta^2,{\cal Q}^2_0)$ belongs to a
given GPD at the input scale ${\cal Q}_0$.

\subsection{Mellin-Barnes representation for amplitudes}
\label{Sec-RepAmp}

We study next  the convolution of a GPD with a given
hard-scattering amplitude. To have a concrete example at hand we
deal here with the so-called Compton form factors  that appear in
the perturbative description of DVCS:
\begin{eqnarray}
\label{Def-ComForFac} {\cal F}(\xi,\Delta^2,{\cal Q}^2)  =\!\!
\sum_{p=u,d,s,g}\int_{-1}^1\!\frac{dx}{\xi} \left[
C_p^{(0)\mp}\left(\frac{x}{\xi}\right)  +
\frac{\alpha_s(\mu)}{2\pi}C_p^{(1)\mp}\left(\frac{x}{\xi},\frac{\mu}{\cal
Q}\right) + {\cal O}(\alpha_s^2) \right]
F_{p}(x,\xi,\Delta^2,\mu^2)\,.
\end{eqnarray}
{F}or this process the skewness parameter is equal to the Bjorken
like scaling variable, i.e., $\eta=\xi$. The same kinematic
constraint appears also in the hard exclusive electroproduction of
mesons. In fact, the result for the Compton factors, we will give
below in Eq.\ (\ref{Res-CFF-LO}), can be adopted for this process,
too. This is trivial to LO and requires some additional work
beyond this order.

Let us consider the partonic Compton form factors to LO accuracy,
\begin{eqnarray}
\label{Def-ComForFac-LO} {\cal F}_p (\xi,\Delta^2,{\cal Q}^2)  =
\int_{-1}^1\!dx \left[\frac{Q^2_p}{\xi-x- i  \epsilon}   \mp
\frac{Q^2_p}{\xi+x- i \epsilon} \right] F_{p}(x,\xi,\Delta^2,{\cal
Q}^2)\,,
\end{eqnarray}
where $Q_p$ is the electric charge of the parton. Employing the
decomposition (\ref{Dec-GPDs}) of GPDs into quark and anti-quark
parts, the partonic Compton form factor (\ref{Def-ComForFac-LO})
reads
\begin{eqnarray}
\label{Def-ParCFF-LO}
{\cal F}_p (\xi,\Delta^2,{\cal Q}^2)  = \int_{-\xi}^1\!dx
\left[\frac{Q^2_p}{\xi-x- i \epsilon}   \mp  \frac{Q^2_p}{\xi+x- i \epsilon}  \right]
\left[q_{p}(x,\xi,\Delta^2,{\cal Q}^2) +
\overline{q}_{p}(x,\xi,\Delta^2,{\cal Q}^2) \right]\,.
\end{eqnarray}
Next we employ the Mellin-Barnes representation (\ref{Rep-GPD})
for GPDs and perform the momentum fraction integration by means of Eq.\
(\ref{Def-IntHarAmp1}). Then the Compton form factors are
expressed in terms of the conformal moments
\begin{eqnarray}
\label{Res-CFF-LO}
{\cal F}_p 
 =
\frac{Q^2_p}{2i}\int_{c-i\infty }^{c+i\infty }\!dj\,
\xi^{-j-1} \frac{2^{j+1}\Gamma(5/2+j)}{\Gamma(3/2)\Gamma(3+j)}
\left(i - \frac{\cos(\pi j) \mp 1}{\sin(\pi j)}  \right)
\left[m_j+ \overline{m}_j\right](\xi,\Delta^2,{\cal Q}^2)\,,
\end{eqnarray}
where the sum of $m_j+ \overline{m}_j$ is given by the analytic
continuation of odd and even conformal moments for the vector and
axial-vector case, respectively. As we realize by comparing with
Eq.\ (\ref{Rep-F-MelBar}) and Eq.\ (\ref{Valpj-Cros}) the
imaginary part of ${\cal F}_p$  is $\pi \left[q+
\overline{q}\right](\xi,\xi,\Delta^2)$ as it must be. The real
part of the amplitude, given by a principal value integral in the
momentum fraction representation, contains in the integrand an
additional factor $\tan(\pi j/2)$ and $-\cot(\pi j/2)$ for the
vector and axial-vector case, respectively.

In a conformal subtraction  scheme the inclusion of perturbative
corrections is straightforward. Higher order corrections can be
written as convolution of the LO hard-scattering amplitude
(\ref{Def-ComForFac-LO}) with certain kernels, which are
conformally covariant. Hence, in analogy to the discussion in
Sect.\ \ref{SubSec-Con}, this yields in the Mellin-Barnes
representation  a multiplication with the corresponding
eigenvalues, which are known from deep inelastic scattering. For
instance, to NLO accuracy the partonic form factor ${\cal H}_p$
for quarks in the parity even sector of DVCS on a nucleon target
reads for $\mu={\cal Q}$
\begin{eqnarray}
\label{Def-ComForFacH}
{\cal H}_p 
 &\!\!\! =\!\!\! &
\frac{Q_p^2}{2 i}\int_{c-i\infty }^{c+i\infty }\!dj\, \xi^{-j-1}
\frac{2^{j+1}\Gamma(5/2+j)}{\Gamma(3/2)\Gamma(3+j)}
\left[i+\tan\left(\frac{\pi j}{2}\right)\right] \left[1 +
\frac{\alpha_s({\cal Q})}{2\pi} {\cal C}^{(1)}_j \right]
\left[h_j+ \overline{h}_j\right](\xi,\Delta^2,{\cal Q}^2)
\nonumber\\
\end{eqnarray}
with the NLO coefficients
\begin{eqnarray}
\label{ResNLO-Coe}
{\cal C}_j^{(1)} \!\!\!&=&\!\!\! C_F \Big[S^2_{1}(1 + j) +
\frac{3}{2} S_{1}(j + 2) - \frac{9}{2}  +
\frac{5-2S_{1}(j)}{2(j + 1)(j + 2)} -   S_{2}(j + 1)
\nonumber\\
&&\quad+\frac{1}{2} \gamma_j^{(0)}
\left\{2S_1(2j+3)-S_1(j+2)-S_1(j+1) \right\} \Big]\,,
\end{eqnarray}
where the analytic continuation of harmonic sums is defined by
derivatives of the $\Gamma$ function
\begin{eqnarray}
S_{1}(z) = \psi(z+1)-\psi(1)\,,\quad
S_{2}(z) = - \frac{d}{dz} \psi(z+1) + \frac{\pi^2}{6}\,,
\quad \psi(z)= \frac{d}{dz} \ln\Gamma(z)\,.
\end{eqnarray}
The first line in Eq.\ (\ref{ResNLO-Coe}) is up to an overall
normalization factor the well-known perturbative correction to the
Wilson coefficients of the structure function $F_1$ in deep
inelastic scattering \cite{ZijNee92}, while the addenda in the
second line is induced by the non-forward kinematics. This result
is verified by a direct rotation from the minimal subtraction
scheme to the conformal one and coincides with the prediction of
the conformal operator product expansion \cite{Mue97a,BelMue97a}.

Another advantage of the Mellin-Barnes representation is that it
might be useful for an analytic approximation of the Compton
factors at smaller values of $\xi$, lets say $\xi \lesssim
10^{-2}$,  which should lead to a rather good approximation of the
scattering amplitude for the kinematics in Collider experiments.
Such an approximation has been already studied within the DD
formalism, see \cite{BelMueKir01} and references therein, however,
it remained restricted to the perturbative LO approximation and
only the term containing the leading power in $\xi$ could be
extracted. The main idea here is to shift the integration path in
Eq. (\ref{Def-ComForFacH}) to the left, so that one picks up the
leading order contribution for $\xi\to 0$. Suppose that for the
vector case the first singularity on the l.h.s.\ is a pole at
$j=\alpha_0$ with $-1 < \alpha_0 < c < 1$, which might depend on
$\Delta^2$. We recall that in this case only the analytic
continuation of odd moments enters and, thus, $c$ can also be
chosen to be positive, however, it must be smaller than one. The
integration path  can then be shifted further to the left such
that $-1 < c^\prime < \alpha_0 $ and all other singularities
remain to the  left of the new integration path, while the leading
pole contribution $(j=\alpha_0)$ is explicitly taken  into
account. To LO accuracy the partonic Compton
form factors thus read
\begin{eqnarray}
\label{Shi-IntPat}
{\cal H}_p 
 &\!\!\! =\!\!\! &
Q_p^2\xi^{-\alpha_0-1}
\frac{2^{\alpha_0+1}\Gamma(5/2+\alpha_0)}{\Gamma(3/2)\Gamma(3+\alpha_0)}
\left[i+\tan\left(\frac{\alpha_0 \pi}{2}\right)\right] \pi\, {\rm
Res}\left[h_{j}+ \overline{h}_{j}\right](\xi,\Delta^2,{\cal
Q}^2)\Bigg|_{j=\alpha_0}
\\
 &&\!\!\! +
\frac{Q_p^2}{2i}\int_{c^\prime-i\infty }^{c^\prime+i\infty }\!dj\,
\xi^{-j-1} \frac{2^{j+1}\Gamma(5/2+j)}{\Gamma(3/2)\Gamma(3+j)}
\left[i+\tan\left(\frac{\pi j}{2}\right)\right] \left[h_j+
\overline{h}_j\right](\xi,\Delta^2,{\cal Q}^2)\,. \nonumber
\end{eqnarray}
This result could even be improved by further shifts of the
integration contour. A more
systematic expansion, requires that also the conformal moments
$m_j(\xi, \Delta^2)$ are expanded in $\xi$ in the
vicinity of $\xi=0$. We remind that $\xi$ is related to the
Bjorken variable by $\xi \sim \Bx/(2-\Bx)$ and is  for present
fixed target experiments certainly not larger than $\sim 0.4$.
Hence, already the inclusion of the $O(\xi^2)$ corrections might be
sufficient to obtain a good approximation of the Compton form factors in
this kinematics.

Unfortunately, the method for a systematic approximation, pointed
out above, must maybe be refined due to the following
complication. Namely, we know that in general the conformal
moments will have a branch point at $\xi=0$ and so a Taylor
expansion around this point is not possible, after the analytic
continuation is performed. Even worse, it turns out that the
appropriate analytic continuation, which guarantees the
correctness of the Mellin-Barnes integral, can induce terms that
lead to a behavior $\propto \eta^j $ or even $ \propto \eta^{2j}
$. Certainly, the latter term requires that we shift the
integration path to the right, while the former require a
consideration of all poles. Corresponding to its behavior at
$j\to \infty$, we might close the integration path so that the
positive or negative axis is included. For specific conformal
moments it is probably still possible to obtain a systematic
expansion of the Compton form factors in powers of $\xi$ up to a
certain order. In general, however, this seems to be a serious
problem.

One might naively expect that this issue can be resolved, if one goes back
to the definition of conformal moments (\ref{Def-ConMom}); expand first
$c_n(x,\xi)$  in $\xi^2$  and take then  moments with
respect to $x$, which in turn can be simply continued to
complex valued $j$. Such an expansion looks like, see Eq.\
(\ref{Exp-2Con2SimMom}),
\begin{eqnarray}
\label{Exp-c}
c_j(x,\xi) = x^j \left[c_{j0} + c_{j2} \frac{\xi^2}{x^2} +
c_{j4} \frac{\xi^4}{x^4} + \cdots   \right]\,.
\end{eqnarray}
The coefficients $c_{jm}$ vanish for integer  value $j=n$ with $m
> n$ and so this expansion degenerates then  into a
polynomial\footnote{Obviously, for non-negative integer $n$ we are
dealing with integrals of the type $\int_{-1}^{1}\! dx\, x^{n-m}
q(x,\eta,\Delta^2)$, well defined for $n \geq m$. To avoid
divergencies for complex valued $j$ with $\Re{\rm e}\, j \leq
m-1$, the integral should be defined as a contour integral in the
complex $x$ plane so that it exist and can be viewed as analytic
continuation with respect to the variable $n-m$}\, \cite{Luk69a}.
in $\eta$ of degree $n$. Hence we can shift from the beginning for
each individual term the integration path in the Mellin-Barnes
integral to the r.h.s., i.e., $c\to c+m$. However, one easily
realizes that this procedure will also introduce new poles in the
complex $j$ plane that are also shifted to the right of the real
axis, remaining, however, to the left of the new integration path.
Finally, it turns out that these poles will remove the power in
$\xi^2$ we gained by the expansion (\ref{Exp-c}). With one word,
taking the limit $\xi\to 0$ in general conformal moments, leads to
the correct leading power behavior of the Compton form factors in
$\xi$, however, the normalization might be wrong.

Let us finally comment on the representation for the scattering
amplitude in the case of the production of a hadron pair by photon
fusion. For the important phenomenological situation that one
photon is on-shell, it is of course related to the DVCS amplitude
by crossing. By means of the crossing relation (\ref{CroRel}) the
amplitude follows to LO accuracy from Eq.\
(\ref{Def-ComForFac-LO}):
\begin{eqnarray}
\label{Def-Com2Pho-LO} {\cal F}_p (\zeta,W^2,{\cal Q}^2)  =
(1-2\zeta)Q^2_p \int_{0}^1\!dz \left[\frac{1}{z}   \mp
\frac{1}{1-z}  \right] \Phi_p(z,\zeta,W^2, {\cal Q}^2)\,.
\end{eqnarray}
Employing the conformal partial wave decomposition
(\ref{ConParWavGDA}) for GDAs, we immediately find the following
series for the scattering amplitude
\begin{eqnarray}
\label{Res-Com2Pho-LO} {\cal F}_p (\zeta,W^2,{\cal Q}^2) =
(1-2\zeta) Q^2_p \sum_{n=0}^\infty
\frac{2^{n+1}\Gamma(5/2+n)}{\Gamma(3/2)\Gamma(3+n)}  \left[1 \mp
(-1)^n\right] M_n(\zeta,W^2)\,.
\end{eqnarray}
As discussed in Sect.\ \ref{SubSec-Cro} the conformal moments
$M_n(\zeta,W^2)$ are related to the GPD ones by crossing, cf.\
Eq.\ (\ref{ConParWavGDA}). Of course, the inclusion of
perturbative corrections and evolution effects is done in an
analogous way as in the case  of DVCS. The only difference is that
we are now dealing with integer $n$.  The sum
(\ref{Res-Com2Pho-LO}) converges, just as the partial wave
decomposition for the GDA itself. However, the  oscillations
caused by the polynomials $p_n(1-2z,1)$ drop out and so its
numerical approximation is easier to handle. This representation
can be directly used in phenomenological studies. Knowing an
appropriate analytic continuation of $M_n(\zeta,W^2)$ would also
allow to rewrite the conformal partial wave expansion
(\ref{Res-Com2Pho-LO}) as Mellin-Barnes integral:
\begin{eqnarray}
\label{Res-Com2Pho-LO-MelBarInt} {\cal F}_p (\zeta,W^2,{\cal Q}^2)
=  (1-2\zeta) \frac{Q_p^2}{2i}\int_{c-i\infty }^{c+i\infty }\!dj\,
 \frac{2^{j+1}\Gamma(5/2+j)}{\Gamma(3/2)\Gamma(3+j)}
 \frac{M_j(1-\zeta,W^2) \mp M_j(\zeta,W^2)}{\sin(\pi j)}\,.
\end{eqnarray}
Here we employed the symmetry relation $M_n(\zeta,W^2)= (-1)^n
M_n(1-\zeta,W^2)$ and that $M_j$ is holomorphic in the first and
forth quadrant.

\section{Evaluation and parameterization of conformal moments}
\label{Sec-ExaGPDs}

This section is devoted to  the parameterization of conformal
moments and the numerical treatment of GPDs and Compton form
factors. We study first the analytic properties of conformal
moments for a simple toy GPD ansatz and give then  examples for
the numerical evaluation. Then  we introduce a simple
parameterization of the conformal moments with respect to the
skewness dependence, i.e., we consider only a reduced GPD and
its crossing analog, and discuss its dependence
on the momentum transfer squared. Finally,
we suggest a simple GPD model, which is rather flexible in its
parameterization.

\subsection{Numerical treatment of the Mellin-Barnes integral}
\label{SubSec-NumTre}

Let us  consider the conformal moments in terms of the Radyushkin
ansatz (\ref{RadAns}) for the reduced GPDs. This toy example can
be treated for integer values of $b$ and $\beta$ in an analytic
manner. To represent the results here in an explicit form we
choose  again $b=0$ and our toy model with  $\beta=1$. In this
case the function $\omega(x, \eta)$ is given in terms of Eq.
(\ref{Dec-GPD-1}) by
\begin{eqnarray}
\label{GPD-toy} q(x,\eta) = \theta(-\eta \le x \le 1)
\frac{2+\alpha}{2\eta}
\left(\frac{x+\eta}{1+\eta}\right)^{1+\alpha} + \theta(\eta \le x
\le 1)\frac{2+\alpha}{-2\eta}
\left(\frac{x-\eta}{1-\eta}\right)^{1+\alpha}.
\end{eqnarray}
The analytic continuation of the function $\mu_n(-\eta)$, see Eq.\
(\ref{DefConMomGPD}), reads
\begin{eqnarray}
\label{cal-muOrg}
\mu_j(-\eta) =
-   \frac{ \Gamma(1/2)  \Gamma(3 + j)}{2^{3+j} \Gamma(3/2 + j)}
( 1 - \eta) \eta^{-1 + j}
{_3F_2}\left({-j, 3+j, 2+\alpha \atop 2, 3 + \alpha}\Bigg|
\frac{-1+\eta }{2 \eta }\right)\,.
\end{eqnarray}
This function has the  behavior needed for the derivation of the
Mellin-Barnes integral. However, it is not appropriate for an
analytic continuation to negative $\eta$, since the argument of
the hypergeometric function would then become larger than one,
i.e., $(-1+\eta)/2 \eta \to (1+\eta)/2 \eta$, and so its value is
given at the branch cut of the hypergeometric function. To analyse
the situation in more detail, we can first express the
hypergeometric function in Eq.\ (\ref{cal-muOrg}) as a sum of two
${_3F_2}$ functions with argument $2 \eta/(-1+\eta)$. This gives
\begin{eqnarray}
\label{cal-mu}
\mu_j(-\eta)
&\!\!\!=\!\!\! &
-\frac{\left( 2 + \alpha  \right)
{\left( 1 -\eta  \right) }^{1 + j}}{2\left( 2 + j + \alpha  \right)\eta }\,
{_3F_2}\left({-j-1, -j, -2-j-\alpha \atop -2j-2, -j-1-\alpha}\Bigg|
\frac{2 \eta }{-1+\eta }\right)
-
\frac{( 2 + \alpha) \eta^{2( 1 + j)} }{2^{2(2 + j)}(1 - \eta)^{2 +j}}
\nonumber
\\
&&
\times
   \frac{\Gamma(1 + j)\Gamma(3 + j) \tan (\pi j )}{
   ( 1 + j - \alpha)\Gamma(3/2 + j)\Gamma(5/2 + j)}\,
{_3F_2}\left({2+j, 3+j, 1+j-\alpha \atop 4+2j, 2+j-\alpha}\Bigg|
\frac{2 \eta }{-1+\eta }\right)
\\
&& - \frac{\eta^{1 + j + \alpha}}{2^{1 + j - \alpha}
(1-\eta)^{1+\alpha}} \frac{\Gamma(1/2)\Gamma(1 + j)}{\Gamma(3/2 +
j)} \frac{\Gamma(3 + \alpha ) \Gamma(1 + j - \alpha )\sin(\pi j
)}{ \Gamma(-\alpha ) \Gamma(3 + j + \alpha )\sin(\pi[j + \alpha]
)} \,. \nonumber
\end{eqnarray}
The first term on the r.h.s.\ contains poles on the real axis at
$j=1/2,3/2,\dots$ and for $\alpha < 0$ at
$j=-\alpha,1-\alpha,\dots$ Both of them are cancelled by the
second and third term, respectively.

The continuation of this function to negative values of $\eta$ is
obviously not unique, since we have  branch points at $\eta= 0$,
which leads to phase factors $e^{\pm i \pi(j+1+\alpha)}$ as well
as $e^{\pm i 2\pi(j+1)}$. Such phase factors in $\mu_j(\eta)$
would, however,  violate the assumptions made for the derivation
of the Mellin-Barnes representation. To ensure that this does not
happen we define the analytic continuation by
 \begin{eqnarray}
 \label{con-mu}
 \mu_j^{\rm AC}(\eta) =  \frac{e^{i\pi j} \mu_j(-\eta e^{-i \pi}) -
e^{-i\pi j} \mu_j(-\eta e^{i \pi})}{2 i\sin(\pi j)}\,.
\end{eqnarray}
were the phase factors $e^{\pm i\pi j}$ compensate the phases
which arise from the $\eta^j$ terms in the definition
(\ref{Def-cHyp}) of conformal moments, while taking the
``discontinuity'' and dividing it by $2i \sin(\pi j)$ ensures that
no new singularities appear on the real axis and that
$\mu_j(\eta)$ is bound for $j\to \infty$. In fact the properties
of Legendre functions imply that the prescription (\ref{con-mu})
provides the contribution from the outer region
\begin{eqnarray}
\label{con-AnaConRes}
 \mu_j^{\rm AC}(\eta)=\int_{\eta}^1\! dx\, c_j (x,\eta)
 \frac{2+\alpha}{2\eta}
 \left(\frac{x+\eta}{1+\eta}\right)^{1+\alpha}
 +  \frac{2+\alpha}{2\eta} \left(\frac{2\eta}{1+\eta}\right)^{1+\alpha}
\left(\frac{\eta}{2} \right)^{j+1}
\frac{\Gamma(1/2)\Gamma(1 + j)}{\Gamma(3/2 + j)}
\,,
\end{eqnarray}
where the second term on the r.h.s.\ arises from a pole at
$x=\eta$ and cancels the contribution from the lower bound of the
integral. Hence, we realize that the central region is still
ignored in Eq.\ (\ref{con-AnaConRes}) and its contribution might
be taken into account within Eq.\ (\ref{Def-mu-cen2}). The term
that appears there from the pole at $x=\eta$ is already contained
in Eq.\ (\ref{con-AnaConRes}) and so only the discontinuity on the
negative axis contributes.

Analytic continuation via Eq.\  (\ref{con-mu}) leads for our toy
ansatz to
\begin{eqnarray}
\label{cal-muAC}
\mu^{\rm AC}_j(\eta) &\!\!\!=\!\!\! &
\frac{\left( 2 + \alpha  \right) {\left( 1 + \eta  \right) }^{1 + j}}{
2\left( 2 + j + \alpha  \right)\eta }\,
{_3F_2}\left({-j-1, -j, -2-j-\alpha \atop -2j-2, -j-1-\alpha}\Bigg|
\frac{2 \eta }{1+\eta }\right)+
\frac{( 2 + \alpha) \eta^{2( 1 + j)} }{2^{2(2 + j)}(1 + \eta)^{2 +j}}
\nonumber
\\
&&\times
   \frac{\Gamma(1 + j)\Gamma(3 + j) \tan (\pi j )}{( 1 + j - \alpha)
   \Gamma(3/2 + j)\Gamma(5/2 + j)}\,
{_3F_2}\left({2+j, 3+j, 1+j-\alpha \atop 4+2j, 2+j-\alpha}\Bigg|
\frac{2 \eta }{1+\eta }\right)
\\
&&
 -
\frac{ \eta^{1 + j + \alpha}}{2^{1 + j - \alpha}
(1+\eta)^{1+\alpha}} \frac{\Gamma(1/2)\Gamma(1 + j)}{\Gamma(3/2 +
j)} \frac{\Gamma(3 + \alpha ) \Gamma(1 + j - \alpha )\sin(\pi
\alpha )}{ \Gamma(-\alpha ) \Gamma(3 + j + \alpha )\sin(\pi[j +
\alpha] )} \,. \nonumber
\end{eqnarray}
As before for $\mu_j(-\eta)$ there are no net singularities in the
first and forth quadrant of the complex $j$ plane, however,
individual terms possesses poles on the real axis. The first term
on the r.h.s.\ is needed to restore the polynomiality in the sum
$\mu_n(\eta)+\mu^{\rm AC}_n(-\eta)$. However, the third term
violates polynomiality and must be cancelled by the contribution
from the central region, still missing. We can restore this
missing term from the polynomiality condition in such a way that
it is free of singularities in the first and forth quadrant of the
complex $j$ plane. The $\sin(\pi\alpha)/\sin(\pi[j+\alpha])$ term
of the third line, however, generates for $j=n=\{0,1,2,\cdots\}$,
a sign alternating series and to get rid of it, we separately
continue even and odd moments:
\begin{eqnarray}
\label{cal-mtoy}
m_j(\eta|\sigma) = \mu_j(\eta) + \mu^{\rm AC}_j(-\eta) +
\sigma
\frac{ \eta^{1 + j + \alpha} }
{2^{1 + j - \alpha} (1+\eta)^{1+\alpha}}
 \frac{ \Gamma(1/2)\Gamma(1 + j)}
{\Gamma(3/2 + j)} \frac{\Gamma(3 + \alpha ) \Gamma(1 + j - \alpha
)}{ \Gamma(-\alpha ) \Gamma(3 + j + \alpha ) )} \,,
\end{eqnarray}
with $\sigma=1$ and $\sigma=-1$ in the  even and odd sector,
respectively. We remark that this result for the conformal moments
of our toy GPD (\ref{GPD-toy}) can be alternatively obtained
within the framework given in Sect.\ \ref{SubSubSec-SomWatTra}.
Especially, the term we restored from the  polynomiality condition
arise from the discontinuity on the negative real axis in Eq.\
(\ref{Def-mu-cen2}).

Now let us come  to the numerical treatment of the Mellin-Barnes
integrals for GPDs (\ref{Rep-F-MelBar},\ref{RepGPD-Even-Odd}) and
Compton form factors (\ref{Res-CFF-LO}). The numerical evaluation
can be easily done once the conformal moments are known in
analytic or numerical form. Corresponding to the behavior of GPDs
at $x\to 1$, they should vanish at large $j$ rather fast
\cite{Yua03}. Hence, the integral converges rather fast, too,  and
so one can in practice perform the integration over a finite
interval. If one includes higher order corrections or the
evolution the numerical treatment remains stable.

\begin{figure}[t]
\begin{center}
\mbox{
\begin{picture}(600,120)(0,0)
\put(55,100){(a)}
\put(10,0){\insertfig{8}{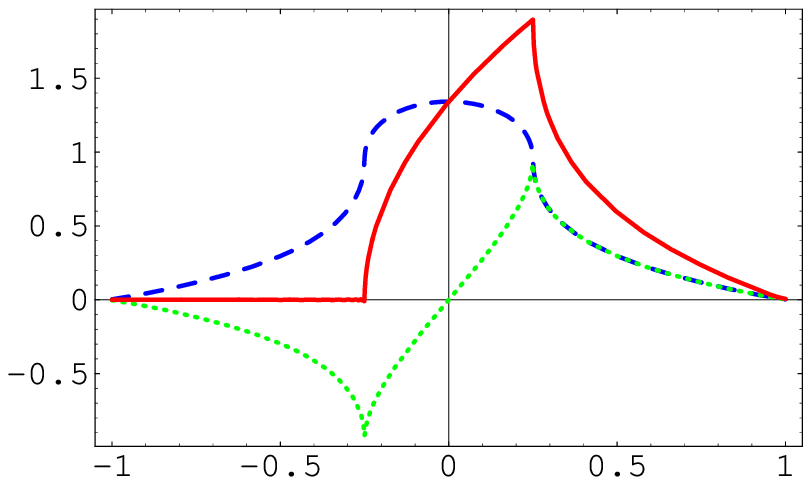}}
\put(0,25){\rotatebox{90}{$q(x,\eta=0.25,{\cal Q}_0^2)$}}
\put(230,-5){$x$}
\put(315,100){(b)}
\put(275,0){\insertfig{7.6}{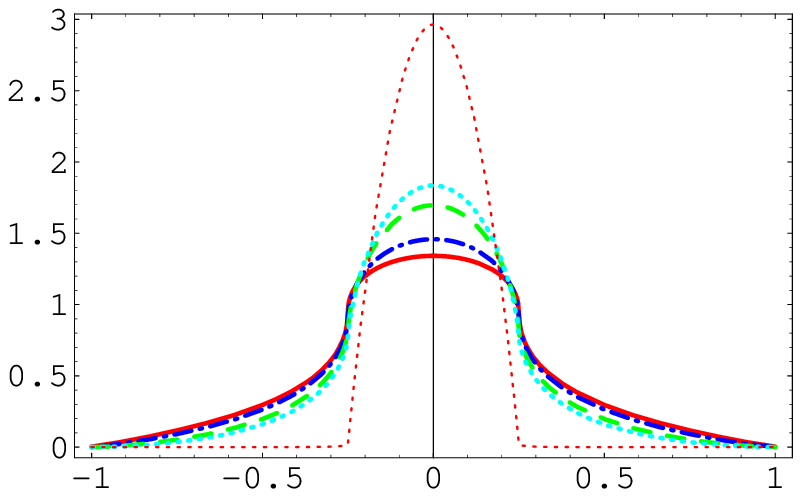}}
\put(258,20){\rotatebox{90}{$q^{\rm even}(x,\eta=0.25,{\cal Q}^2)$}}
\put(485,-5){$x$}
\end{picture}
}
\end{center}
\caption{
\label{FigGPD1}
In panel (a) the toy GPD (\ref{GPD-toy}) is displayed for
$\alpha=-1/2$ and $\eta=0.25$ as solid line and its symmetric and
antisymmetric part as dashed and dotted line, respectively. Panel
(b) shows the evolution of the symmetric part from the input scale
${\cal Q}_0^2=0.5\ \GeV^2$ (solid) to the scales ${\cal Q}^2=1\
\GeV^2$ (dash-dotted), ${\cal Q}^2=10\ \GeV^2$ (dashed),  ${\cal
Q}^2=100\ \GeV^2$ (dotted), and (nearly) asymptotic limit (thin dotted).
}
\end{figure}
For our toy model we display in Fig.\ \ref{FigGPD1}(a) for
$\alpha=-1/2$ and $\eta=0.25$ the GPDs arising from odd and even
conformal moments, while the sum of them (solid line) provides the
original support.
 The evolution with respect to the renormalization/factorization
scale $\mu^2={\cal Q}^2$ is taken into account simply by including
the evolution operator in the Mellin-Barnes integral. For the
vector case the even quark moments evolve separately, i.e., they
do not mix with the gluonic ones, and so we can employ  Eq.\
(\ref{Sol-EvoEqu-LO}). For $\alpha_s({\cal Q})$ we take its LO
approximation for three quark flavors and $\Lambda=0.22$ GeV. That
the numerics is completely unproblematic is demonstrated in Fig.\
\ref{FigGPD1}(b), where the asymptotic case ${\cal Q}^2 \to
\infty$ is nearly reached by setting ${\cal Q}^2 = 10^{1000}
\GeV^2$. We remind that in this limit the outer region will die
out and only the lowest conformal moment contributes leading to
the asymptotic GPD
\begin{eqnarray}
 q^{\rm asy}(x,\eta) =
 \theta(\eta-|x|) \frac{3}{4|\eta|} \frac{\eta^2-x^2}{\eta^2}\,.
\end{eqnarray}

\begin{figure}[t]
\begin{center}
\mbox{
\begin{picture}(600,110)(0,0)
\put(10,0){\insertfig{7.3}{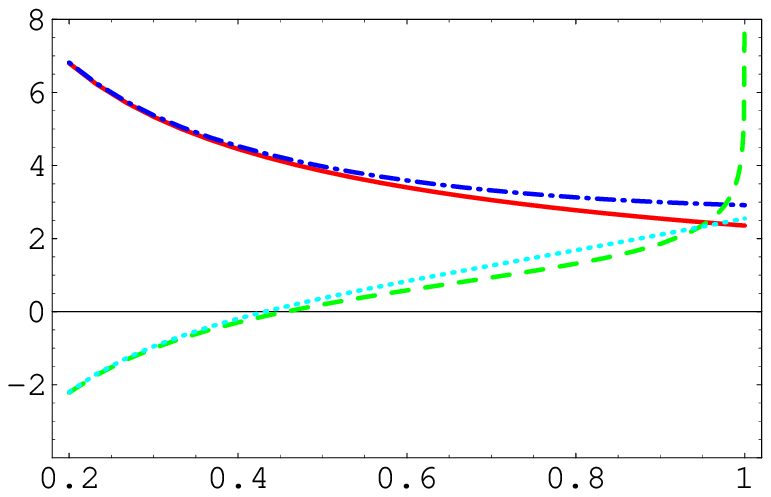}}
\put(-2,30){\rotatebox{90}{${\cal H}(\xi,\Delta^2=0,{\cal
Q}_0^2)$}} \put(207,-8){$\xi$} \psfrag{0.02}[cc][cc]{}
\psfrag{0.05}[cc][cc]{} \psfrag{0.002}[cc][cc]{}
\psfrag{0.005}[cc][cc]{} \put(280,-3){\insertfig{7.3}{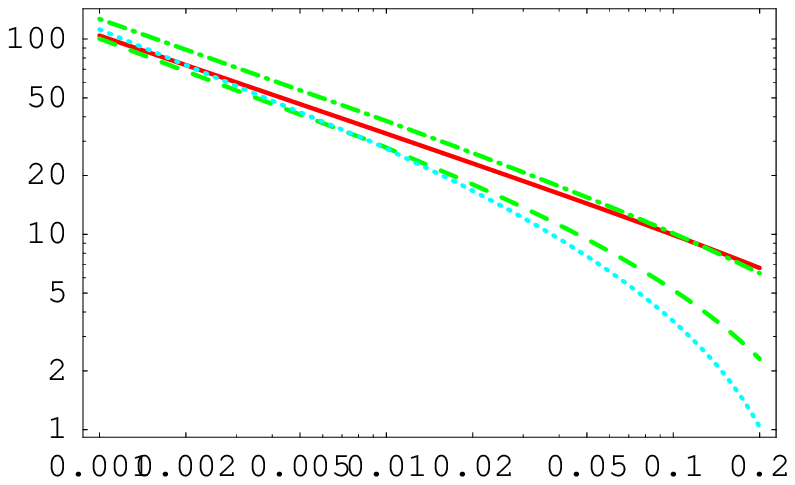}}
\put(259,30){\rotatebox{90}{${\cal H}(\xi,\Delta^2=0,{\cal
Q}^2)$}} \put(476,-8){$\xi$}
\end{picture}
}
\end{center}
\caption{ \label{Fig-AppCFF}  The Compton form factor ${\cal
H}(\xi,\Delta^2=0,{\cal Q}^2)$, arising from the toy GPD
(\ref{GPD-toy}), to LO accuracy. Left: Exact imaginary (solid) and
real (dashed) part of the Compton form factor and their
approximation by Eq. (\ref{App-CFF-toy}) (dash-dotted and dotted
lines, respectively). Right: The Compton form factor evaluated
numerically  at the input scale ${\cal Q}_0^2= 0.5\, \GeV^2 $
(solid and dashed) and evolved with the flavor non-singlet
evolution equation to ${\cal Q}^2= 4\, \GeV^2 $ (dash-dotted and
dotted). The imaginary part is shown as solid and dash-dotted
lines and the absolute value of the real part as dashed and dotted
lines. }
\end{figure}

The numerical procedure for the calculation of scattering
amplitudes is even easier to handle than that for GPDs themselves.
As mentioned, a nice feature of the Mellin-Barnes integral is that
it can be used to derive an expansion in powers of $\xi$. Let us
suppose that we like to evaluate the Compton form factor
(\ref{Def-ComForFacH}) to LO accuracy, where the charge $Q_p$ is
set to one. Here only the analytic continuation of the odd
conformal moments is needed. Guided by the small $x$ behavior of
parton densities, we choose the parameter $\alpha$ to be negative
and larger than -3/2. To ensure that all singularities are on the
l.h.s.\ of the integration path, we take $- {\rm
Max}(1/2,-1-\alpha) <c < {\rm Min}(1/2,-\alpha)$. The first pole
which appear in the integrand  on the negative axis arises from
$\tan(\pi j/2)$ and is at $j=-1$. We remark that the pole which we
would have expected in the forward case, namely, at $j=-1-\alpha$
is absent for $\xi> 0$. Rather there appears a new one at
$j=-1+\alpha$, which is associated with the behavior of the GPD at
the cross-over point $x=\eta$. {F}rom our explicit expression for
the conformal moments, given in Eqs.\ (\ref{cal-mu}),
(\ref{cal-muAC}), and (\ref{cal-mtoy}) it is obvious that the
integrand contains three different pieces,  proportional to
$\xi^{-1-j}$, $\xi^{1+j}$, and $\xi^\alpha$. For the term
proportional to  $\xi^{-1-j}$ we can arrange a systematic
expansion in $\xi$ by a shift of the integration path to the left.
The poles appear here at $j=\{-1-\alpha,-2-\alpha,\cdots\}$,
$j=\{-1,-2,\cdots\}$, and at $j=\{-1/2,-3/2,\cdots\}$. In the latter
case the contribution will be  cancelled by those from the  piece
proportional to  $\xi^{1+j}$, which has  poles on the positive
axis only,  at $j=\{1/2,3/2,\cdots\}$. This is established by a
shift of the integration path to the right. What remains are the
terms proportional to $\xi^\alpha$. Here we can close the
integration path so that the first and forth quadrant is included
and employ then the residue theorem, where the only poles are at
$j=\{-\alpha,1-\alpha,\cdots\}$. All of these poles contribute to
the leading power behavior and they must be resummed. Finally,
collecting the results and neglecting power corrections of order
${\cal O}(\xi^3)$ leads to the approximation
\begin{eqnarray}
\label{App-CFF-toy}
{\cal H} (\xi) &\!\!\! =\!\!\! &
2^{\alpha }\pi( 2 + \alpha)
\left[i  + \cot\left(\frac{\pi\alpha}{2}\right) \right] \xi^{\alpha}
\Big\{ 1 -
  i ( 1 + \alpha)  \xi \tan\left(\frac{\pi \alpha }{2}\right)
\nonumber\\
&&+ \frac{( 1 + \alpha)(2 + \alpha)}{2} \xi^2 +
  {\cal O}(\xi^3) \Big\}
-\frac{2(2 + \alpha)}{\alpha }
\left\{ 1 + \frac{( 2 + \alpha  + {\alpha }^2)}
       {( 1 - \alpha)  ( 2 - \alpha) } {\xi }^2 +
       {\cal O}(\xi^3)  \right\}\,.
\end{eqnarray}
This expansion coincides with that for the exact Compton form
factor, which for our toy model (\ref{GPD-toy}) is exactly
calculable. It is remarkable that this  expansion does contain odd
powers of $\xi$. In fact, the $\xi$ term in the first line should
be especially important in the small $\xi$ region. Numerically,
this approximation works quite well for the region that is of
phenomenological interest. The deviation  from the exact
expression is about $3\%$ for $\alpha=-1/2$ and $\xi=0.4$ for the
imaginary part. For the real part the approximation induces a
small shift of the zero of the exact expression from $\xi\sim
0.456$ to $\xi\sim 0.431$. The accuracy of the approximation
(\ref{App-CFF-toy}) grows with increasing $\alpha$ and rapidly
with decreasing $\xi$. It is amazing that this approximation
remains qualitatively correct as long as $\xi$ does not approaches
one, see left panel in Fig.\ \ref{Fig-AppCFF}. In the right panel
of this figure we show  the imaginary and real part of the
Compton form factor, evaluated numerically with the Mellin-Barnes
integral, for smaller values of $\xi$. Here the difference between
the approximate and exact result is even  invisible. Let us stress
specifically, that we do not encounter any numerical problems in
the small $\xi$ region.

\subsection{Ans\"atze for conformal moments}

The conformal moments for complex valued conformal spin, appearing
in the GPD representation (\ref{Rep-F-MelBar}), can be expanded in
terms of $c_k(1,\eta)$ using  a suitable integral transformation.
Such an integral transformation can be viewed as the analytic
continuation of the conformal expansion (\ref{Def-Exp-Coe})  for
non-negative integer conformal spin. In such an representation the
$\eta$ and $\Delta^2$ dependence in the conformal moments
separates and only an ansatz for the form factors $F_{j
k}(\Delta^2)$ is required.

Lets have a closer look at these form factors. The GPD
Mellin--Barnes integral representation (\ref{Rep-F-MelBar} can be
interpreted as describing the effective summation of all particle
exchanges in the $t$-channel. These are labelled by their
conformal spin. Of course, conformal symmetry is broken in the
non-perturbative sector. However, the conformal spin can still be
used for the classification of excitations, just like in quantum
mechanics for a  non-spherical potential, the partial wave
expansion in terms of spherical harmonics can still be employed to
solve the Schr\"odinger equation. For vanishing skewness, i.e.,
$\eta=0$, the quantum number of the conformal spin can be replaced
by the common spin $J=j+1$ and the conformal moments are given by
the ``diagonal'' form factor $F_{jj}(\Delta^2)$
\begin{eqnarray}
m_j(\eta=0,\Delta^2) =  F_{jj}(\Delta^2)\,,
\end{eqnarray}
cf.\ Eq.\ (\ref{Def-Exp-Coe}).  Having in mind that the conformal
partial waves $p_{j}(x,\eta=0)$ are then given by $ (1/x)^J$,
where $1/x$ plays the role of a (rescaled) energy, the
Mellin--Barnes integral for GPDs looks similar to the $t$-channel
scattering amplitude at large energies. This suggest that there
exists a connection with Regge theory or at least with Regge
phenomenology. Let us remind that in deep inelastic scattering,
i.e., for $\Delta^2=0$, this connection shows up in the small $x$
behavior of structure functions, which is governed by the
intercept of the corresponding leading Regge trajectories
\cite{LanPol70}. Note, however, that the small $x$ behavior of
parton densities depends on their conventions, i.e., on the
factorization scheme and scale. A more recent analysis for the
unpolarized valence quark densities can be found in Ref.\
\cite{DieFelJakKro04}. Below it will be demonstrated that for
$\eta=0$ there exist phenomenological indications that the
$\Delta^2$ dependence in $F_{jj}(\Delta^2)$ is related to Regge
trajectories.

Also for general kinematics, i.e., $\eta\neq 0$, one can take this
duality between $t$- and $s$-channel serious, see the discussion
in Sect. \ref{SubSec-AnaGPDs}. A description of DVCS in the
high-energy limit is given in \cite{BalKuc00}, where the leading
Regge trajectory arises from the pomeron exchange in the
t-channel. Moreover, the $t$-channel description provides
arguments for the so-called $D$-term \cite{PolWei99}, appearing in
the nucleon GPDs $H(x,\eta,\Delta)$ and $E(x,\eta,\Delta)$, and
for the so-called pion pole term \cite{PenPolGoe99}, appearing in
$\widetilde E(x,\eta,\Delta)$. Note that these effects are taken
into account by a modification of GPDs that affects only the
central region and so their $s$-channel counterpart is absent. On
the basis of the conformal partial wave expansion, applied to LO
accuracy, and the crossing relation between GPDs and GDAs a dual
description of the former ones in terms of $t$-channel exchanges
has been suggested in Ref.\ \cite{PolShu02}. Here the conformal
moments have been decomposed into contributions with definite
angular momentum. Since the concept of conformal spin has to the
best of our knowledge not been worked out for applications in
hadron spectroscopy, one should focus on this more appropriate
quantum number. Nevertheless, let us point out that the form
factors $F_{j k}(\Delta^2)$ for $j\neq k$, defined in Eq.\
(\ref{Def-Exp-Coe}), measure the strength of (non-perturbative)
conformal symmetry breaking.  In this paper, however, we do not
proceed with the spectroscopic interpretation of these form
factors (or some rotated version of them) and  the search of an
appropriate representation for the conformal moments with complex
valued conformal spin in terms of them.

Instead, our aim is now more pragmatic, namely we will introduce
and study specific ans\"atze for the conformal moments. Therefore,
we now introduce and explore ans\"atze that are simpler to handle.
We start in Sect.\ \ref{SubSubSec-AnsConMomRed} with a reduced
variable dependence by setting $\Delta^2 =0$. Then we implement in
Sect.\ \ref{SubSubSec-AnsConMomDep-t} the $\Delta^2$ dependence.
In Sect. \ref{SubSubSec-AnsConMomEta0} we especially consider this
dependence for the case $\eta=0$ and discuss the resulting GPD
$H(x,\eta=0,\Delta^2)$, especially, its interpretation as three
dimensional parton density.

\subsubsection{Ans\"atze for reduced conformal moments}
\label{SubSubSec-AnsConMomRed}

For our simple toy GPD example (\ref{GPD-toy}), studied in Sect.
\ref{SubSec-NumTre}, the conformal moments appear to be rather
complicated functions. Indeed, it is not clear at all whether the
complicated structure of $m_j(\eta,\Delta^2=0)$ in our toy model,
is an artifact of the analytic continuation procedure or indicates
some physics, related to the skewness dependence. So let us
explore several ``minimal'' ans\"atze for the conformal moments.

To ensure that the forward limit is correctly reproduced, we  factorize
the poles in the complex $j$-plane that survive the limit $\eta\to 0$:
\begin{eqnarray}
\label{Ans-mj}
m_j(\eta,\Delta^2=0|\alpha,\beta) =
\frac{\Gamma(\alpha+1+j) \Gamma(\alpha+\beta+2)}{
\Gamma(\alpha+1)\Gamma(\alpha+\beta+2+j)} b_j(\eta^2)\,.
\end{eqnarray}
Here the normalization of the function  is $b_j(\eta^2=0)=1$ and,
moreover, we considered it as convenient to normalize the lowest
moment according to $m_0(\eta,\Delta^2=0|\alpha,\beta) =1$. In the
forward limit we arrive after an inverse Mellin transform at
\begin{eqnarray}
\label{Ans-mj-Mel}
m_j(\eta=0,\Delta^2=0 |\alpha,\beta)\quad \to \quad
\frac{\Gamma(\alpha+\beta+2)}{\Gamma(\alpha+1)\Gamma(\beta+1)}
 x^\alpha (1-x)^\beta\,,
\end{eqnarray}
which is well defined for all realistic values of $\alpha$, except
for $\alpha=-1$.  For this special value one can first multiply
the expression with an appropriate normalization factor that
cancels the factor $1/\Gamma(\alpha+1)$. The standard
parameterization of a parton density, i.e., $N x^\alpha \left(1 + A
\sqrt{x} + B x \right)(1-x)^\beta $, can be obtained in Mellin
space by the linear combination
\begin{eqnarray}
\label{Par-ParDen}
q_j=N^\prime
\left[m_j(0,0 |\alpha,\beta) + A^\prime\, m_j(0,0 |\alpha+1/2,\beta)+
B^\prime\, m_j(0,0 |\alpha+1,\beta)\right]\,,
\end{eqnarray}
where
\begin{eqnarray}
N^\prime =
N \frac{\Gamma(1+\alpha)\Gamma(1+\beta)}{\Gamma(2+\alpha+\beta)}
\,,\quad A^\prime =
\frac{\Gamma(3/2+\alpha)\Gamma(2+\alpha+\beta)}{
\Gamma(1+\alpha)\Gamma(5/2+\alpha+\beta)} A \,,
\quad  B^\prime =  \frac{1+\alpha}{2+\alpha+\beta} B\,.
\end{eqnarray}
It is required that the function $b_j(\eta^2)$ for non-negative
integer values of $j=n$ reduces to a polynomial of order $n/2$ and
$(n\pm 1)/2$ for even and odd $n$, respectively. Here the order of
the odd moments depends on the specific GPD. Moreover, these
functions have only singularities in the second and third quadrant
of the complex $j$ plane. We can allow that these functions grow
with the real part of $j$, however, they must be bounded for large
imaginary parts of $j$,   $|{\rm arg}(j)| \leq \pi/2$.

One can classify the conformal moments $b_n(\eta^2)$ in general
and their analytic continuation with respect to their dependence
on both variables $\eta$ and $j$, which leads to a classification
of reduced GPDs. (We do not know whether this mathematical fact is
known already from some other context.) For the time being, we
concentrate on functions $b_j(\eta^2)$ that can be expanded around
the point $\eta^2=0$. Several simple examples can be
given in terms of  hypergeometric functions
\begin{eqnarray}
\label{Def-b}
b_j\left(\eta^2\Big|\{a,b\},\{r,s\},\{\sigma,p\}\right) =
\frac{
 {_2\!F}_1\!\left({ -j/2 + (1-2 p)(1-\sigma)/4 ,a \atop b}\Big| r+s\,
 \eta^2\right)}{
 {_2\!F}_1\!\left({ -j/2 + (1-2 p)(1-\sigma)/4 ,a \atop b}\Big| r\right)}\,,
\end{eqnarray}
where the normalization condition at $\eta=0$ is satisfied. Here
the parameters $a$ and $b$ may depend on $j$. For the analytic
continuation of even moments we set as above $\sigma=1$ in the
case of odd ones $\sigma=-1$ and the choice $p=1$ leads for $j=n$
to polynomials of order $(\eta^2)^{(n+1)/2}$, while $p=0$ gives
polynomials of order $(\eta^2)^{(n-1)/2}$.  To ensure that no
branch cut appears in the interval $0\leq \eta^2\leq 1$, the
parameters $r$ and $s$ must fulfill the inequalities $r\leq 1$ and
$r+s\leq 1$.

The definition (\ref{Def-b}) is quite general and contains  a
number of special cases. For instance, if we set $\sigma=1,
a=(1-j)/2, b=2,$ and $r=1$,
\begin{eqnarray}
\label{Ans-Fun-b}
b_j\left(\eta^2\Big|\{(1-j)/2 ,2\},\{1,s\},\{1,0\}\right) =
\frac{\Gamma(3/2)\Gamma(3+j)}{2^{j+1}\Gamma(3/2+j)}
{_2\!F}_1\!\left({ -j/2 ,(1-j)/2 \atop 2}\Big| 1+s\, \eta^2\right)\,,
\end{eqnarray}
we recover for $s=-1$ the definition of conformal moments
(\ref{Def-cHyp}) with $x=1$. Here they are expressed in terms of
hypergeometric functions that arise from the original ones by a
so-called quadratic transformation. These moments can be used for
even and odd values of $j$. If one needs odd moments that are of
order $(\eta^2)^{(n+1)/2}$, one should set $a=(-1-j)/2$.   We
remark that the expansion in the vicinity of $\eta^2=0$ exists
only as linear combination of two power series in $\eta^2$, one of
them containing the overall factor $(\eta^2)^{3/2+j}$.

Besides the other parameters, $s$ controls the strength of the
skewness dependence. If we set it to zero, the function
$b_j(\eta^2)$ is simply one.

If $a=b$, $b_j(\eta^2)$ reduces to the simple function
\begin{eqnarray}
b_j\left(\eta^2\Big|\{a,a\},\{r,s\},p\right) =
\left(1-\frac{s}{1-r}\, \eta^2\right)^{(j+p)/2}\,.
\end{eqnarray}
Here $1-r$ appears as a scaling factor and so in the following $r$
can be set to zero, when simultaneously $s$ is restricted to
$s\leq 1$. For negative values of $s$ the function (\ref{Def-b})
is even analytic  for $1\leq \eta$. To get a clue which values of
$a(j)$ and $b(j)$ are allowed in the ansatz (\ref{Ans-Fun-b}), we
generate ``associated'' conformal moments of
$b_j\left(\eta^2\Big|\{a,a\},\{0,s\},p\right)$ by the convolution
integral
\begin{eqnarray}
\label{Gen-ConMom-Int}
\int_{0}^1\! dz\, f_j(z) (1-z\, s\,\eta^2)^{(j+p)/2}\,,
\quad \mbox{with}\quad \int_{0}^1\! dz\, f_j(z) =1\,.
\end{eqnarray}
Here it is required that $f_j(z)$ as function of $j$ is bound for
$j\to \infty$ with $|{\rm arg}(j)| \leq \pi/2$ for $0\leq z\leq
1$. To arrive at  the parameterization in terms of hypergeometric
functions we choose
\begin{eqnarray}
\label{Gen-ConMom-Ans}
f_j(z)   =   \frac{\Gamma(b(j))}{\Gamma(a(j))\Gamma(b(j)-a(j))}\,
z^{a(j)-1}  (1-z)^{b(j)-a(j)-1}\,.
\end{eqnarray}
The requirement for the bound of $f_j(z)$ is certainly satisfied
for $\Re{\rm e}\, a(j)> 0$ and $\Re{\rm e}(b(j)-a(j))> 0$.
Especially, if $a(j)$ tends to infinity for $j \to \infty$, $b(j)$
has to grow as $a(j)$ or even faster. The case $a=(1-j)/2$ and
$b=2$, mentioned above, can be obtained within  an analogous
treatment. Here, however, one has to choose an integration contour
in Eq.\ (\ref{Gen-ConMom-Int}) in the complex plane that encircles
the point $z=0$ and so the convergence condition for the integral
can be relaxed. We will skip this issue here and refer, for
instance, to Ref.\ \cite{Luk69a}.

We now explore the resulting GPDs and GDAs. As example we take the
reduced valence quark GPDs and GDAs in the vector case with the
rather realistic values $\alpha=-1/2$ and $\beta=3$. The
unpolarized parton density is normalized to one and reads, cf.\
Eqs.\ (\ref{Ans-mj}) and (\ref{Ans-mj-Mel}),
\begin{eqnarray}
q(x) = \frac{35}{32}\;  x^{-1/2} (1-x)^3
\end{eqnarray}
The function $b_j(\eta^2)$ is the analytic continuation of
conformal moments with even $n$ and the parameters are
specified in Table \ref{Tab-Cases}.
\begin{table}[t]
\begin{tabular}{|c||c|c||c|}
\hline cases & \multicolumn{2}{c|}{parameters of
$b_j\left(\eta^2\Big|\{a ,b\},\{r,s\},\{1,0\}\right)$} & explicit
expression
\\
\hline
    & $\{a ,b\}$ & \{r,s\} &
\\
\hline\hline
(a)  & $\{(1-j)/2 ,2\}$ &  $\{1,-1\}$ &
$\frac{\Gamma(3/2)\Gamma(3+j)}{2^{j+1}\Gamma(3/2+j)}
{_2\!F}_1\!\left({ -j/2 ,(1-j)/2 \atop 2}\Big| 1-\eta^2\right)$
\phantom{\Bigg|}\\
\hline
(b) & $\{a ,b\}$ &  $ \{1,0\}$ & $1$
\phantom{\Bigg|}\\
\hline
(c)  &  $\{-199/4+ j, 2/3+2j\}$ &  $\{0,-1/4\}$ &
$
 {_2\!F}_1\!\left({ -j/2 ,-199/4+ j \atop 2/3+2j}\Big|-
 \frac{\eta^2}{4}\right)$
\phantom{\Bigg|}\\
\hline
(d) &  $\{-199/4+ j, 2/3+2j\}$ &  $\{0,1/4\}$ &
${_2\!F}_1\!\left({ -j/2 ,-199/4+ j \atop 2/3+2j}\Big|
\frac{\eta^2}{4}\right)$
 \phantom{\Bigg|}\\
\hline
(e) & $\{a,a\}$ &  $\{0,\pm 1/4\}$ &
$ \left(1\pm\frac{\eta^2}{4}\right)^{j/2}$
\phantom{\Bigg|}\\
\hline
\end{tabular}
\caption{ The parameters and resulting functions for our ans\"atze
concerning the analytic continuation of even conformal moments
(\ref{Def-b}), classified by case (a) -(e).
\label{Tab-Cases} }
\end{table}
A few comments are in order. In case (a) we took the conformal
moments itself, here in a more appropriate representation that is
for $\eta>0$ equivalent to $c_j(1,\eta)$. They possess the
properties we required in the derivation of the Mellin--Barnes
integral and of course, they reduce to polynomials for both even
and odd non-integer values of $n$. We will use them in the
Mellin-Barnes integral (\ref{Rep-F-MelBarSym}) within even
conformal partial waves  (\ref{Def-pSym}). Crossing symmetry , see
Eq.\ (\ref{ConParWavGDA}), requires the existence of an analytic
continuation to $\eta>1$ and this is achieved here by the change
of arguments   $c_j(1,\eta) \to c_j(1-2\zeta,1)$, see Eq.\
(\ref{Def-Exp-Coe}). For the numerical evaluation of the GDAs the
conformal partial wave series (\ref{ConParWavGDA}) or
alternatively the Mellin-Barnes integral (\ref{Rep-GDA-MelBar})
can be used. The projection on the even moments can be achieved by
symmetrization:
\begin{eqnarray}
\Phi(z,\zeta) \to  \frac{1}{2} \left[\Phi(z,\zeta) +
\Phi(1-z,\zeta)\right]\,.
\end{eqnarray}

The case (b) seems to be trivial. But after crossing it results in
$(1-2\zeta)^j$, which leads for odd moments with $\zeta> 1/2$ to
an alternating series with an  absolute value that converges to
one in the end-points. For the convergence of the conformal
partial wave series  (\ref{Def-Exp-Coe}) in terms of polynomials
an exponential suppression factor $2^{-n}$ is required. Hence,
this series only converges in the  interval $1/4 < \zeta < 3/4$.
The analytic continuation with respect to $\zeta$ can be achieved
by the Mellin-Barnes integral (\ref{Rep-GDA-MelBar-uni}).
However, outside of the convergence region, i.e., $\zeta  < 1/4$
or $3/4< \zeta$,  we will find a GDA that does not vanish at the
end-points $z=\{0,1\}$.

The cases (c) and (d) differ by the sign of the argument in the
hypergeometric function and follows from (e) by an integral
transformation (\ref{Gen-ConMom-Int}). They reduce to polynomials
for even $n$ only. The factor $1/4$ in the argument improves the
convergence property after crossing. The second parameter in the
upper line of the hypergeometric function, i.e., $-199/4+j$, can
for certain non-integer values of $j=n <  49$ be a negative
integer. As explained above, this does not generate problems. The
choice of this ``big'' constant $-199/4$ induces a numerical
enhancement of ${\cal O}(\eta^2)$ terms. The parameter $3/2+2 j$
in the lower line compensates for rather large values of $j$ the
growing of the second argument in the first line. We did not
include any poles on the positive $j$ axis. As a consequence, we
have new poles on the negative one. They appear at
$j=-1/3,-4/3,\cdots$ and die out in the limit $\eta\to 0$. With
our choice $\alpha=-1/2$, which determines the small $x$ behavior
of the parton densities and is associated with a pole at $j=-1/2$,
a new ``leading'' pole at $j=-1/3$ arises for non-zero skewness.
Its contribution, however, should be suppressed as ${\cal O}
(\eta^2)$. So for instance, in the Compton form factors it should
only give rise to a $\eta^2 \eta^{-2/3}=\eta^{4/3}$ term.
Moreover, we changed the normalization of the other poles for $0 <
\eta$. For not too large values of $\eta$ this should produce only
a numerically small  effect. Since we included a numerical
enhancement, this might induce some sizeable changes of the  GPD
and GDA shapes for not too small values of $\eta$.

\begin{figure}[t]
\begin{center}
\mbox{
\begin{picture}(600,110)(0,0)
\put(-5,60){\rotatebox{90}{$ b_{2}(\eta^2)$}}
\put(10,-2){\insertfig{7.3}{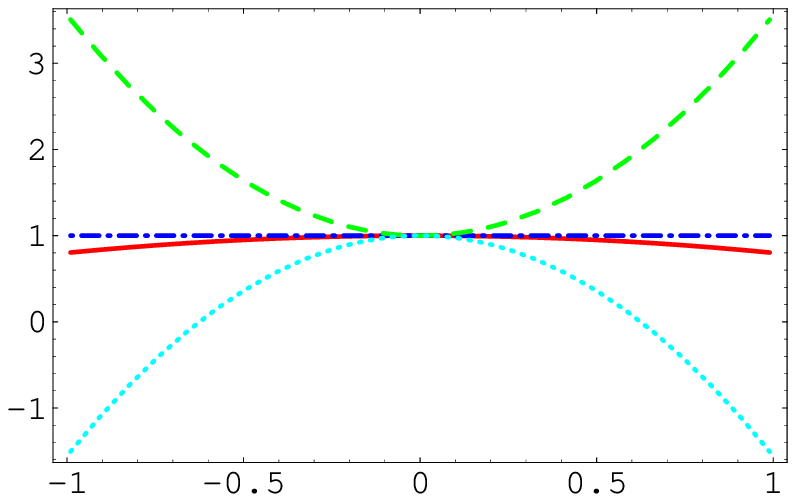}} \put(210,-8){$\eta$}
\put(265,30){\rotatebox{90}{$\scriptstyle(1-2\zeta)^2\,
b_{2}(1/(1-2\zeta)^2)$}} \put(280,0){\insertfig{7}{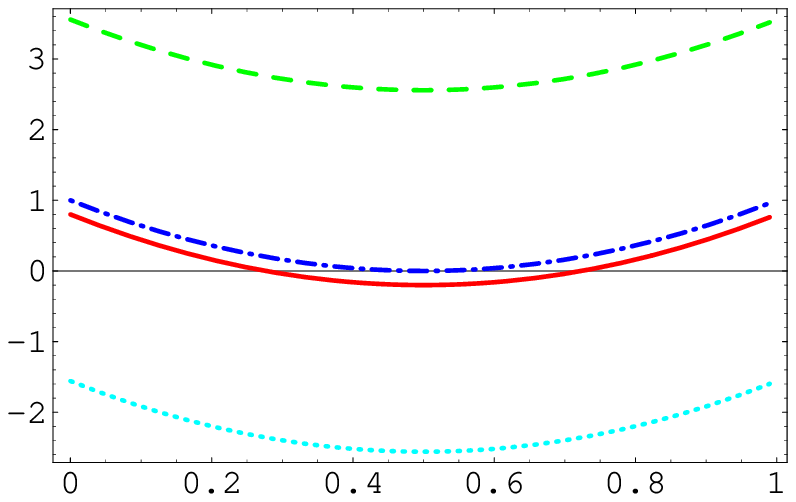}}
\put(470,-8){$\zeta$}
\end{picture}
}
\end{center}
\caption{ \label{FigConMom2} The $\eta$ (left) and $\zeta$ (right)
dependence of $b_n(\eta^2)$ and $(1-2\zeta)^2\,
b_{n}(1/(1-2\zeta)^2)$, respectively,  for $n=2$ are shown for the
parameterizations given in Table \ref{Tab-Cases}: (a) solid,  (b)
dash-dotted, (c) dashed, and (d) dotted lines. }
\end{figure}
In Fig.\ \ref{FigConMom2} we display the $\eta$ and $\zeta$
dependence for the second moment, i.e, of $b_2(\eta^2)$.
The $\eta$ dependence of the conformal moment in case (a) [(solid line)],
i.e.,
\begin{eqnarray}
c_2(\eta,1) = 1 - \frac{{\eta }^2}{5}
\quad\Rightarrow\quad c_2(1-2\zeta,1) = (1-2\zeta)^2-\frac{1}{5}\,,
\end{eqnarray}
is rater weak and so it is ``similar'' as in  case (b) [constant
or $(1-2\zeta)^2$ (dash-dotted line)]. In contrast we find for
cases (c) [dashed line] and (d) [dotted line] a large deviation in
opposite directions. For GPD moments it will die out for $\eta\to
0$, while for the GDA moments the difference  is nearly $\zeta$
independent and large. We will skip here the detailed discussion
of the dependence in higher moments, which are suppressed by the
factor ${\rm B}(1/2+n,4+n) \sim 1/n^4 $ for $n\to \infty$. Note,
however, that only the conformal moments (a) for $\eta^2> 0$
continuously tend to zero with increasing $n$. Also the rescaled
conformal moments (a) for all values of $\zeta$ possess this
behaviour. We remark that the conformal moments (e) will mostly
not give quantitatively different results as (b), so we will in
the following not present them.

\begin{figure}[t]
\begin{center}
\mbox{
\begin{picture}(600,140)(0,0)
\put(15,-1){\insertfig{7.9}{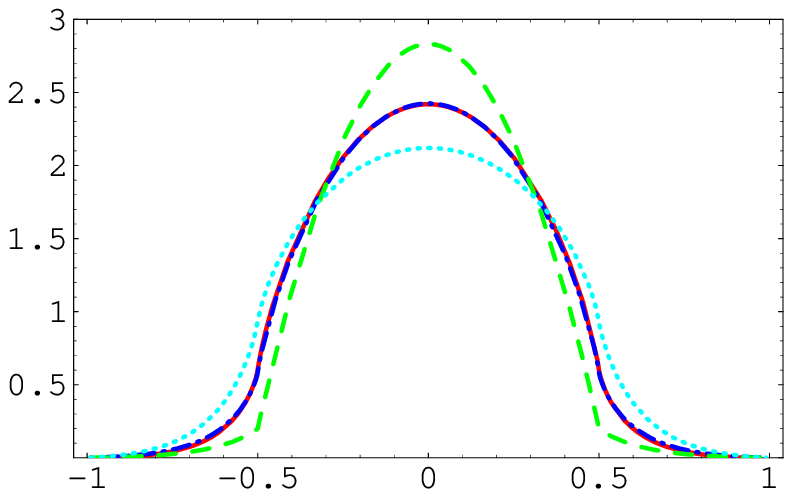}}
\put(225,-8){$x$}
\put(0,50){\rotatebox{90}{$q(x,\eta=0.5)$}}
\put(280,0){\insertfig{7.5}{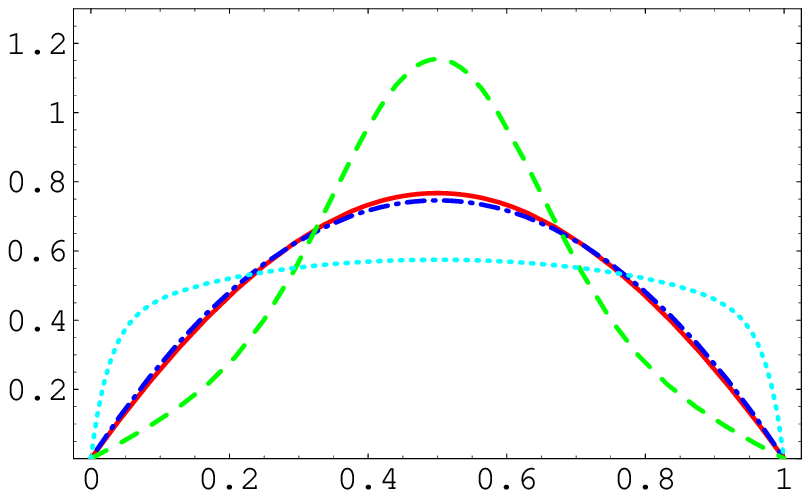}}
\put(485,-8){$z$}
\put(265,50){\rotatebox{90}{$ \Phi(z,\zeta=0.4)$}}
\end{picture}
}
\end{center}
\caption{ \label{FigFor2} The momentum fraction shape of reduced
GPDs with $\eta=0.5$ (left) and GDAs  with $\zeta=0.4$ (right),
same labelling as in Fig.\ \ref{FigConMom2}. }
\end{figure}
In Fig.\ \ref{FigFor2} we show the resulting GPDs for $\eta=0.5$
and GDAs for $\zeta = 0.4$. As has already been discussed the
skewness dependence of the conformal moments in Fig.\
\ref{FigConMom2}, cases (a) and (b) can  be hardly distinguished,
for GPDs they are quite the same.  In case (c) the area in the
central region and the magnitude of the maximum at $\eta=0$ are
enhanced, while the area of the outer region shrinks (the lowest
moment of all functions is normalized to one). Consequently, also
the magnitude at the cross-over points $x=\pm \eta$ decreases. The
reverse situation is observed for case (d). This is caused by the
analytic properties of the conformal moments with respect to the
variable $j$, which we explained above. Comparing both panels in
Fig.\ \ref{FigFor2}, one immediately sees that the ``width'' of
GPDs shows up in the end-point behavior of GDAs. It is remarkable
that the GDAs  (a) and (b) possess shapes that are governed mainly
by the first partial wave of the conformal expansion. Indeed, for
case (a) it turns out that in the region $0.05\leq \zeta \leq 0.95
$ the higher partial waves give at most a $10\%$ percent
correction. In the much narrower region $0.3\leq \zeta \leq 0.7 $
this is within the same accuracy also true for the GDA (b). This
region starts slightly about or below the value, where the series
turns over  to be divergent, i.e., for $\zeta \leq 0.25$ and  $
0.75\leq \zeta$. A slightly smaller convergency radius holds for
the GDA (c) and (d).

\begin{figure}[t]
\begin{center}
\mbox{
\begin{picture}(600,120)(0,0)
\put(15,0){\insertfig{7.5}{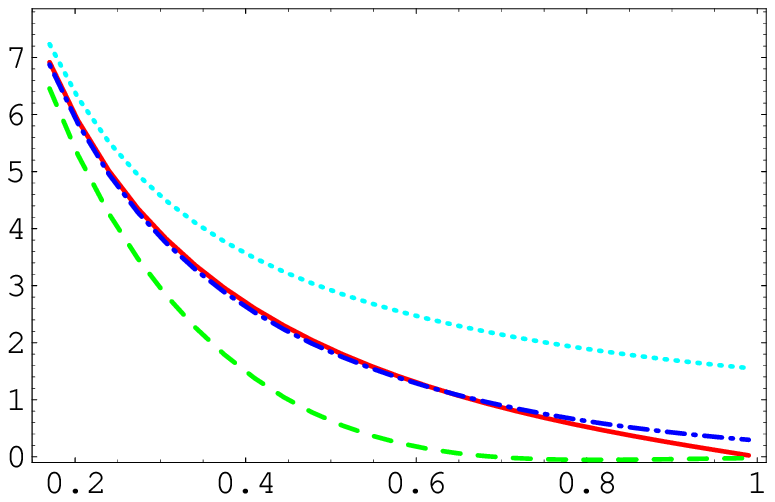}}
\put(215,-8){$\eta$}
\put(0,40){\rotatebox{90}{$\pi q(x=\eta,\eta)$}}
\put(277,0){\insertfig{7.7}{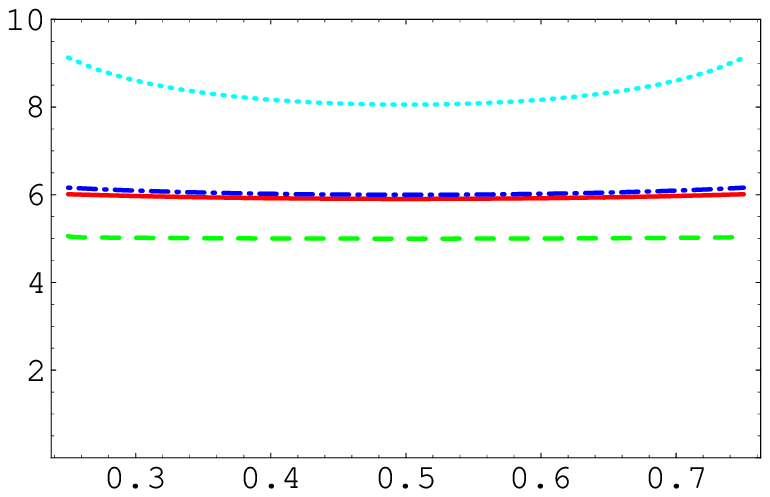}}
\put(485,-8){$\zeta$}
\put(260,40){\rotatebox{90}{$\int_0^1\!dz\, \Phi(z,\zeta)/z$}}
\end{picture}
}
\end{center}
\caption{ \label{FigAmp2} The reduced GPD at the cross-over point
$x=\eta$, multiplied with the factor $\pi$, (left) versus $\eta$
and the convolution of the GDA with  $1/z$ (right) versus $\zeta$,
same as in Fig.\ (\ref{FigConMom2}).}
\end{figure}
In Fig.\ \ref{FigAmp2} we show the quantities that enter the
scattering amplitude for a hard exclusive process, in which only
the virtuality of the incoming  photon  can be varied. Under such
circumstances the Compton form factors, i.e., Eq.\
(\ref{Def-ComForFac-LO}) or their crossing analogues Eq.\
(\ref{Def-Com2Pho-LO}), are measurable. The non-perturbative distributions
can not be directly determined by deconvolution, but at least within  our toy
ans\"atze the  correspondence between momentum fraction dependence
and the amplitudes is clear-cut.  In the left panel
of Fig.\ \ref{FigAmp2} we depict the GPD at the point $x=\eta$
multiplied by $\pi$. This cross-over point trajectory is
accessible in single spin asymmetry measurements. Within our
ans\"atze both the normalization and the slope encodes information
about the GPD shape. Compared with the left panel in Fig.\
\ref{FigFor2}, one realizes that the narrowest (widest) GPD has
the smallest (largest) value at the cross over point $x=\eta$. The
differences of the trajectories diminish with decreasing $\eta$.
Cases (a) and (b) are only distinguishable when $\eta$ tends to
one. Certainly, the slope of the  cross-over point trajectories is
dictated by the strength of the skewness dependence of the
conformal moments.

In the right panel of Fig.\ \ref{FigAmp2} we show the $\zeta$
dependence of the scattering amplitude, given as convolution of a
GDA  with a LO hard-scattering amplitude, see Eq.\
(\ref{Def-Com2Pho-LO}). It has been evaluated within the partial
wave expansion (\ref{Res-Com2Pho-LO}), taking the first thirty
terms into account.  The result is only displayed for the region
in which also the series for the conformal moments (b)-(d)
converges, i.e., for  $0.25\lesssim \zeta \lesssim 0.75$,  The
end-point behavior of GDAs determines  the normalization of the
scattering amplitude while the $\zeta$ dependence is in all cases
(rather) flat. As already mentioned, these convergency problems
arise from an exponential growth caused by the factor $2^n$ that
is contained in the normalized conformal partial waves
(\ref{Def-pn}). One would expect that the conformal partial wave
expansion of GDAs converges for all physical values of $\zeta$ and
$z$. To ensure this, the conformal moments must be exponentially
suppressed by a factor $2^{-n}$ for large $n$. We remind that the
normalization has been adopted from  parton densities and so this
factor will drop out in the limit $\eta\to 0$.

We come now to the $\eta$ dependence of the cross-over point
trajectory for smaller values of $\eta$. The trajectory is
represented by the Mellin-Barnes integral
\begin{eqnarray}
\label{Def-TraToy}
q(\eta,\eta)=
\frac{1}{2i}\int_{c-i\infty }^{c+i\infty }\!dj\,
\eta^{-j-1} \frac{2^{j+1}\Gamma(5/2+j)}{\Gamma(3/2)\Gamma(3+j)}\,
\frac{\Gamma(1/2+j) \Gamma(9/2)}{\Gamma(1/2)\Gamma(9/2+j)} b_j(\eta^2)\,,
\end{eqnarray}
which  can be used to derive, as explained in Sect.\
\ref{Sec-RepAmp}, an analytic expansion in terms of powers in
$\eta$. The leading terms
\begin{eqnarray}
\label{App-TraToy}
q(\eta,\eta)= \frac{35}{6 {\sqrt{2}\pi}\,{\sqrt{\eta}}} -
\frac{105\,{\sqrt{\eta }}}{8{\sqrt{2}\pi}}
 +\cdots
\end{eqnarray}
are ``universal'' in all cases and arise from the poles at
$j=-1/2$ and $j=-3/2$. The accuracy of this approximation for the
cross-over point trajectory is for all cases about  $1\%$ for
$\eta=0.05$ and, of course, starts to be much better with
decreasing $\eta$. Corrections to the approximation
(\ref{App-TraToy}) depend on the specific model and will not be
discussed here. For large values of $\eta$ they are smaller for
ansatz (a) than for (c) or (d).  Remarkably, in  case (b) the
integral (\ref{Def-TraToy}) can be exactly calculated:
\begin{eqnarray}
q(\eta,\eta)=
\frac{( 2 - \eta)^{7/2} }{2 {\sqrt{2}}\pi\, \eta }\,
{_2F}_1\!\left({1/2,5/2 \atop 9/2}\Big|
     1 -\frac{2}{\eta}\right)\,.
\end{eqnarray}

Let us summarize the lessons from this investigation. Unluckily,
after crossing it turns out that three of our models, (b)-(d),
lead to convergence problems for the conformal partial waves
series. Since these series should converges, the conformal moments
must for $\eta >1$ exponentially decrease as $2^{-n}$ with
increasing $n$. Viable ans\"atze for the conformal moments are
holomorphic functions of the conformal spin $j$ that respect the
following requirements:
\begin{itemize}
\item  they should be  bound for
$j \to \infty$ with $|{\rm \arg}(j)| \leq \pi/2 $ for $|\eta| \leq 1$
\item  An expansion of these functions  in $\eta^2$ should exist
\item  they should show an exponential suppression for $\eta^2 >1$  by a
factor $2^{-n}$ (for integer $n$)
\end{itemize}
In addition they should be flexible enough to be able to generate
a large variety of GPD shapes. In our examples only case (a),
i.e., the analytic continuation of Gegenbauer polynomials, satisfy
the three requirements. Applying suitable integral transformations
to them allows to generate ``associated'' polynomials. We did not
study in detail how flexible the  GPD shapes can be parameterized
within this method but got the impression that only small changes
for large $\xi$ are possible. We are award that these examples do
not cover all possible types of conformal moments. One important
lesson is that the analytic properties of conformal moments
determine the qualitative features of GPDs. Certainly, our
examples here give only a first insight into this connection. For
completeness, we mention the existence of a symmetry relation that
arises from the definition (\ref{Def-cHyp}), namely, the
invariance under the replacement $j\to -j-3$ \cite{ManKirSch05}:
\begin{equation}
\frac{2^{1+j}
\Gamma(3/2 + j)}{\Gamma(3/2) \Gamma(3 + j)} \eta^{-j}c_{j}(x,\eta) =
\frac{2^{1+j}
\Gamma(3/2 + j)}{\Gamma(3/2) \Gamma(3 + j)}\;
\eta^{-j}  c_j(x,\eta)\Big|_{\textstyle j\to -j-3}\,.
\end{equation}
Since, however, perturbative QCD, e.g., anomalous dimensions,
already violate this symmetry, there is no need to implement it in
a non-perturbative ansatz for conformal moments.

\subsubsection{Implementation of momentum squared dependence}
\label{SubSubSec-AnsConMomDep-t}

Now we address the implementation of the $\Delta^2$ dependence. We
might  include it by introducing $\Delta^2$ depended parameters in
the ansatz (\ref{Ans-mj}) in such a way that no singularities in
the first and forth quadrant of the complex $j-$plane appear. For
the sake of simplicity we write here a factorized ansatz
\begin{eqnarray}
\label{Ans-mj-t}
m_j(\eta,\Delta^2|\alpha,\beta) =
\frac{\Gamma(\alpha+1+j) \Gamma(\alpha+\beta+2)}{\Gamma(\alpha+1)
\Gamma(\alpha+\beta+2+j)} b_j(\eta^2) F_j(\Delta^2)\,,
\end{eqnarray}
with the normalization condition $F_j(\Delta^2=0)=1$.  In the case
that the form factors  $F_j(\Delta^2)$ are $j$ independent the $x$
and $\Delta^2$ dependence is obviously factorized, too. Lattice
calculations for the first moments of $u$ and $d$ quark GPDs
suggest that such a factorization is actually not correct.
According to these results, the cut-off mass squared in a dipole
fit increases with $j=n$. Unluckily, the systematic theoretical
uncertainties of these results, especially, those associated with
the chiral extrapolation are still kind of large,  but  the
preferred fits suggest a linear growth with $(j+1)$ \footnote{The
$m^2_{d,n}= (n+1) m^2_{d,0}$ dependence for the squared dipole
masses $m^2_{d,n}$, taken from Table I of Ref.\ \cite{Hag04a}
gives a very good  fit for the generalized form factors
$A_{n,0}^{u-d}$ ($m_\pi=897$ MeV ) and  $A_{n,0}^{u+d}$
($m_\pi=744$ MeV ) with $n=0,1,2$. For the large pion mass
$m_\pi=897$ MeV  one might have the impression that the growth is
stronger (weaker) for the (axial-)vector case, which might be
caused by a different intercept in the power behavior. Certainly,
a definite conclusion can not be drawn from present lattice
measurements.} \cite{Hag04a}. A second constraint on the
$\Delta^2$ dependence arises from the lowest moment of GPDs, which
is related to partonic form factors, e.g., for the GPD $H$:
\begin{eqnarray}
F^{p,{\rm val}}_1(\Delta^2) =
\int_{-\eta}^1\!\, dx H_{p_{\rm val}}(x,\eta,\Delta^2)\,.
\end{eqnarray}
Where $ F^{p,{\rm val}}_1(\Delta^2)$ can be expressed in terms of
the proton ($p$) and neutron ($n$) electromagnetic Dirac form
factors according to
 \begin{eqnarray}
2 F^{u,{\rm val}}_1(\Delta^2) =
2 F^{p}_1(\Delta^2) +  2 F^{n}_1(\Delta^2)\,, \quad
F^{d,{\rm val}}_1(\Delta^2) = F^{p}_1(\Delta^2) +  2 F^{n}_1(\Delta^2)\,.
 \end{eqnarray}
We remark that the lowest moment of a GPD is simply obtained by
setting $j=0$ in the ansatz (\ref{Ans-mj-t}).

Now we are in the position to build a minimal GPD ansatz for the
valence quark GPDs $H$ which satisfy the theoretical constraints
in the forward case and provide for the lowest moment the correct
$\Delta^2$ dependence. Suppose we use the parameterization of the
forward parton distribution at a given input scale in the form
(\ref{Par-ParDen}), the conformal GPD moments read then, e.g., for
the $u$ quark
\begin{eqnarray}
\label{Par-ConMomGPD}
m^{u_{\rm val}}_j(\eta, \Delta^2) &\!\!\!=\!\!\! &
\frac{2}{1+A^\prime + B^\prime}
\left[m_j(\eta, \Delta^2 |\alpha,\beta) +
A^\prime\, m_j(\eta,\Delta^2 |\alpha+1/2,\beta)+
B^\prime\, m_j(\eta,\Delta^2|\alpha+1,\beta)\right]\,,
\nonumber\\
m_j(\eta, \Delta^2 |\alpha,\beta) &\!\!\!=\!\!\! &
\frac{\Gamma(\alpha+1+j)
\Gamma(\alpha+\beta+2)}{\Gamma(\alpha+1)\Gamma(\alpha+\beta+2+j)}
b_j(\eta^2) F^{u,{\rm val}}_1(\Delta^2/(j+1))\,,
\end{eqnarray}
normalized to $2$ for $j=0$ and $\Delta^2=0$. Guided by the
lattice results, we rescaled here the dipole masses in the form
factor $F^{u,{\rm val}}_1$ by $j+1$, which results in a dependence
on the ratio $\Delta^2/(j+1)$. The remaining free parameters
$A^\prime,B^\prime, \alpha,\beta $ are taken from the parton
density parameterization and for $b_j(\eta^2)$ one can use one of
the functions suggested above, where preference is given to case
(a). Analogously, one can deal with other species of GPDs that
reduce in the forward limit to parton densities. However, for
these the $\Delta^2$ dependence is typically far less known. In
particular, unpolarized sea quark GPDs are not constraint by
elastic form factor measurements.

Let us have a closer look at the parameterization
(\ref{Par-ConMomGPD}). For $j=0$ we obviously find from this
parameterization
\begin{eqnarray}
\int_{-\eta}^1\!\, dx H_{u_{\rm val}}(x,\eta,\Delta^2) =
2 b_0(\eta^2) F^{u,{\rm val}}_1(\Delta^2)\,,
\end{eqnarray}
where the $\eta$-dependence drops out, $b_0(\eta^2)=1$ and the
correct normalization is ensured. Setting $\Delta^2=0$, it is also
clear from Eq.\ (\ref{Ans-mj-t})  that we arrive by an inverse
Mellin transform at the parton density. It has been mentioned
above that the GPD for  $\eta\to 0$ and small $x$ is dominated by
the leading Regge trajectory. In the following we argue that after
a small modification this trajectory is  already present in our
parameterization. In our case  we are dealing with the $\rho^0$ and
$\omega$ trajectories, which according  to the analysis of Ref.\
\cite{DieFelJakKro04} are parameterized by a  linear
$\Delta^2$-dependence
 \begin{eqnarray}
 \label{RegTraDieJakKro}
 \alpha_\omega(\Delta^2) = 0.42 +  \Delta^2\, 0.95\, \GeV^{-2},\quad
  \alpha_\rho(\Delta^2) = 0.48 + \Delta^2 \, 0.88\, \GeV^{-2}\,.
 \end{eqnarray}
If we take for the elastic nucleon form factor the standard dipole
parameterization and consider only the first pole in the complex
$j$-plane of the  parton density Mellin--moments, i.e., the pole
that appears on the negative axis at the largest value of $j$, our
parameterization for the conformal moments reads
\begin{eqnarray}
\label{Res-Reg}
m^{{\rm val}}_j(\eta, \Delta^2) &\!\!\!\propto\!\!\!&
\frac{1}{\alpha+1+ j }\,
\frac{(j+1+\alpha)}{m^2_d (j+1+\alpha) - \Delta^2 (1+\alpha)}
\\
&&\times
\left( \frac{A}{4 M^2_N  -(1+c)\Delta^2/(j+1+c)} +
\frac{B}{m^2_d - (1+d) \Delta^2/(j+1+d)}\right)\,.
\nonumber
\end{eqnarray}
Here $M_N$ is the  nucleon mass, $m_d$ is the dipole mass
appearing the parameterization of the Sachs form factors, e.g., in
the electric one of the proton  $G^p_E(\Delta^2) =
1/(1-\Delta^2/m^2_d)^2$. $A$ and $B$ are two constants,  the
values of which can be read off, e.g., from  Ref.\
\cite{BelMueKir01}. Moreover, we modified here the rescaling of
the $\Delta^2$ dependence in such a way that the pole at
$j+1+\alpha=0$ cancels against the scaling factor $(j+1+\alpha)$.
The constants $c$ and $d$, appearing in the scaling factors of the
remaining $\Delta^2$ dependence, are chosen to be positive so that
they do not interfere with the leading Regge trajectory. This
linear trajectory is given by
\begin{eqnarray}
\label{AnsRegAns1}
\alpha(\Delta^2) = \alpha(0) + \alpha^\prime(0)\Delta^2 \,,
\qquad \alpha(0)= -\alpha\,,
\quad \alpha^\prime(0)=\frac{1+\alpha}{m_d^2}
\end{eqnarray}
and so we find with Eqs.\ (\ref{Par-ConMomGPD}) and (\ref{Res-Reg})
\begin{eqnarray}
\label{AnsRegAns2}
m^{{\rm val}}_j(\eta, \Delta^2) =
\frac{2}{1+A^\prime + B^\prime}
\frac{\Gamma(\alpha+2+j)
\Gamma(\alpha+\beta+2)}{\Gamma(\alpha+1)\Gamma(\alpha+\beta+2+j)}
\frac{\beta_j(\Delta^2) b_j(\eta^2)}{(j+1)-\alpha(\Delta^2)}
\left(1 + A^\prime \cdots + B^\prime \cdots \right) \,,
\nonumber\\
\end{eqnarray}
where the remaining $\Delta^2$ dependence is accumulated in the function
\begin{eqnarray}
\label{AnsRegAns3}
\beta_j(\Delta^2) =  \frac{A}{4 M^2_N  -(1+c)\Delta^2/(j+1+c)} +
\frac{B}{m^2_d - (1+d) \Delta^2/(j+1+d)}\,.
\end{eqnarray}

As we saw, after rescaling of the dipole masses the leading pole
at $j=-1-\alpha$ in Eq.\ (\ref{Par-ConMomGPD}) is in the
parameterization (\ref{AnsRegAns2}) shifted to
$j=-1+\alpha(\Delta^2)$ by an amount proportional to  $\Delta^2$.
The numerical value of the dipole mass  is $m_d=840$ MeV and that
of $\alpha$ depends on the factorization scale. When we choose the
scale to be the intercept of the Regge trajectories
(\ref{RegTraDieJakKro}) we find  the following slopes
\begin{eqnarray}
\alpha_\omega^\prime(0)= 0.82\, \GeV^{-2}\,,
\quad \alpha_\rho^\prime(0)= 0.74\, \GeV^{-2}\,.
\end{eqnarray}
These values are only about $15\%$ smaller than the ones given in
Eq.\ (\ref{RegTraDieJakKro}). In view of the fact that we used just
the standard dipole parameterization of the elastic form factors
and neglected all non-leading Regge trajectories, this agreement
is quite astonishing. We interpret it as further evidence that
Regge theory or at least Regge phenomenology is indeed applicable
to GPDs. A deeper understanding of this issue could provide the
key to a dual description of GPDs in terms of hadronic degrees of
freedom and certainly warrants dedicated efforts.
\begin{figure}[t]
\begin{center}
\mbox{
\begin{picture}(600,120)(0,0)
\put(50,60){(a)}
\put(10,-4){\insertfig{8.2}{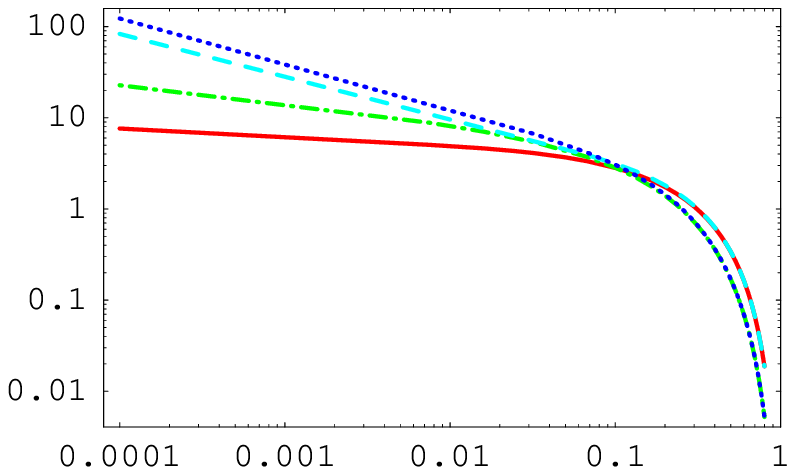}}
\put(0,25){\rotatebox{90}{$H_u(x,\eta=0,\Delta^2,{\cal Q}^2)$}}
\put(240,-5){$x$}
\put(310,60){(b)}
\put(275,0){\insertfig{7.7}{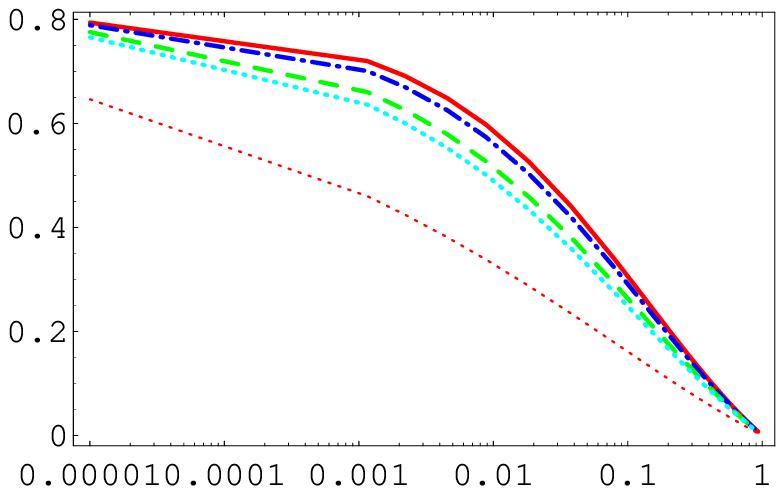}}
\put(260,35){\rotatebox{90}{$\langle \vec{b}^2 \rangle_u(x)\;
[{\rm fm}^{2}]$}}
\put(485,-5){$x$}
\end{picture}
}
\end{center}
\caption{ \label{Figavb} The $u$-valence quark GPD
$H_u(x,\eta=0,\Delta^2,{\cal Q}^2)$ of the proton is displayed in
panel (a) versus $x$ for fixed $\Delta^2=-0.25\, \GeV^2$ with a
$\Delta^2$-dependence arising from a Regge trajectory via Eqs.\
(\ref{AnsRegAns1}), (\ref{AnsRegAns2}), and (\ref{AnsRegAns3})
(solid   and dash-dotted) and with a rescaled $1/(j+1)$
$\Delta^2$-dependence via Eq.\ (\ref{Par-ConMomGPD}) (dashed and
dotted). The GPD is evolved from the input scale ${\cal Q}^2=0.5\,
\GeV^2$ (solid and dashed) to ${\cal Q}^2= 10\, \GeV^2$
(dash-dotted and dotted). In panel (b) we show for the
$\Delta^2/(j+1)$-depended GPD $H_u$ the expectation value for the
square of  the impact parameter (\ref{Res-aveb}). The different
lines correspond to  ${\cal Q}^2=0.5\, \GeV^2$ (solid), ${\cal
Q}^2=1\, \GeV^2$ (dash-dotted), ${\cal Q}^2=10\, \GeV^2$ (dashed),
${\cal Q}^2=100\, \GeV^2$ (dotted) and  ${\cal Q}^2=10^{100}\,
\GeV^2$ (thin dotted). We used the following parameter set:
$A^\prime=B^\prime=0, \alpha=-1/2, \beta=3$ and $c=d=1$. }
\end{figure}

\subsubsection{Numerical consequences for the probabilistic
interpretation of GPD $H$}
\label{SubSubSec-AnsConMomEta0}

As mentioned in  the introduction, GPDs  possesses for $\eta=0$ a
probabilistic interpretation in the infinite momentum frame. In
particular the Fourier transform\footnote{To avoid a confusion
with the definition of the GPD $\widetilde{H}_q(x,\eta,\Delta^2)$
in the axial-vector case, we omit the tilde symbol  for the
Fourier transform of $H_q(x,\eta,-\vec{\Delta}^2)$. The quantities
in the two-dimensional impact space $\vec{b}$ are indicated by
their argument $\vec{b}$.}
\begin{eqnarray}
\label{Def-Hq-eta0}
H_q(x,\vec{b}) =
\int\!\frac{d^2\vec{\Delta}}{(2\pi)^2}\,
e^{-i \vec{b}\cdot \vec{\Delta}} H_q(x,\eta=0,\Delta^2=-\vec{\Delta}^2)
\end{eqnarray}
is the probability to find a quark species $q$ inside the nucleon
with  momentum fraction $x$ at impact parameter $\vec{b}$. The
latter is defined relative to the center of momentum of the
hadron, i.e., $\vec{b}$ is the distance of the active parton in the
transversal direction from this center. It is worth mentioning
that the definition of such a center is based on the existence of
a Galilean subgroup of transverse boosts in the infinite moment
frame \cite{KogSop70}. Within  light-cone quantization such
transverse boosts have a field theoretical definition  in terms of
two conserved charges, expressed by the plus components of the
energy momentum tensor. Their eigenvalues are good quantum number
that labels the states of the hadron. Conveniently, they are
chosen to be zero for the center.

This probabilistic interpretation of the GPD (\ref{Def-Hq-eta0})
has inspired several authors to build GPD ans\"atze to get a first
glimpse of the three-dimensional tomographic picture of the
nucleon. This has often been done using an exponential ansatz for
the $\Delta^2$-dependence, which might serve it purpose in the
space like region, however, violates the analytic properties of
scattering amplitudes etc.\ such that  crossing relations
become meaningless. In the following we study our conformal
moments for $\eta =0$ under this aspect. The only uncertainty
which is left in our GPD representation is the $j$-dependence of
the form factor. In Fig.\ (\ref{Figavb}) we present the momentum
fraction dependence for fixed $\Delta^2= -0.25\, \GeV^2$ and two
different resolution scales. Obviously, the implementation of the
Regge trajectory (solid and dash-dotted lines) results in a
flatter $x$-dependence compared to a simple rescaling  of the
dipole mass squared with $j+1$ (dashed and dotted lines).

As a side remark, we comment on the factorized $\Delta^2$ ansatz
for GPDs. Although, it is wrong in principle, this does not
necessarily imply that all estimates for  observables fail
completely, at least not for smaller values of $\Delta^2$. For the
ans\"atze of conformal moments, we discussed in the both previous
sections, the normalization of the resulting GPDs between the
different versions of $\Delta^2$-dependence varies for $0.08
\lesssim x\lesssim 0.2$ and $|\Delta^2| < 0.3\, \GeV^2$, i.e., in
the fixed target kinematics, not larger than $20\%$. Certainly,
for lower or larger values of $x$ the differences especially in
the overall size can increase drastically, compare, e.g., the solid
and dashed line in Fig.\ \ref{Figavb} (a). On the other hand the
suppression introduced by the Regge motivated ansatz (see solid
and dash-dotted lines) is welcome to suppress sea quark and
gluonic contributions, which notoriously are overestimated in the
factorized $\Delta^2$ ansatz for hard exclusive electroproduction
processes.

The average distance from an active parton to the center of
the nucleon is defined as \cite{Bur00,Bur02}
\begin{eqnarray}
\label{Def-aveb}
\langle \vec{b}^2 \rangle_q(x,{\cal Q}^2) =
\frac{\int\!d\vec{b}\,\vec{b}^2 H_q(x,\vec{b},{\cal Q}^2)}{\int\!d\vec{b}\,
H_q(x,\vec{b},{\cal Q}^2)} =
4 \frac{\partial}{\partial \Delta^2}
\ln H_q(x,\eta=0,\Delta^2,{\cal Q}^2)\Big|_{\Delta^2=0}\,.
\end{eqnarray}
Within the parameterization (\ref{Par-ConMomGPD}), where for
simplicity we again rescale  the $\Delta^2$ dependence by
$1/(j+1)$,  this average distance can be exactly calculated and
expressed in terms of forward parton distributions
\begin{eqnarray}
\label{Res-aveb}
\langle \vec{b}^2 \rangle_q(x,{\cal Q}^2) =
\frac{\int_x^1\!\frac{dy}{y}\, q(y,{\cal Q}^2)}{q(x,{\cal Q}^2)}
\,4 \frac{\partial}{\partial \Delta^2} \ln
F^{q}_1(\Delta^2)\Big|_{\Delta^2=0}\,.
\end{eqnarray}
The  $1/(j+1)$ factor that arises in Mellin space from the
differentiation with respect to $\Delta^2$ gives in $x$ space rise
to the integral $\int_x^1\! dy/y\cdots$. Obviously, by a more
refined rescaling of the $\Delta^2$ dependence, see Eq.\
(\ref{Res-Reg}), we can express the resulting average by a more
complex integral. For small  $x$ this quantity tends to a
constant, depending only on the resolution scale, while at large
$x$ it vanishes as $(1-x)$. The latter behavior is a consequence
of the linear $j$-dependence of the dipole masses. Such a behavior
has been rejected in Ref.\ \cite{Bur04}. Namely, the quantity
\begin{eqnarray}
d(x)= \frac{\langle \vec{b}^2 \rangle_q(x)}{(1-x)^2}
\end{eqnarray}
is interpreted, based on a partonic picture, as the distance of
the active parton from the center of momentum of the spectators
and should therefore be finite for $x\to 1$. Consequently,
$\langle \vec{b}^2 \rangle_q(x)$ should vanishes at least with
$(1-x)^2$ for $x\to 1$. Although this argumentation is further
supported by perturbative QCD arguments \cite{Yua03}, it is in our
opinion not excluded that the simple parton picture can  be
misleading for quantities, which have no well-defined
field-theoretical analog.  Note also that a $(1-x)^2$ behavior of
$\langle \vec{b}^2 \rangle_q(x)$ for $x\to1$ requires a $(1+j)^2$
growth of the squared dipole masses. This means in terms of Regge
phenomenology that the trajectories should possesses a (small)
non-linear term.

Let us come back  to our ansatz. To include the scale dependence
we use for the GPD $H_q(x,\eta=0,\Delta^2,{\cal Q}^2)$  in Eq.\
(\ref{Def-aveb})  the Mellin-Barnes integral and insert the
evolution operator, compare with Eq.\ (\ref{Sol-EvoEqu-LO}),
\begin{eqnarray}
\langle \vec{b}^2 \rangle_q(x,{\cal Q}^2) =\frac{
\int_{c-i
\infty}^{c+i \infty}\!dj\, \frac{x^{-j-1}
\Gamma(1+j+\alpha)}{\Gamma(1+j+\alpha+\beta) (j+1)}\,
e^{\left\{-\frac{\gamma_j^{(0)}}{2}
\int_{{\cal Q}_0^2}^{{\cal Q}^2}\frac{d\sigma}{\sigma}
\frac{\alpha_s(\sigma)}{2\pi} \right\}}}{\int_{c-i
\infty}^{c+i \infty}\!dj\, \frac{x^{-j-1}
\Gamma(1+j+\alpha)}{\Gamma(1+j+\alpha+\beta)}\,
e^{\left\{-\frac{\gamma_j^{(0)}}{2}
\int_{{\cal Q}_0^2}^{{\cal Q}^2}\frac{d\sigma}{\sigma}
\frac{\alpha_s(\sigma)}{2\pi} \right\}}}\,
4 \frac{\partial}{\partial \Delta^2}
\ln F^{q_{\rm val}}_1(\Delta^2)\Big|_{\Delta^2=0}\,.
 \end{eqnarray}
For experimental accessible values  of ${\cal Q}^2$ and for small
$x$, both the  numerator and denominator are dominated by the
leading pole at $j= -1-\alpha$. Shifting the integration path to
the left we find for  $x\to 0$ the constant value
\begin{eqnarray}
\label{Res-aveb1}
\langle \vec{b}^2 \rangle_q(x=0,{\cal Q}^2)  =
-\frac{4}{\alpha} \frac{\partial}{\partial \Delta^2}
\ln F^{q_{\rm val}}_1(\Delta^2)\Big|_{\Delta^2=0} \,.
\end{eqnarray}
We remark that a Regge motivated ansatz would induce a double pole
in the numerator and, consequently,  a logarithmic modification
with respect to both the $x$- and ${\cal Q}^2$-dependencies.
Within the dipole ansatz for the partonic form factors, taken from
Ref.\ \cite{BelMueKir01}, we have
\begin{eqnarray}
\label{Lim-aveb}
\langle \vec{b}^2 \rangle_u(x=0,{\cal Q}^2)  =
-\frac{1}{\alpha} 0.4  {\rm fm}^2
~ ~ &~ ~ ~&  ~ ~
\langle \vec{b}^2 \rangle_d(x=0,{\cal Q}^2)  =
-\frac{1}{\alpha} 0.53  {\rm fm}^2
\end{eqnarray}
for the  $u$  and $d$ valence quarks. From Fig.\ \ref{Figavb}(b),
where we used the generic value $\alpha=-1/2$, we can read off the
qualitative $x$-dependence of $\langle \vec{b}^2 \rangle_{q_{\rm
val}}(x,{\cal Q}^2)$ for given resolution scale ${\cal Q}^2$. It
can be roughly approximated by a logarithmic growth with
decreasing $x$ which changes slope  at some ``cusp'' point $x_{\rm
cusp}({\cal Q}^2) \sim 10^{-3}$. With increasing ${\cal Q}^2$ this
cusp is washed out. The increase with $\ln(1/x)$ visible in Fig.\
\ref{Figavb} (b), will not continue in the limit $x\to 0$ but will
saturate at a finite value (\ref{Lim-aveb}). Numerically, we
checked this qualitative feature up to ${\cal Q}^2 = 10^{10000}\,
\GeV^2$ and $1/x = 10^{40}$.

So within our simple ansatz, which can be easily refined, we
reproduce following well-known grand picture. At large $x$ the
valence quarks are in the center of the nucleon. With decreasing
$x$ they become ever more delocalized in the transverse direction
and will move away in transversal direction with decreasing
momentum fraction $x$. In the valence quark region with $x\sim
0.3$ the average transverse distance is of the order $\sim 0.4\,
{\rm fm}$. With increasing resolution scale this value is slowly
decreasing. In the small $x$ region their transverse distance
grows, reaches (for the ansatz we used) $\sim 0.9\, {\rm fm}$  in
the limit $x\to 0$  for $u$ valence quarks and a slightly higher
value for $d$ quarks. Hence, the charge of the proton, which is
probed in a scattering experiment, has a non-trivial transverse
distribution.

\section{Summary and conclusions}

In this paper we have derived a new representation for leading
twist-two GPDs and GDAs in terms of Mellin-Barnes integrals over
the complex conformal spin. This representation is rather
analogous to the partial wave expansion of scattering amplitudes
with respect to complex angular momentum, given in terms of
Legendre polynomials or functions, respectively. Indeed, also our
partial waves can be  expressed in terms of associated Legendre
functions of the first and second kind. Mathematically, there
should exist a one-to-one relation to other representations that
are based on  conformal symmetry. The transformations between
different  representations, however, still have to be worked out
in detail. For instance, taking the Mellin transform of the
Mellin-Barnes integral must lead to  the integral kernel that maps
the ``effective forward parton distribution'' to GPDs
\cite{Shu99,Nor00} or employing the Fourier transform one must
arrive at the light-cone position space representation
\cite{BalBra89,KivMan99b,ManKirSch05}. We confirm and generalize
the results for the gluonic sector given in Ref.\
\cite{ManKirSch05}. The advantage of our new representation is
that the central and outer region are obtained from the same
conformal moments of a GPD and polynomiality is manifestly
implemented. This has not been done so far in the approaches, we
mentioned. Sometimes,  the central region was even treated
incorrectly, for comments see Ref.\ \cite{Nor00}. We derived a
``spectral" representation of conformal kernels with complex
valued conformal spin. However, we are not completely satisfied
with this representation, since some support restriction have to
be fixed explicitly by step-functions. So far an orthogonal
eigenfunction basis for these kernels can only be given as series
of mathematical distributions, which are labelled by non-negative
integer conformal spin.

To leading order the kernels and hard-scattering amplitudes
respect conformal symmetry and so we could provide the solution of
the evolution equation and the Compton form factors for DVCS as
Mellin-Barnes integral.  This representation allows a simple and
stable numerical evaluation of these quantities. This has some
practical advantages, especially, having efficient numerical
routines at hand for the evaluation of the Mellin-Barnes integral,
one might be able to evaluate this quantities in ``real time''
rather than using a database of Compton form factors or GPDs.  We
even demonstrated for the Compton form factors that by means of
the Mellin--Barnes integral a systematic analytic approximation in
powers of $\xi$ is feasible.

Beyond LO order the conformal symmetry is broken in a subtle way
by the minimal subtraction scheme, applied to the divergencies of
composite operators, and the trace anomaly of the energy momentum
tensor. The first effect can be cured by a finite renormalization,
while the latter one can be absorbed into either the
hard-scattering amplitudes or the evolution of GPDs. Such symmetry
breaking effects have been studied for $\eta=1$ in connection with
the pion distribution amplitude and the pion elastic and $\gamma
\gamma^\ast \to \pi^0$ transition form factor. Depending on the
model for the pion distribution amplitude one finds a  $10\%$
variation of the NLO corrections \cite{Mue98}. For skewness $\eta
< 1$ one would expect an even smaller effect because the conformal
symmetry breaking is  suppressed by a factor $\eta^2$. For
realistic experiments typically  $\eta \lesssim 0.4$. Further
studies are desirable to clarify this issue and, hopefully might
provide a systematic expansion of the conformal symmetry breaking
effects in powers of $\eta^2$.

Certainly, the evaluation of the analytic continued  conformal
moments from a given GPD ansatz is a rather non-trivial task, the
difficulty of which depends on the analytic properties of the
corresponding GPD. Since, however, GPDs are almost unknown
non-perturbative functions, one should rather  model the conformal
moments directly instead of the GPDs or DDs. This is a non-trivial
task with respect to skewness dependence, since certain analytic
properties of the conformal moments must be respected. Some effort
is required to tune the internal skewness dependence, arising from
the conformal partial waves,  accordingly. In the end, however, it
turned out that our ans\"atze generated GDAs with non-vanishing
end-point contributions. This illustrates again that known
constraints for GPDs are very difficult to fulfill. An important
observation one can draw from our  examples is that ``strong
skewness'' effects  already show up in the first few conformal
moments for non-negative integer conformal spin. So it is
worthwhile to calculate them on the lattice to get a clue for the
strength of the skewness dependence of  GPDs.

A partial wave expansion of the conformal moments itself avoids
the problems described above  yields a most flexible
parameterization of GPDs in terms of form factors.  These form
factors are related to particle exchanges in the $t$-channel, but
this connection still has to be worked out in detail. For $\eta=0$
there is no difference between  conformal spin and ordinary spin
and the conformal partial wave expansion turns over into one  with
respect to spin. In this kinematical domain  we found  some
evidence that Regge phenomenology can be used as a reliable guide
for modelling the conformal moments. Especially, for unpolarized
valence quark GPDs the parameterization of parton densities and
elastic electromagnetic form factors can be unified and
interpreted as a leading Regge trajectory. Since GPDs depend on
the factorization conventions, such a connection can only hold
approximatively. Moreover, Regge theory is only applicable for
physical amplitudes and going beyond the leading trajectory has
its own difficulties. Nevertheless, the Regge analogy provides
some guidance for the modelling of conformal moments. This is
especially important for those GPDs that are difficult to extract
from experiments.

The advantage of the Mellin-Barnes representation has been
demonstrated for several analytic and numerical examples,
especially, for unpolarized valence quarks in the $\eta=0$ case.
The only unknown in this limit is the spin, i.e., $j$, dependence
of the form factors, which for $j=0$, are measured in elastic
electron proton scattering. The momentum fraction dependence  of
the GPD follows then from the $t$- and ${\cal Q}^2$-dependencies,
where the boundary condition  at the input scale ${\cal Q}_0^2$
and at $\Delta^2=0$ can be simply taken from the parameterization
of Mellin moments for parton densities. Remarkably, no additional
fitting procedure is needed to satisfy the GPD constraints.
However, different parameterization of the form factors with
respect to $j$, can lead to a different holographic picture of the
nucleon. Certainly, the important task here is to pin down the
remaining degrees of freedom for the $j$ dependence. No question,
improved lattice calculation with a realistic pion mass can
provide at least a partial answer.

We would like to add a speculation concerning  the  experimental
access to this dependence in hard exclusive reactions. Suppose it
turns out from lattice measurements that the skewness dependence
of the conformal moments is weak, the $\xi$-dependence for the
scattering amplitude is (approximately) known and the only degree
of freedom left is the unknown $j$-dependence of these form
factors, which determines the shape of the trajectory of
the cross-over point of a GPD as function of  $\xi$, in dependence
of $\Delta^2$. Such a trajectory can be explored in single beam
spin experiments. Taking Mellin moments of this trajectory, which
requires of course some interpolation, one directly gets the form
factors, up to some normalization factor, in dependence of the
(conformal) spin.

Let us finally stress that the crossing relation between GPDs and
GDAs is very simple for conformal moments. Apart from a trivial
rescaling procedure of the skewness dependence it involves only
the $\Delta^2$ or $W^2$ dependence. One has to perform an analytic
continuation of the form factors from the space- to the time-like
region, which requires only a suitable parameterization in terms of
rational functions (linear combination of monopole or dipole
forms), which scale for $\Delta^2\to-\infty$ such as predicted by
dimensional counting rules. As there exist  also non-perturbative
scales, like the hadron masses themselves or $\Lambda_{\rm QCD}$,
anomalous, i.e., logarithmical deviations, from the canonical
scaling should be present to some extend. Although, we were not
able to give for all examples of conformal moments  an unified
representation of GPDs and GDAs in terms of  a Mellin-Barnes
integral this is not a restriction in practice. If one has an
ansatz for the dependence of the conformal moments on the complex
conformal spin, one can employ for GDAs the partial wave expansion
with integer conformal spin.

In conclusion, we have introduced a representation that makes it
easier to include GPDs and GDAs in phenomenological studies and
offers new theoretical possibilities for the investigation of
perturbative and non-perturbative aspects.

{\ }\\

\noindent
{\bf Acknowledgement}

\noindent For discussions on  mathematical aspects  we are
indebted to A.\ Manashov and for a general discussion on GPDs we
like to thank M.~Diehl, which inspired us to include several
remarks. This project has been supported by the DFG.

\appendix

\section{Integrals}
\label{App-Int}

In this appendix we collect several integrals, which appear in the
evaluation of moments or the convolution of GPDs  with the hard
scattering amplitude. The partial waves $p_j(x,\eta)$, given in
Eqs.\ (\ref{Def-p-all}), (\ref{Def-p-P}), and (\ref{Def-p-Q}),
might be expressed by associated Legendre functions of the first
and second kind, i.e., by $\sqrt(1-x^2) P_{j+1}^{-1}(x)$ and
$\sqrt(1-x^2) Q_{j+1}^{-1}(x)$. The conformal moments
$c_j(x,\eta)$, see Eq.\ (\ref{Def-cHyp}), can be represented
within the same basis, however, divided by the weight $(1-x^2)$
\cite{AbrSte}. The  integrals, presented in the following, can
then be read off from diverse integral tables. Moreover, we give
the relation between conformal and ordinary Mellin moments.

The conformal moments of the partial waves for complex valued
conformal spin read in the outer region
\begin{eqnarray}
\label{Def-Ort-Out} \int^{\infty}_{\eta}\! dx\, c_k(x,\eta)
p_j(x,\eta)= \frac{\sin(\pi j)}{\pi}  \frac{{\cal N}_{kj}(\eta)
}{k-j}\,, \quad \Re{\rm e}\,j>\Re{\rm e}\,k\,,
\end{eqnarray}
where the normalization factor is given by
\begin{eqnarray}
{\cal N}_{kj}(\eta)=\frac{2 \Gamma(3 + k)\Gamma(5/2 + j)}
{\Gamma(3 + j)\Gamma(3/2 + k)(j+k+3)}
\left(\frac{\eta}{2}\right)^{k-j}\,, \quad {\cal N}_{jj}(\eta) =
1\,.
\end{eqnarray}
 Note that this integral
(\ref{Def-Ort-Out}) only converges for $\Re{\rm e}\,j>\Re{\rm
e}\,k$. In the central region we have the following integral
\begin{eqnarray}
\label{Def-Ort-Cen} \int^{\eta}_{-\eta}\!dx\, c_k(x,\eta)
p_j(x,\eta)= \left( \frac{\sin(\pi j)}{\pi}  -
\frac{(j+1)(j+2)}{(k+1)(k+2)} \frac{\sin(k \pi)}{\pi}  \right)
 \frac{{\cal N}_{kj}(\eta) }{j-k}\,.
\end{eqnarray}
On the r.h.s.\ two terms appear in the brackets. The
sum of the former one and the integral
(\ref{Def-Ort-Out}) is proportional to the identity
$\delta(j-k)$, however, the latter one gives an
addendum. Fortunately, for integer value $k=m=
0,1,2,\cdots$ it  will not contribute. Remarkably, if
the support of $c_k(x,\eta)$ is extended to $x\leq
-\eta$ by means of
\begin{eqnarray}
\label{Def-c-ext} c_k(x,\eta) = \left(\frac{\eta}{2}\right)^{2k+2}
\frac{\sin(k \pi)
\Gamma(1+k)\Gamma(3+k)}{\Gamma(3/2+k)\Gamma(5/2+k)} x^{-k-3}\;
{_2\!F}_1\!\left({(k+3)/2,(k+4)/2\atop
5/2+k}\Big|\frac{\eta^2}{x^2}\right),\; x\leq-\eta
\end{eqnarray}
and for $p_j(x,\eta)$ by analytic continuation into the region $x < 0$,
we have the following integral
\begin{eqnarray}
\label{Def-Ort-Out2} \int_{-\infty}^{-\eta}\! dx\, c_k(x,\eta)
p_j(x,\eta)= \frac{(j+1)(j+2)}{(k+1)(k+2)} \frac{\sin(k \pi)}{\pi}
\frac{{\cal N}_{kj}(\eta) }{j-k}\,, \quad \Re{\rm e}\,k>\Re{\rm
e}\,j\,.
\end{eqnarray}
Up to the sign it is equal to the $\sin(k \pi)$ proportional
contribution on the r.h.s.\ in Eq.\ (\ref{Def-Ort-Cen}). Hence,
the sum of integrals (\ref{Def-Ort-Out}), (\ref{Def-Ort-Cen}) and
(\ref{Def-Ort-Out2}) might be understand as a limit $\Re{\rm
e}\,k\to\Re{\rm e}\,j$ that yields the identity, formally written as
\begin{eqnarray}
\int^{\infty}_{-\infty}\!dx\, c_k(x,\eta)
p_j(x,\eta) = -2 i \sin(\pi j) \delta(k-j)\,.
\end{eqnarray}
We add that for non-negative integer values of the conformal spin
the integral (\ref{Def-Ort-Cen}) establishes the orthogonality
relation (\ref{OrtRel}) for Gegenbauer polynomials.

Let us also give here the Mellin moments of the partial waves
$p_j(x,\eta)$ for integer value $n =0,1,2,\cdots $. For the
integration in the outer region we have
\begin{eqnarray}
\label{Res-x-mon-out} \int_{\eta}^\infty dx\, x^n p_j(x,\eta) =
\frac{\eta^{n-j} \sin(\pi j)}{(n-j)\pi }\; {_3F_2}\left({1/2+j/2,
1+j/2, j/2-n/2 \atop 5/2+j, 1+j/2-n/2}\Bigg| 1\right)\,, \quad
\Re{\rm e}\,j>\Re{\rm e}\,n\,,
\end{eqnarray}
while from the central region we find, up to the overall sign, the
same expression
\begin{eqnarray}
\label{Res-x-mon-cen}
\int_{-\eta}^\eta dx\, x^n p_j(x,\eta) =
\frac{\eta^{n-j} \sin(\pi j)}{(j-n)\pi }\;
{_3F_2}\left({1/2+j/2, 1+j/2, j/2-n/2 \atop 5/2+j, 1+j/2-n/2}\Bigg|
1\right)\,.
\end{eqnarray}
Neglecting the $\sin(\pi j)$ term, these expressions will contain
single poles at $j=\{0,2,\cdots, n\}$ and $j=\{1,3,\cdots, n\}$
for even and odd $n$, respectively. The values for the two lowest
moments are:
\begin{eqnarray}
\int_{-\eta}^\eta dx\,  p_j(x,\eta) &\!\!\! =\!\!\! &
\left(  \frac{2}{\eta}\right)^j \frac{2^3\,\Gamma(5/2 + j) }
    {j\; \Gamma(1/2)\Gamma(4 + j)} \frac{\sin (\pi j )}{\pi}\,,
\\
\int_{-\eta}^\eta dx\,  x\, p_j(x,\eta) &\!\!\! =\!\!\!&
\left(  \frac{2}{\eta}\right)^{j-1} \frac{2^4\,\Gamma(5/2 + j) }
    {(j-1)(4+j)\; \Gamma(1/2)\Gamma(3 + j)} \frac{\sin (\pi j )}{\pi}\,.
\end{eqnarray}
Analogous as discussed above, the sum of both integrals
(\ref{Res-x-mon-out}) and (\ref{Res-x-mon-cen}) should be
understood as a limit that results in a linear combination of
``$\delta$-functions'', which are concentrated in
$j=\{0,1,\cdots,n\}$. {F}rom the Mellin-Barnes representation of
GPDs we explicitly find then the usual Mellin moments, expressed
in terms of conformal ones:
\begin{eqnarray}
\label{Exp-Sim2ConMom}
\int_{-1}^{1}\!dx\, x^n q(x,\eta,\Delta^2) = \sum_{i=0}^n
\left(\frac{\eta}{2}\right)^{n-i}
\frac{\left( 1 + {\left( -1 \right) }^{n-i} \right) n!\, \Gamma(5/2 + i)}
  {2\, i!\, \Gamma(1+n/2 - i/2 )
  \Gamma(5/2 + i/2 + n/2)} m_i(\eta,\Delta^2)\,.
\end{eqnarray}
The inverse relation can be brought in the form
\begin{eqnarray}
\label{Exp-2Con2SimMom}
m_n(\eta,\Delta^2)  =
\sum_{i=0}^{[n/2]} \left(\frac{\eta}{2}\right)^{2i} \frac{(-1)^i  n!
\Gamma(3/2 +n- i)}
  { i!\,(n - 2 i )!\, \Gamma(3/2 + n) }
  \int_{-1}^{1}\!dx\, x^{n-2i} q(x,\eta,\Delta^2)\,.
\end{eqnarray}

In the convolution of the hard-scattering amplitude with
generalized parton distributions the following two  integrals
appear:
\begin{eqnarray}
\label{Def-IntHarAmp1} \int_{-\xi}^{\infty} \frac{dx}{\xi+x}
p_j(x,\xi) &\!\!\!=&\!\!\! \left(\frac{2}{\xi}\right)^{1+j}
\frac{\Gamma(5/2+j)}{\Gamma(3/2)\Gamma(3+j)}\,,
\\
\int_{-\xi}^{\infty} \frac{dx}{\xi-x- i\epsilon} p_j(x,\xi)
&\!\!\!=&\!\!\! e^{-i \pi j} \left(\frac{2}{\xi}\right)^{1+j}
\frac{\Gamma(5/2+j)}{\Gamma(3/2)\Gamma(3+j)}\,.
\end{eqnarray}
We note that the reduction to non-negative integer values of the
conformal spin is simply done by the replacement $j\to n$. In this
case only the central region contributes in these integrals,
especially, the imaginary part will drop out

\section{Gluonic sector}
\label{App-GluCas}

Here we present the Mellin-Barnes integral for gluon GPDs, defined
in  Eq.\ (\ref{Def-GPD-g}). Let us first note that the index of
Gegenbauer polynomials, appearing in the definition of conformal
moments, is determined by  group theory. To be more general, we
consider the light-ray operator
\begin{eqnarray}
\label{Def-GenOpe}
{\cal O}(\kappa_1,\kappa_2) =      \phi(\kappa_2 n) \phi(\kappa_1 n)
\end{eqnarray}
that contains two quantum fields, which live on the light-cone
$n^2=0$. For gluons the field is build by the field strength
tensor. We assume that these fields have definite spin projection
$s$ on the light-cone, i.e.,
 \begin{eqnarray}
n^\mu \Sigma_{\mu \nu} \widetilde{n}^\nu \phi(\kappa n) =
s\, \phi(\kappa n)\,.
 \end{eqnarray}
Here $\Sigma_{\mu \nu}$ is the usual generator of Lorentz
transformation, acting on a field $\phi(x)$ at $x=0$. Moreover,
the canonical dimension of the field $\phi(x)$ is denoted as
$\ell$. The conformal spin of the field is defined as
\begin{eqnarray}
j= \frac{1}{2}(\ell+s)\,.
\end{eqnarray}
It characterizes the behavior of the field under collinear
conformal transformation, which can be viewed as the projective
transformations on a line:
\begin{eqnarray}
\kappa \to \kappa^\prime &\!\!\!=\!\!\!&
\frac{a \kappa + b}{c\kappa +d}\,,\quad a d- b c=1\,,
\nonumber\\
\phi(\kappa n)\to \phi^\prime(\kappa n) &\!\!\!=\!\!\!&
(c \kappa + d)^{-2 j} \phi\left(\frac{a \kappa + b}{c\kappa +d}\, n\right)\,.
\end{eqnarray}
In four dimensional space the quantum numbers are
\begin{eqnarray}
\ell &\!\!\!=\!\!\! & \frac{3}{2}\,,\qquad  s =
\pm \frac{1}{2} \qquad\;\; \mbox{for quark fields}\\
\ell &\!\!\!=\!\!\! & 2\,, \qquad\,
s = 0,\pm 1 \qquad \mbox{for gluon field strength tensor}
\end{eqnarray}
The conformal operators are obtained by the group theoretical
decomposition of the light-ray operator (\ref{Def-GenOpe}) into
irreducible representations. If both quantum fields have the same
conformal spin, then the (local) conformal operators which have
definite conformal spin are characterized by Gegenbauer
polynomials with index $\nu=2j-1/2$:
\begin{eqnarray}
\label{Def-GenConOpe}
{\cal O}_{nl}= i^l \left(\partial_{\kappa_1}+\partial_{\kappa_2}\right)^l
C_n^{\nu}
\left(\frac{\partial_{\kappa_1}-\partial_{\kappa_2}}{\partial_{\kappa_1}+
\partial_{\kappa_2}}\right)
{\cal O}(\kappa_1,\kappa_2)\Big|_{\kappa_1=\kappa_2=0}\,,\quad n\leq l\,,
\end{eqnarray}
where the conformal spin of these operators is $2 j+n$. For each
given conformal spin, there appears an infinite tower of operators
which are labelled by the quantum number $l$, related to their
spin or if one likes to their canonical dimension. For leading
twist operators the spin projection $s$ must be maximal so that
the twist of the fields $t=\ell-s$ is minimal. Consequently, to
leading twist the quark fields have conformal spin $j=1$ and gluon
ones $j=3/2$. The index of Gegenbauer polynomials is in the former
and latter case $\nu=3/2$ and $\nu=5/2$. We remark that the gauge
link factor connecting the fields along the light-cone does not
change the construction via Eq.\ (\ref{Def-GenConOpe}).

For gluonic GPDs (\ref{Def-GPD-g}) we define the conformal moments
for $n=1,2,\cdots$
\begin{eqnarray}
\label{Def-cG} {^G\! c}_n(x,\eta) =  \eta^{n-1}\, {^G\!
c}_n\!\left(\!\frac{x}{\eta}\!\right) \quad\mbox{with}\quad {^G\!
c}_n(x) = \frac{\Gamma(5/2)\Gamma(n)}{2^{n-1} \Gamma(3/2+n)}
C_{n-1}^{5/2}\left(x\right)\,.
\end{eqnarray}
in such a way that in the forward limit the usual normalization of
Mellin moments appear: $\lim_{\eta\to 0}{^G\! c}_n(x,\eta) =
x^{n-1}$. Moreover, as in the quark sector, they project on
operators with conformal spin $2+n$. Compared to the Mellin
moments of parton densities within our convention, one power in
$x$ seems to be missed, however, it is included in the definition
of GPDs. For instance, in the vector case we have
\begin{eqnarray}
\label{ForLim-gGPD}
\lim_{\Delta\to 0} {^{G}\!F}^{V}(x,\eta,\Delta^2) =
x g(x)\quad\mbox{for}\quad x\geq0\,,
\end{eqnarray}
where $g(x)$ is the unpolarized gluon parton density. Analogous
convention holds for the axial-vector case. The analytic
continuation of the conformal spin $j$ in the conformal moments is
again done in terms of hypergeometric functions
\begin{eqnarray}
{^G\! c}_j(x,\eta) = \frac{\Gamma(3/2)\Gamma(j+4)}{2^4
\Gamma(3/2+j)} \; \left(\frac{\eta}{2}\right)^{j-1}
{_2\!F}_1\!\left({-j+1,j+4\atop
3}\Big|\frac{\eta-x}{2\eta}\right)\,.
\end{eqnarray}

The conformal partial waves are analogously constructed as
described in Sect.\ \ref{SubSubSec-ConParWavExp}. Namely, by
including the weight $(1-x^2/\eta^2)^2$ with a suitable
normalization and from the requirement that the partial waves are
vanishing at $x=-\eta$ and are continuous at the cross-over point
$x=\eta$:
\begin{eqnarray}
\label{Def-p-all-G}
 {^G\! p}_j(x,\eta)  = \theta(\eta-|x|) \eta^{-j}\,
{^G\!{\cal P}}_j\left(\frac{x}{\eta}\right) +\theta(x-\eta)
\eta^{-j}\, {^G\!{\cal Q}}_j\left(\frac{x}{\eta}\right)\,
\end{eqnarray}
where
\begin{eqnarray}
\label{Def-p-P-G} {^G\!{\cal P}}_j(x)&\!\!\!=\!\!\!&
 \frac{2^{j}\,\Gamma(5/2 + j)}{\Gamma(1/2)\Gamma(j)}
(1+x)^2 \, {_2\!F}_1\!\left({-j-1,j+2\atop
3}\Big|\frac{1+x}{2}\right)\,,
\\
\label{Def-p-Q-G} {^G\!{\cal Q}}_j(x) &\!\!\!=\!\!\!&
\frac{\sin(\pi j)}{\pi}\; x^{-j}\;
{_2\!F}_1\!\left({j/2,(j+1)/2\atop 5/2+j}\Big|\frac{1}{x^2}\right)
\,.
\end{eqnarray}
Moreover, Bose symmetry implies definite symmetry of the  gluonic
GPDs (\ref{Def-GPD-g}) under the transformation $x\to -x$, i.e.,
in the (axial-)vector case they are always (anti-)symmetric. This
property is simply restored by forming symmetric or antisymmetric
partial waves, see  Eq.\ (\ref{Def-pSym}). Corresponding to our
normalization (\ref{ForLim-gGPD}), we write
\begin{eqnarray}
{^G\!q}(x,\eta,\Delta^2) = \frac{i}{2}\int_{c-i \infty}^{c+i
\infty}\!dj\, \frac{(-1)}{\sin(\pi j)} \left[ {^G\! p}_j(x,\eta)
\pm {^G\! p}_j(-x,\eta)\right] m_j(\eta,\Delta^2)\,,
\end{eqnarray}
where only the analytic continuation of conformal moments for even
(odd) integer values of $n$ is needed for (anti-)symmetric gluon
GPDs.

Let us comment on the properties of the gluonic conformal partial
waves (\ref{Def-p-all-G}). They are related to the quark ones by a
derivation with respect to $x$:
\begin{eqnarray}
p_j(x,\eta) = \frac{1}{j} \frac{d}{dx}\! {^G\!p}_j(x,\eta)\,.
\end{eqnarray}
As a simple consequence, the first derivative of the gluonic
partial wave is smooth at the cross-over point, while the second
one has a jump as it is the case in the quark sector. Also at the
point $x=-\eta$ the gluonic partial waves vanish as $(x+\eta)^2$
rather than $(x+\eta)$ as for the quark ones. Consequently, the
convolution with the following hard-scattering amplitudes, which
appear in the electroproduction of transversely polarized vector
mesons \cite{ManPil99}, exist
\begin{eqnarray}
\label{Def-IntHarAmpG2} \int_{-\xi}^{\infty} \frac{dx}{(\xi+x)^2}
{^G\! p}_j(x,\xi) &\!\!\!=&\!\!\! \left(\frac{2}{\xi}\right)^{1+j}
\frac{j\, \Gamma(5/2+j)}{\Gamma(3/2)\Gamma(3+j)}\,,
\\
\int_{-\xi}^{\infty} \frac{dx}{(\xi-x- i\epsilon)^2} {^G\!
p}_j(x,\xi) &\!\!\!=&\!\!\! e^{-i \pi (j-1)}
\left(\frac{2}{\xi}\right)^{1+j} \frac{j\,
\Gamma(5/2+j)}{\Gamma(3/2)\Gamma(3+j)}\,.
\end{eqnarray}
It has been argued already in Ref.\ \cite{ManPil99} that this
integrals without any further regularization exist, which is shown
here from a more general point of view. We add that in  the
convolution of the gluonic GPDs with the hard-scattering amplitude
at leading twist-two the integrals appear
\begin{eqnarray}
\label{Def-IntHarAmpG1} \int_{-\xi}^{\infty} \frac{dx}{\xi+x}
{^G\!p}_j(x,\xi) &\!\!\!=&\!\!\! \left(\frac{2}{\xi}\right)^{j}
\frac{4\, \Gamma(5/2+j)}{\Gamma(3/2)\Gamma(4+j)}\,,
\\
\int_{-\xi}^{\infty} \frac{dx}{\xi-x-i\epsilon} {^G\!p}_j(x,\xi)
&\!\!\!=&\!\!\! e^{-i \pi (j-1)} \left(\frac{2}{\xi}\right)^{j}
\frac{4\, \Gamma(5/2+j)}{\Gamma(3/2)\Gamma(4+j)}\,.
\end{eqnarray}

Finally, the convolution with conformal kernels is analogous done
as outlined in Sect.\ \ref{Sec-CS-Sch} and yields in the
Mellin-Barnes integral representation of GPDs to a multiplication
of the conformal moments with its eigenvalues. Here we only
mention that in correspondence with our normalization these
eigenvalues follows from the Mellin transform of the forward
limit, i.e., from the Mellin moments of the usual DGLAP kernels.

\section{Mellin--Barnes representation of conformal kernels}
\label{App-MelBar-Ker}

In the following  we derive the Mellin--Barnes integral for a
generic kernel (\ref{SpeRepERBLK}), which is conformal covariant
and possess only non-negative eigenvalues $k_n$. This is
analogously done  as for GPDs in Sect.\ \ref{SubSec-ConColSpi}.
Let us first remind on the support extension of the spectral
representation (\ref{SpeRepERBLK}). The procedure for the kernel
$K(x,y)$ is well-known and arise from the representation
\begin{eqnarray}
\label{RepKerDD-K}
K(x,y) =  \int_0^1\! dw_+\! \int_{-1+w_+}^{1-w_+}\! dw_- \;
\delta(x- y w_+ -w_-)\;  \kappa(w_+,w_-)
\,.
\end{eqnarray}
Here $\kappa(w_+,w_-)$ is the analog of the DD in the case of GPD,
see Eqs.\ (\ref{RepKerDDq}) and (\ref{Def-resc-GPD}). Such a
kernel appears in the convolution with a light-ray operator
(\ref{Def-GenOpe}) and it is in the mathematical sense a
distribution. The support in the whole $(x,y)$-plane can be read
off from Eq.\ (\ref{DefCroGPD}) and is written here as
\begin{eqnarray}
\label{Str-Ker-K} K(x,y) = \theta(y-x) \theta(x+1) \left[  k(x,y)
- \theta(x-1) k(-x,-y) \right] +\left\{ {x\to -x \atop y\to -y
}\right\} \,,
\end{eqnarray}
where the distribution $k(x,y)$ has the integral representation
\begin{eqnarray}
k(x,y)=  \int_0^\frac{1+x}{1+y}\! dw_+\,  \kappa(w_+,x- y w_+)\,.
\end{eqnarray}
Analogous as in the case of a GPD, see discussion in Sec.\
\ref{SubSec-AnaGPDs}, this integral is only uniquely defined in
the central region, i.e., $|x|\leq 1$, while in the outer region
$1 \leq |x|$ only the difference $k(x,y)-k(-x,-y)$ enters. Here
the ambiguities in the support extension of $\kappa(w_+,x- y w_+)$
drop out. The support extension from the region $|x|,|y|\leq 1$ to
the whole support is unique \cite{MueRobGeyDitHor94}, see
analogous discussion as in the last paragraph of Sect.\
\ref{SubSec-Cro}.

Now we are in the position to derive the
Mellin-Barnes integral representation for the
kernel $K(x,y)$. To do so, we represent the series
(\ref{SpeRepERBLK}) in the region $|x|,|y| < 1$ as
integral in the complex plane that includes the
positive real axis
\begin{eqnarray}
K(x,y)= \frac{1}{2 i}\oint_{(0)}^{(\infty)}\!dj\,
\frac{1}{\sin(\pi j)}  p_j(x) k_j  c_j(y) \,.
\end{eqnarray}
Here the functions  $c_j(x) = c_j(x,1)$ and
$p_j(x)=p_j(x,1)$  are defined in Eqs.\
(\ref{Def-cHyp}) and (\ref{Def-pn-SchInt}),
respectively, and $k_j$ is the analytic continuation
of the eigenvalues $k_n$. They coincide with the
Mellin moments of the corresponding DGLAP kernel and
might possesses a logarithmical growing for $j\to
\infty$. The integrand has simple poles at
$j=n=0,1,2,\cdots$ and so the residue theorem leads
to the series (\ref{SpeRepERBLK}). Next, we deform
the integration contour in such a way that it
includes the imaginary axis and is closed by an arc
with infinite radius, see Fig.\ \ref{FigCon1} (a).
It remains to show that this latter contribution
vanishes. The behavior of the integrand for large
$j$ with $|{\rm arg}(j)| \leq \pi/2 $ can be
estimated from the behavior of hypergeometric
functions  and is in our case given by \cite{Luk69}
\begin{eqnarray}
\label{Est-Cir} \frac{1}{\sin(\pi j)}p_j(x+i \epsilon) c_j(y-i
\epsilon) \sim \frac{e^{\pm j \left\{{\rm arccosh}(-x-i \epsilon)
\pm {\rm arccosh} (y-i \epsilon)\right\}} }{\sin(\pi j)} \,.
\end{eqnarray}
Here the analytic continuation in the complex plane by the
$i\epsilon$ prescription determines the branch of the ${\rm
arccosh}$ function, not uniquely defined for real valued $x,y <
1$. It is done in such a way that the estimate (\ref{Est-Cir}) is
applicable   for  $ -1 < y$ and $x < 1$ and $ |{\rm arg} (j)| \leq
\pi/2 $. Obviously, on the infinite arc, both the denominator and
numerator in Eq.\ (\ref{Est-Cir}) will exponentially grow and the
integrand will vanish as long as the condition $|{\rm
arccosh}(-x-i \epsilon)\pm {\rm arccosh}(y-i  \epsilon)|\leq \pi$
is satisfied. This is the case for $ -1< x < y < 1$. Thus, for
this region we arrive at the Mellin--Barnes representation for the
Kernel $K(x,y)$, i.e., for the   function
\begin{eqnarray}
\label{Rep-k-MelBar} k(x,y) = \frac{i}{2}\int_{c-i \infty}^{c+i
\infty}\!dj\, \frac{1}{\sin(\pi j)} p_j(x) k_j  c_j(y)\,.
\end{eqnarray}
Form the spectral representation (\ref{RepKerDD-K}) it follows
that the convolution of the kernel with a holomorphic test
function $\tau(x)$ yield a holomorphic function, depending on $y$.
Thus, we might now employ analytic continuation to extend the
representation (\ref{Rep-k-MelBar}) into the region $1\leq  y$.
This will not alter the convergency properties of the integral,
since ${\rm \arg}({\rm arccosh}(y)) = 0$ for $1\leq y$. We remind
that $p_j(x)$ coincides for $x \leq 1$ with the holomorphic
function ${\cal P}_j(x)$, defined in Eq.\ (\ref{Def-p-P}).
${\cal P}_j(x)$ has a branch cut, starting at $x=1$,
along the positive real axis and we might define its value on the
this cut by $({\cal P}_j(x+i \epsilon  ) + {\cal P}_j(x- i
\epsilon ))/2$. Hence, within this procedure the function $k(x,y)$
is uniquely continued into the region $1\leq y$ for all values $-1
\leq x$, cf.\ with the prescription \ (\ref{Pre-Con}).

The Mellin-Barnes integral for the kernel $K(x,y)$ follows now in
an unique way from the result (\ref{Rep-k-MelBar}) and its support, cf.\
Eq.\ (\ref{Str-Ker-K}). For $1\leq x$ we can write $K(x,y)$ as the
difference
\begin{eqnarray}
K(x,y) = k(x,y) -k(-x,-y)\quad\mbox{for}\quad  1 \leq x\leq y\,.
\end{eqnarray}
We represent $k(-x,-y)$ via the Mellin--Barnes integral
(\ref{Rep-k-MelBar}), where  $p_j(-x)$ is  obtained by analytic
continuation while $c_j(-y)$ has now for $1\leq y$ a branch cut
and a  single pole at $y=1$ on the real axis. We have now to
employ  the principal value prescription,  as above for $ {\cal
P}_j(x)$, and so its value on the cut is
\begin{eqnarray}
\frac{1}{2}\left[c_j(-y+ i \epsilon) + c_j(-y- i \epsilon)\right]
= \cos(\pi j) c_j(y) + \sin(\pi j) d_j(y)\,,
\end{eqnarray}
where the function $d_j(y)$ is defined in Eq.\ (\ref{Def-d-com}).
Using the asymptotic behavior of hypergeometric functions for
large $j$, it can be shown that in  the region $ 1 \leq x < y$ the
contribution $ {\cal P}_j(-x) k_j d_j(y) $, appearing in the
integral (\ref{Rep-k-MelBar}), exponentially vanishes for $j\to
\infty$ with $|{\rm arg}(j)| \leq \pi/2$. Hence,  we can close the
integration path so that the positive real axis is encircled.
Since it is a holomorphic function, the corresponding integral
vanishes. So we can drop this contribution and find that for
$ 1\leq x < y$ the extension of the integral kernel,
\begin{eqnarray}
\label{Def-Ker-MelBar}
 \frac{k_j}{\sin(\pi j)}  p_j(x)  c_{j}(y)=
\frac{k_j}{\sin(\pi j)}\left[\frac{1}{2} {\cal P}_j(x+ i \epsilon)
+ \frac{1}{2} {\cal P}_j(x- i  \epsilon)  - \cos(\pi j) {\cal
P}_j(-x) \right] c_j(y)\,,
\end{eqnarray}
of the Mellin--Barnes integral for $K(x,y)$
\begin{eqnarray}
\label{Res-Ker-MelBar}
K(x,y) =  \frac{i}{2}\int_{c-i \infty}^{c+i \infty}\!dj\,
\frac{k_j}{\sin(\pi j)}  p_j(x)  c_{j}(y)\quad\mbox{for}\quad x < y\,.
\end{eqnarray}
The formula (\ref{Def-Ker-MelBar}) defines the continuation of
$p_j(x)$ for $1\leq x$, which coincides with Eqs.\
(\ref{Def-p-all})-(\ref{Def-p-Q}) for $\eta=1$. The missing part
of the kernel, i.e., for $y<x$, follows by the symmetry
transformation $x,y\to -x,-y$ from Eq.\ (\ref{Res-Ker-MelBar}).
Hence, the integral kernel can be written as
\begin{eqnarray}
K(x,y) =   \frac{i}{2 }\int_{c-i \infty}^{c+i \infty}\!dj\,
\frac{k_j}{\sin(\pi j)} \left[ \theta(y-x) p_j(x)  c_j(y) +
\theta(x-y) p_j(-x)   c_j(-y)  \right].
\end{eqnarray}
We excluded in our analyze here the line $x=y$. As long $j k_j$ is
vanishing for $j\to\infty$, we can use analytic continuation to
approach it.  In all other cases a more advanced analyze is
required. However, this is in fact not necessary here, since it is
obvious that a constant or logarithmic behavior of $k_j$ for
$j\to\infty$ is associated with $\delta$-functions and
$+$-prescriptions for singularities at $x=y$. It is easy to check
by forming the lowest moment that the Mellin-Barnes integral is
correct in that case, too.


\end{document}